\documentclass[fleqn,usenatbib]{mnras}
\usepackage{amsmath,amstext}
\usepackage{mathptmx}
\usepackage[T1]{fontenc}
\usepackage{ae,aecompl}

\usepackage{amsmath}	
\usepackage{amssymb}	
\usepackage{natbib}
\usepackage[colorinlistoftodos]{todonotes}
\usepackage{color}
\usepackage{rotating}
\usepackage{txfonts}

\title[On the Nature of HorI, PicI, GruI and PheII]{On the Nature of Ultra-faint Dwarf Galaxy Candidates. III.\\Horologium I, Pictor I, Grus I, and Phoenix II}

\author[H. Jerjen et al.]{Helmut Jerjen$^{1}$\thanks{e-mail: helmut.jerjen@anu.edu.au},
Blair C. Conn$^{1}$,
Dongwon Kim$^{2}$
and Mischa Schirmer$^{3,4}$
\\
$^{1}$Research School of Astronomy and Astrophysics, Australian National University, Canberra, ACT 2611, Australia\\
$^{2}$Dept. of Astronomy, University of California, Berkeley, CA 94720\\
$^{3}$Max-Planck-Institut f\"ur Astronomie, K\"onigstuhl 17, D-69117 Heidelberg, Germany\\
$^{4}$Gemini Observatory, Casilla 603, La Serena, Chile\\
}

\pubyear{2018}

\begin{document}
\label{firstpage}
\pagerange{\pageref{firstpage}--\pageref{lastpage}}
\maketitle

\begin{abstract}
We use deep Gemini/GMOS-S $g,r$ photometry to study the stellar populations of the recently discovered Milky Way satellite candidates Horologium I, Pictor I, Grus I,  and Phoenix II. Horologium I is most likely an ultra-faint dwarf galaxy at  $D_\odot = 68\pm3$ kpc, with $r_h = 23^{+4}_{-3}$pc and $\langle $[Fe/H]$ \rangle = -2.40^{+0.10}_{-0.35}$\,dex. It's color-magnitude diagram shows evidence of a split sub-giant branch similar to that seen in some globular clusters. Additionally, Gaia DR2 data suggests it is, or was, a member of the Magellanic Cloud group. Pictor I with its compact size ($r_h = 12.9^{+0.3}_{-0.2}$pc) and metal-poor stellar population ($\langle $[Fe/H]$ \rangle = -2.28^{+0.30}_{-0.25}$) closely resembles confirmed star clusters. Grus I lacks a well-defined centre, but has two stellar concentrations within the reported half-light radius ($r_h = 1.77^{+0.85}_{-0.39}$ arcmin) and has a mean metallicity of $\langle $[Fe/H]$ \rangle = -2.5\pm0.3$. Phoenix II has a half-light radius of $r_h = 12.6\pm2.5$pc and an $\langle $[Fe/H]$ \rangle = -2.10^{+0.25}_{-0.20}$ and exhibits S-shaped tidal arms extending from its compact core. Great circles through each of these substructures intersect at the Large Magellanic Cloud (LMC). This suggests that these objects are, or once were, satellites of the LMC. 
\end{abstract}




\section{Introduction}
In recent years, many new Milky Way satellite stellar systems have been reported in the literature, e.g. \citet{Balbinot2013, Belokurov2014,Laevens2014, Bechtol2015, Drlica-Wagner2015, Kim2, Kim2015b, KimJerjen2015a, KimJerjen2015b, Koposov2015, Laevens2015a, Laevens2015b, Martin2015, Kim2016, Luque2016, Martin2016b, Torrealba2016a, Torrealba2016b, Koposov2017}. The majority of these findings are based on relatively shallow photometry from the SDSS\footnote{Sloan Digital Sky Survey}, Pan-STARRS1\footnote{Panoramic Survey Telescope and Rapid Response System, \citet{2016arXiv161205560C}} and DES\footnote{Dark Energy Survey, http://des.ncsa.illinois.edu/releases/sva1D} imaging surveys, and thus we have just started to understand these objects in terms of their stellar population, structural properties, distance and luminosity. In order to make them also valuable for testing near-field cosmology predictions and verifying Milky Way formation scenarios, including the role of the Magellanic Clouds, the true nature of each ultra-faint stellar system has to be unambiguously determined. 

In this paper, the third in our series, we establish deep Gemini/GMOS-S $g,r$ stellar photometry to better constrain the properties of four recently discovered southern dwarf galaxy candidates: Horologium I (Hor I; DES J0222.7-5217), Pictor I (Pic I; DES J0443.8-5017), Grus I (Gru I) and Phoenix II (Phe II; DES J2339.9-5424). We deliberately use both designations here for each object to reflect the fact that it still remains unclear whether they are in fact dark matter dominated dwarf galaxies, baryonic star clusters, or something else. They were detected in the first-year Dark Energy Survey (DES) data, although the three with alternate designations are from \citet{Bechtol2015} and \citet{Koposov2015}, while Gru I is only reported in the latter. The preliminary fundamental parameters derived for these objects are listed in separate columns in Table \ref{table:oridata}. The third column for Phe II contains the values determined by \citet{MutluPakdil2018}.

\begin{table*}
\centering
\begin{tabular}{llccclclccc}
\hline
 & \multicolumn{2}{c}{Horologium I} & \multicolumn{2}{c}{Pictor I} && Grus I && \multicolumn{3}{c}{Phoenix II}\\
 & \multicolumn{2}{c}{(DES J0255.4-5406)} & \multicolumn{2}{c}{(DES J0443.8-5017)} & &&  & \multicolumn{3}{c}{(DES J2339.9-5424)}\\
 \hline
R.A.(J2000) (deg) & 43.8820 & 43.87 & 70.9475 & 70.95 && 344.1765 && 354.9975& 354.99 & 354.9929\\
DEC (J2000) (deg)& $-54.1188$ & $-54.11$ & $-50.2830$ & $-50.28$ && $-50.1633$ && $-54.4060$& $-54.41$ & $-54.4049$\\
significance ($\sigma$)& 11.6 & 8.2 & 17.3 & 7.1 && 10.1 && 13.9& 5.1 &n/a\\
D$_\odot$ (kpc) & 79 & 87 &  114 & 126 && 120 && 83& 95 & $84.3\pm4.0$\\
$r_h$ (arcmin)& $1.31^{+0.19}_{-0.14}$& $0.24^{+0.30}_{-0.12}$ & $0.88^{+0.27}_{-0.13}$ & $1.2^{+3.5}_{-0.60}$ && $1.77^{+0.85}_{-0.39}$ && $1.09^{+0.26}_{-0.16}$& $1.2^{+0.60}_{-0.60}$ & $1.5\pm0.3$\\
$r_h$ (pc)& $30^{+4.4}_{-3.3}$& $60^{+76}_{-30}$& $29^{+9.1}_{-3.3}$& $43^{+153}_{-21}$& &$62^{+29.8}_{-13.6}$&& $26^{+6.2}_{-3.9}$& $33^{+20}_{-11}$ & $37\pm6$\\
ellipticity ($1-\frac{b}{a}$)& $<0.28$ & ...&$0.47^{+0.12}_{-0.29}$& ...&& $0.41^{+0.20}_{-0.28}$&& $0.47^{+0.08}_{-0.29}$& ... & $0.4\pm0.1$\\
P.A.(deg) & ...& ...& $78\pm23$& ...&& $4\pm60$ && $164\pm54$& ... & $156\pm13$\\
$M_V $ (mag) & $-3.4\pm0.1$& $-3.5\pm0.3$ & $-3.1\pm0.3$ & $-3.7\pm0.4$ &&$-3.4\pm0.3$ &&$-2.8\pm0.2$& $-3.7\pm0.4$ & $-2.7\pm0.4$\\
\hline
\end{tabular}
\caption{Parameters for the four ultra-faint dwarf galaxy candidates available in the literature prior our study. First column from \citet{Koposov2015}, second column from \citet{Bechtol2015}, third column from \citet{MutluPakdil2018}. 
\label{table:oridata}}
\end{table*}

\begin{figure*}
\begin{center}
\includegraphics[width=1.0\hsize]{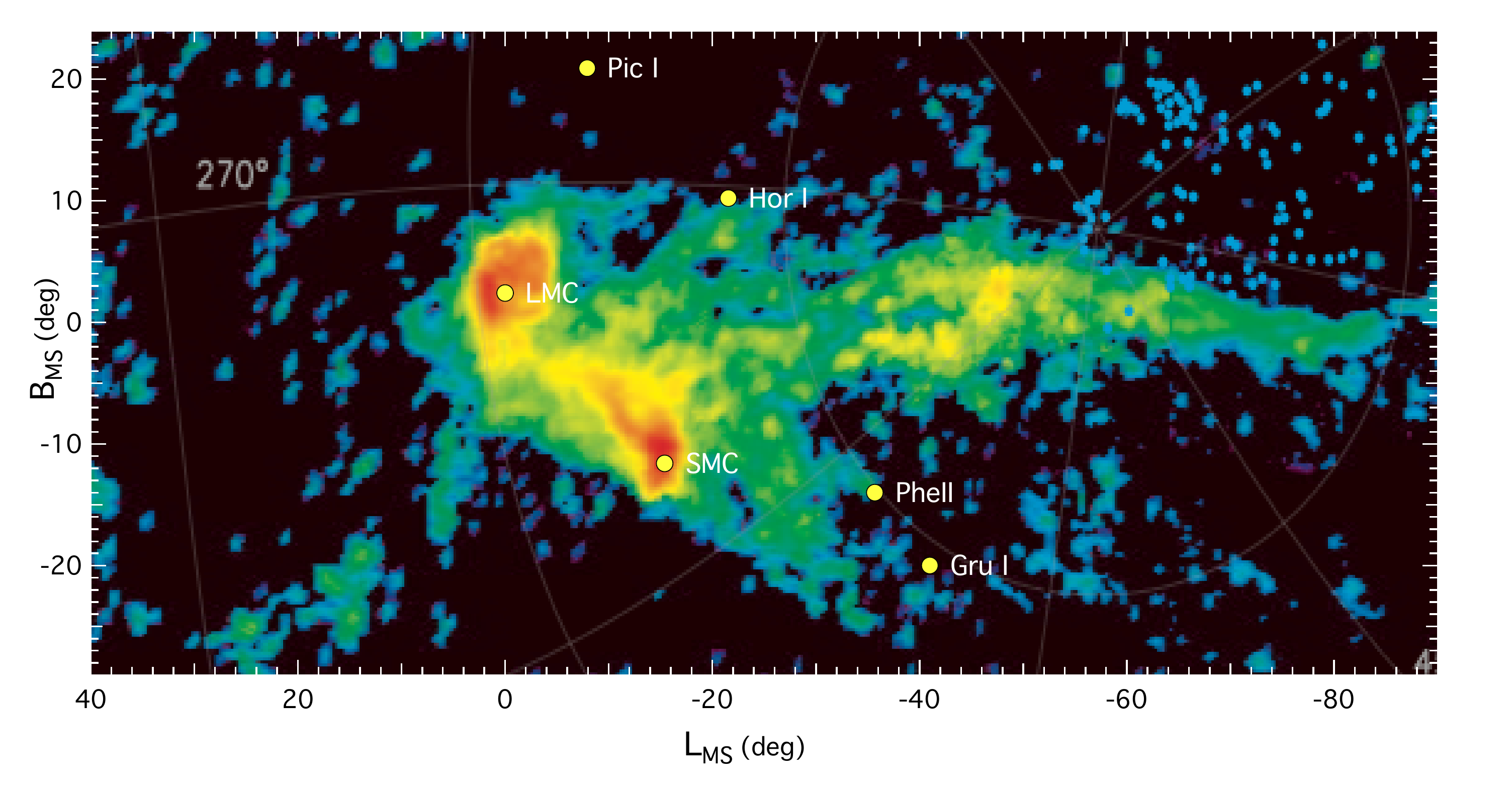}\vspace{-0.5cm}
\caption{On sky distribution of the four Milky Way satellite candidates with respect to the Magellanic Clouds and the neutral hydrogen gas of the Magellanic Stream. The HI column density ($\log(N_{HI})$ in units of cm$^{-2}$) is shown over six orders of magnitudes, ranging from $\log(N_{HI})=16$ (black) to 22 (red). For more details we refer to \citet{2010ApJ...723.1618N}.}
\label{fig:MCs}
\end{center}
\end{figure*}

Like many other ultra-faint dwarf galaxy candidates in the southern hemisphere, these four stellar overdensities are relatively close in projection to the Magellanic Clouds. This further raises the question of whether some of them might be, or were part of the Magellanic Clouds' own satellite system. Figure~\ref{fig:MCs} shows their locations with respect to the Large and Small Magellanic Clouds and the gaseous Magellanic Stream. \citet{Koposov2015} finds that Hor I, Pic I and Phe II have similar half-light radii ($r_{h,Hor I}= 30^{+4.4}_{-3.3}$\,pc, $r_{h,Pic I}= 29^{+9.1}_{-3.3}$\,pc, $r_{h,Phe II}= 26^{+6.2}_{-3.9}$\,pc), with Hor I and Phe II residing at approximately the same heliocentric distance (79 and 83 kpc, respectively). Gru I is roughly double the physical size of the other objects ($r_{h, Gru I}= 62^{+29.8}_{-13.6}$\,pc) and along with Pic I they are both about twice as distant as the Magellanic Clouds ($D_\odot=120$ and 114 kpc, respectively). 

With the recent release of the Gaia DR2 catalogue, some of the ultra-faint dwarf galaxy candidates now also have proper motions associated with them. For our sample, Hor I and Gru I both have five member stars each in the catalogue published by \citet{Fritz2018} and Phe II has six stars \citep{Fritz2018b}. The authors concluded that Hor I and Phe II are co-orbiting relative to the majority of classical Milky Way satellites, while Gru I is counter-orbiting with respect to that group. There are no data available on Pic I yet. 

As can be seen in \citet{Bechtol2015} and \citet{Koposov2015}, the discovery data drawn from DES has a typical  limiting magnitude of $g \sim 23$ reaching, at best, the subgiant branch of the stellar population. Given the low total luminosity of these ultra-faint stellar systems they statistically have very few stars brighter than this limit. The relatively shallow depth of the photometry then impacts on the accuracy to which the primary properties of these objects can be measured. Here we present the deeper photometry necessary to trace the low-mass stellar members several magnitudes below the main sequence turn-off (MSTO). Achieving high photometric accuracy with these stars significantly improves our ability to constrain the size, shape and chemistry of the ultra-faint systems, and search for signs of tidal disturbance. As with \citet[][hereafter Paper I, II]{Conn2018a,Conn2018b}, this paper seeks to provide the best possible constraints on these ultra-faint systems and to shed more light on their true nature.

In Section~\ref{sec:obs}, we introduce the data, discuss the reduction procedures, photometric calibration and artificial star experiments. In Section~\ref{sec:param_analysis}, we describe the process by which we determine the parameters for each object and in Sections~\ref{sec:HorIprop}, \ref{sec:PicIprop}, \ref{sec:GruIprop} and \ref{sec:PheIIProp} we present the results. Section~\ref{sec:discussion} is the discussion and Section~\ref{sec:conclusion}, the summary. Appendix~\ref{App:AppendixA} contains additional information related to the artificial star experiment.

\section{Observations and Data Reduction}\label{sec:obs}
\begin{figure*}
\begin{center}
\includegraphics[width=0.35\hsize]{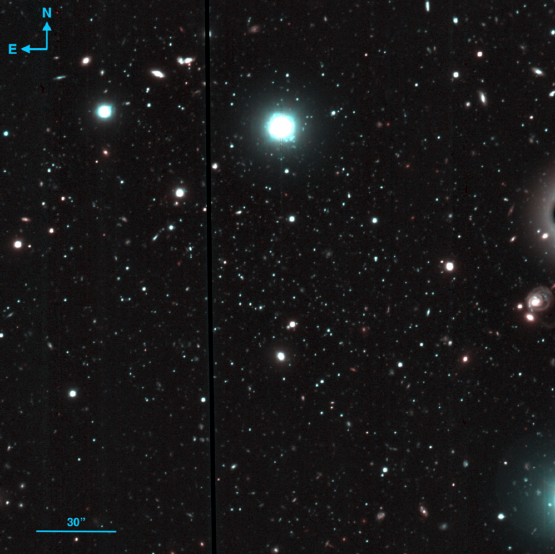} 
\includegraphics[width=0.35\hsize]{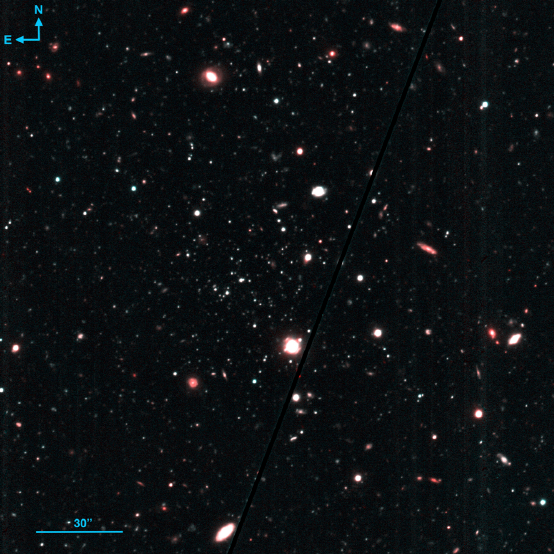}\\ 
\includegraphics[width=0.35\hsize]{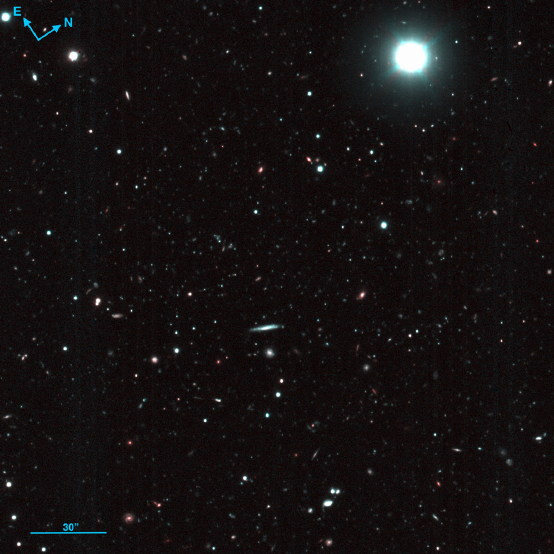}  
\includegraphics[width=0.35\hsize]{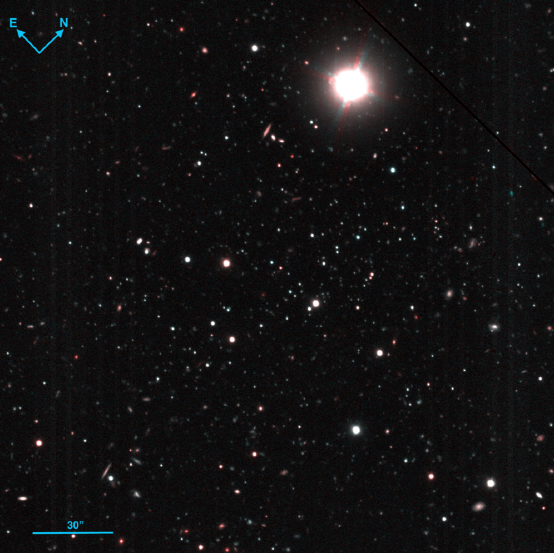}
\caption{False colour RGB images of the ultra-faint dwarf galaxy candidates Hor I, Pic I, Gru I, and Phe II (left to right, top to bottom) produced from deep GMOS-S images. The contrast has been stretched in an attempt to highlight the stellar populations of each system. The $g$-band co-added image was used for the blue and the $r$-band co-added image for the red channel. Each field is approximately 3$\farcm$5 $\times$ 3$\farcm$5, the ruler in the lower left corner has a length of 30 arcsec and the orientation of the CCDs can be found in the top left corner. Image Credit: Jennifer Miller, Gemini Observatory}
\label{fig:images}
\end{center}
\end{figure*}

\begin{table*}
\centering
\begin{tabular}{lrccccccc}
\hline
Field & R.A. & DEC & Position Angle & Filter & Observation & Airmass & Exposure & Seeing\\
 &  (deg, J2000) & (deg, J2000) & (deg) & & Date& & (s) & (\arcsec)\\
 \hline\hline
Horologium I  & 43.8820   &       $-$54.1188    &    0   &   $g\_G0325$   &      2016 Oct 6  &    1.093 - 1.095 &  3$\times$600  &   0.57     \\
(DES J0255.4-5406)      &     &          &    0   &  $r\_G0326$   &      2016 Oct 6  &    1.094 - 1.111 &  3$\times$600   &   0.48    \\\hline
Pictor I    &  70.9475  &       $-$50.2830     &    180 &   $g\_G0325$   &      2016 Oct 6  &    1.068 - 1.077 &  3$\times$600  &    0.58      \\
(DES J0443.8-5017)      &   &           &    180 &  $r\_G0326$   &      2016 Oct 6  &    1.064 - 1.076 &  3$\times$600  &  0.51   \\\hline
Grus I        & 344.1765 &       $-$50.1633    &    60  &  $g\_G0325$   &      2016 Sep 27  &    1.094 - 1.128 &  3$\times$600   &  0.46   \\ 
      &  &           &    60  &    $r\_G0326$   &      2016 Sep 27  &    1.091 - 1.183 &  3$\times$600   &   0.38      \\\hline
Phoenix II    &  354.9975 &       $-$54.4060     &    225 &   $g\_G0325$   &      2016 Aug 31  &    1.213 - 1.276 &  3$\times$600   &   0.52    \\
(DES J2339.9-5424)      &  &       	    &    225 &    $r\_G0326$   &      2016 Aug 31  &    1.208 - 1.372 &  3$\times$600   &   0.48   \\
\hline
\end{tabular}
\caption{Observing Log of GS-2016B-Q-7.}
\label{table:data}
\end{table*}

The imaging data were obtained using the Gemini Multi-Object Spectrograph South (GMOS-S) mounted on the 8m Gemini South Telescope through Program ID: GS-2016B-Q-7. In accordance with the Gemini Observatory standards, the observing conditions under which they were taken were dark, clear skies (SB50\footnote{SB50 - Sky Brightness 50$^{th}$ percentile}/CC50\footnote{CC50 - Cloud Cover 50$^{th}$ percentile}) combined with atmospheric seeing that was typically better than 0.6 arcsecond at zenith (IQ20\footnote{IQ20 - Image Quality 20$^{th}$ percentile}). The observations took place on the nights of 2016 August 31, September 27 and October 6 (see Table~\ref{table:data}). IQ20 conditions allow us to utilise the $1\times1$ binning mode of GMOS-S and to thereby take advantage of its finest pixel scale ($0\farcs08$ per pixel). The field of view for GMOS-S is $5\farcm5\times 5\farcm5$ and each object was observed using the $g$-band ($g\_G0325$) and $r$-band ($r\_G0326$) consisting of a short 60s exposure centred on the target followed by three dithered exposures of 600s each. The panels in Figure~\ref{fig:images} present the false-colour images of the co-added frames for HorI, PicI, GruI and PheII, respectively.

The data reduction proceeded in two phases with the generating master biases and master twilight flats, bias subtraction and flat fielding, astrometry and co-addition performed using the {\sc theli} pipeline \citep{Erben2005,2013ApJS..209...21S}. While the point spread function (PSF) photometry was undertaken on the co-added frames using {\sc dolphot} \citep{2000PASP..112.1383D}.

\subsection{Photometric Calibration}\label{sec:photcalib}

\begin{table}
\centering
\begin{tabular}{lcc}
\hline
 & $g$ band & $r$ band \\ \hline\hline
Colour term $(g -r)$ & $+0.026^{+0.045}_{-0.046}$& $-0.059^{+0.042}_{-0.041}$\\
Horologium\,I offsets& $-3.132^{+0.053}_{-0.054}$ & $-2.850^{+0.049}_{-0.051}$\\
Pictor\,I offsets& $-3.159^{+0.057}_{-0.059}$ & $-2.846^{+0.052}_{-0.053}$ \\
Grus\,I offsets& $-3.316^{+0.047}_{-0.047}$& $-3.028^{+0.042}_{-0.043}$  \\
Phoenix\,II offsets& $-3.124^{+0.053}_{-0.055}$& $-2.856^{+0.050}_{-0.050}$ \\\hline
\end{tabular}
\caption{Colour terms and photometric offsets derived from comparison with APASS calibrated DES photometry. All instrumental photometry assumed a zero point of 30.00 for both filters prior to the offsets being applied.The offset values listed are a combination of the true zero point correction and the atmospheric extinction correction.}
\label{table:calib}
\end{table}

To calibrate the {\sc dolphot} photometry we cross-matched the instrumental magnitudes with APASS\footnote{The AAVSO Photometric All-Sky Survey}~\citep{2015AAS...22533616H} calibrated DECam photometry\footnote{DECam photometry generated using the procedures outlined in \citet{KimJerjen2015b}.}. After removing hot pixels, extended objects and those with large photometric errors, the resulting subset was fit with a linear function to determine the colour term, zero point offset and atmospheric extinction correction for calibration. As per Papers I and II of this series, the colour term has been fixed for all fields along with a nominal zero point of 30.0 magnitudes leaving only a offset needed for each filter to correct the photometry. Table~\ref{table:calib} lists the colour terms and offsets, along with their corresponding errors, used to calibrate the data.

\subsection{Artificial Star Experiments}\label{sec:completeness}
To determine the photometric completeness of our data, we performed an artificial star experiment on each field. {\sc dolphot} injects each artificial star individually into a randomised pixel position and then attempts to recover that star using the same parameters as those used in generating the science catalog. In total we used around 500,000 artificial stars in each field which yields approximately 5000 stars per 0.1 magnitude bin. The histogram generated from the ratio of recovered stars to input stars is then fit with a logistic function to compute the 50\% completeness level. The logistic function has the following form:
\begin{eqnarray}\label{eqn:logistic}
Completeness = \frac{1}{1 + e^{(m - mc)/\lambda}}\\
Error = \sqrt{N C(1 - C)}/N
\end{eqnarray}

\begin{figure*}
\begin{center}
\includegraphics[width=0.4\hsize]{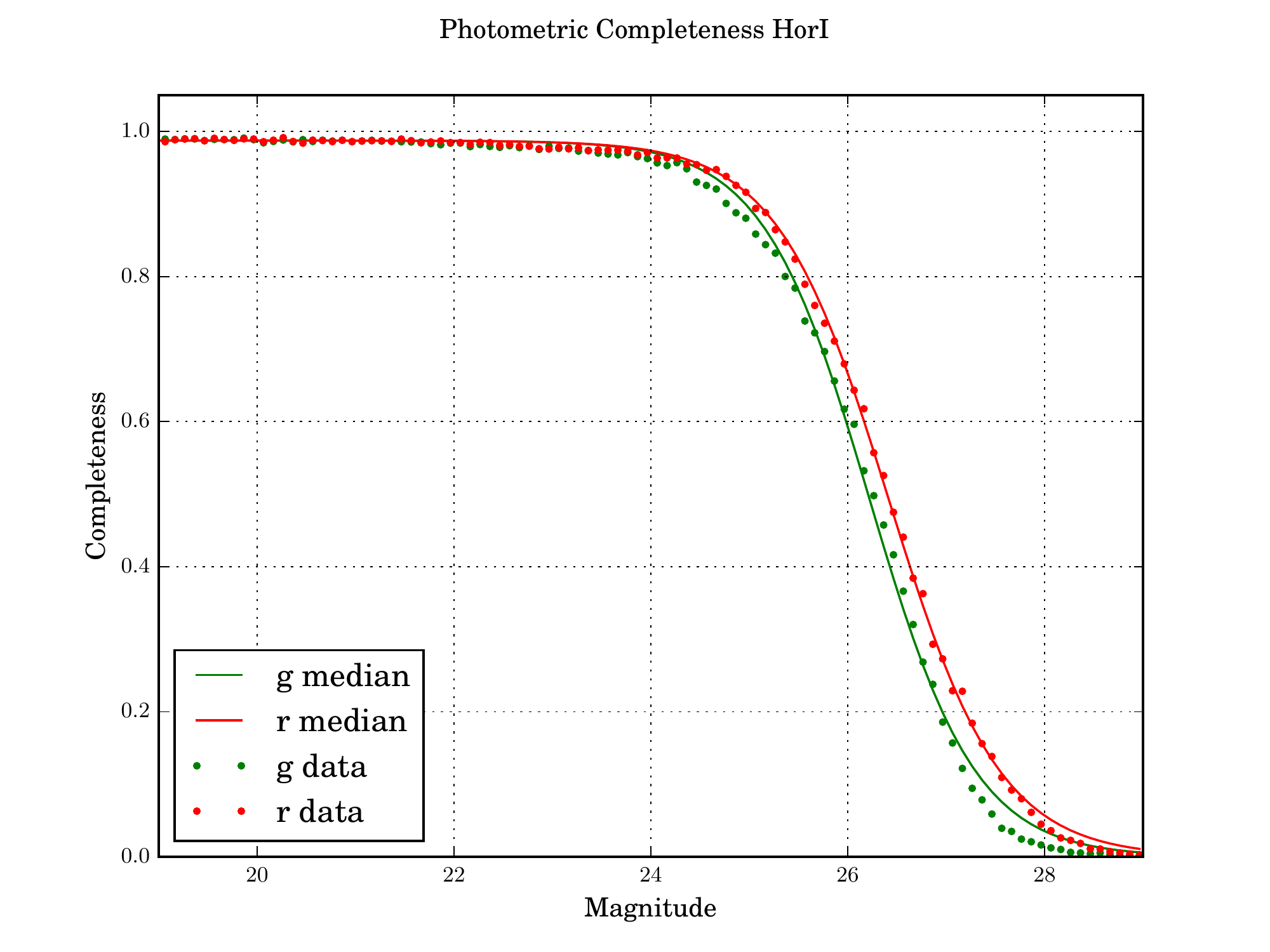}\hspace{-0.2cm}
\includegraphics[width=0.4\hsize]{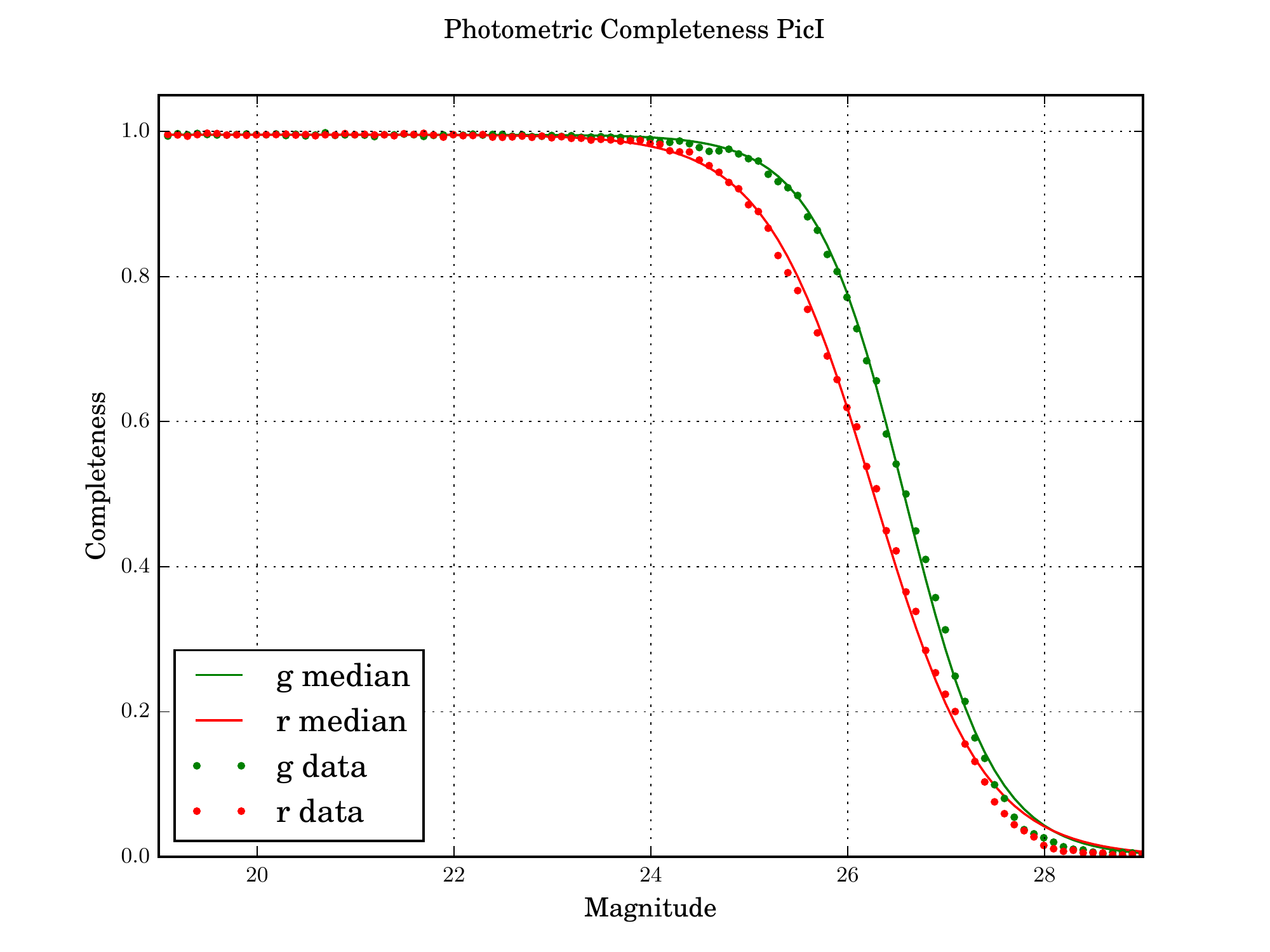}\\
\includegraphics[width=0.4\hsize]{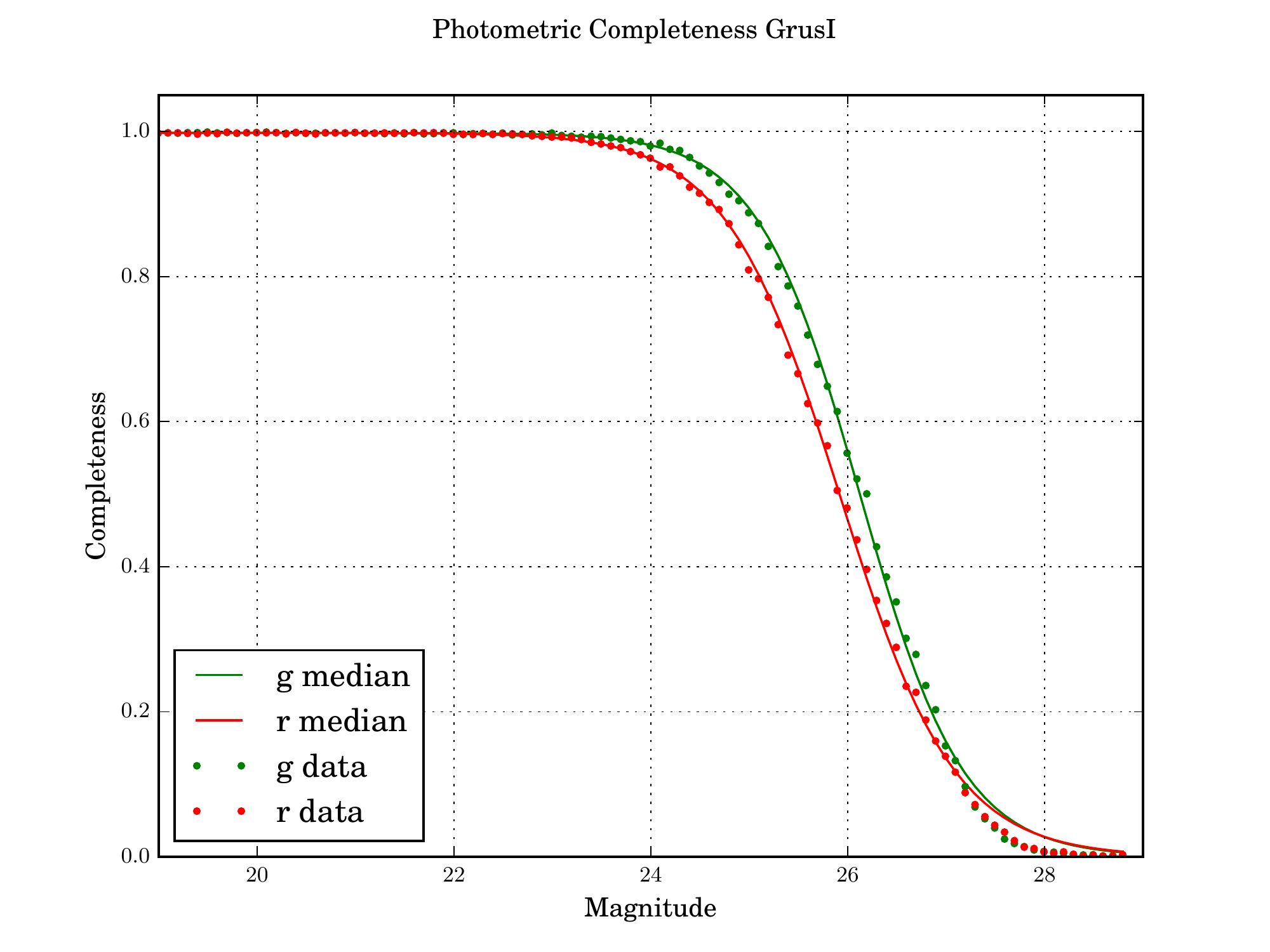}\hspace{-0.2cm}
\includegraphics[width=0.4\hsize]{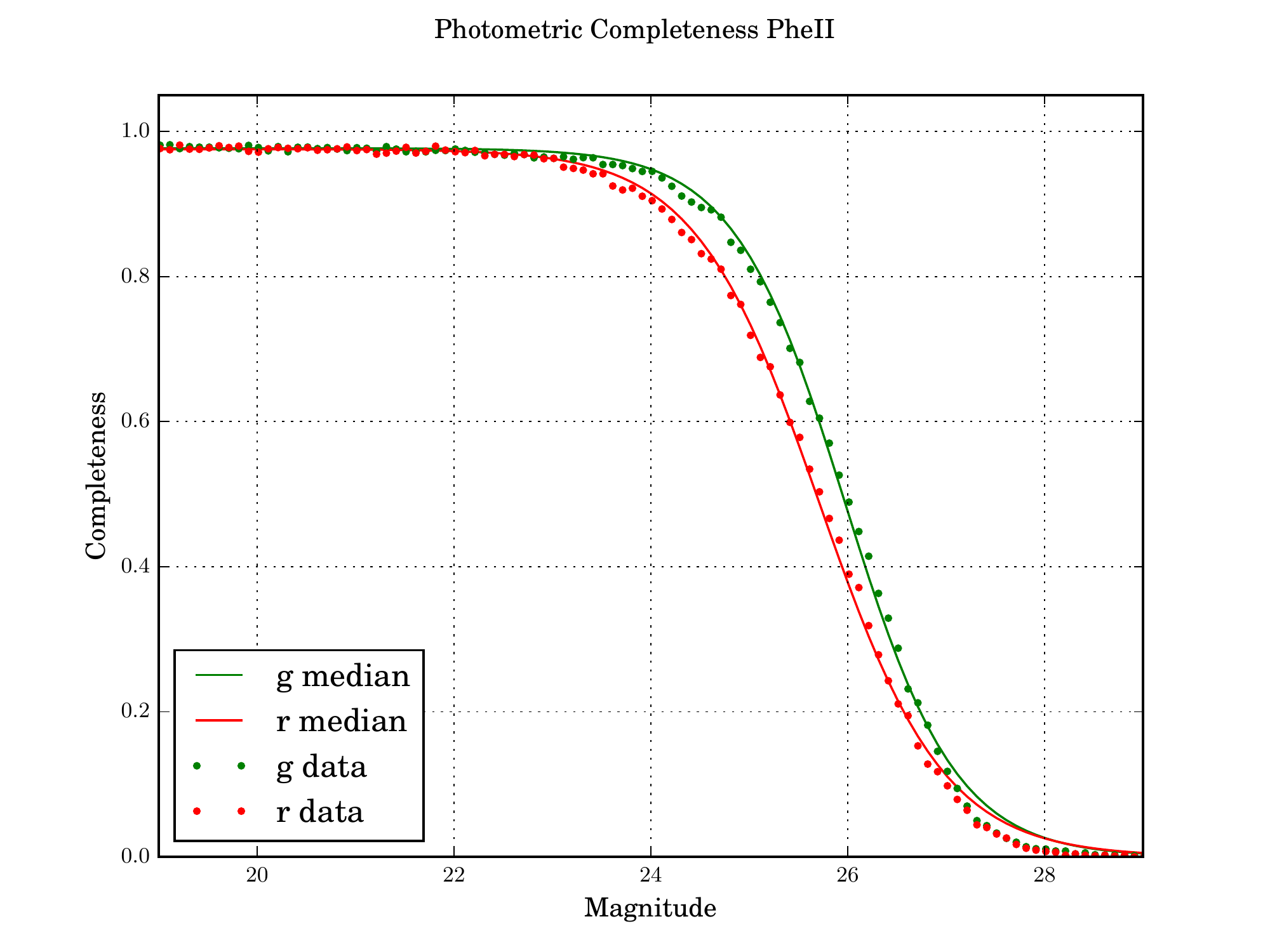}
\caption{Photometric completeness plots for the Hor I, Pic I, Gru I, and Phe II fields (left to right, top to bottom): fitting the Logistic function (Equation 1) to the results of the artificial star tests outlined in Section~\ref{sec:completeness}. The statistical error of each data point (Equation\,2) is roughly the same size as the points and have not been plotted.}
\label{fig:CompletenessPlots}
\end{center}
\end{figure*}

\begin{table}
\caption{50\% Photometric Completeness Estimates.}
\centering
\begin{tabular}{lcccc}
\hline
Field & $mc_g$ & $\lambda_g$&$mc_r$& $\lambda_r$ \\\hline\hline
Hor I &$26.22\pm{0.01}$ & 0.425$\pm{0.005}$ & $26.49\pm{0.01}$ & $0.465\pm{0.005}$ \\
Pic I &$26.58\pm{0.01}$ & $0.459\pm{0.005}$ & $26.27\pm{0.01}$ & $0.553\pm{0.006}$ \\
Grus I &$26.13\pm{0.01}$ & $0.524\pm{0.005}$ & $25.92\pm{0.01}$ & $0.584\pm{0.005}$ \\
Phe II &$25.97\pm{0.01}$ & $0.564\pm{0.005}$ & $25.71\pm{0.01}$ & $0.632\pm{0.006}$ \\
\hline
\end{tabular}
 \label{table:completeness}
\end{table}

\noindent where $m$ is the magnitude, $mc$ is the 50\% completeness value and $\lambda$ is roughly the width of the rollover. For the error: $N$ is the number of artificial stars per bin and $C$ is the completeness in that bin. Figure~\ref{fig:CompletenessPlots} shows the photometric completeness plots and 
Table~\ref{table:completeness} lists the best-fitting logistic function parameters for each filter and field.

\subsection{Colour-Magnitude Diagrams}
\begin{figure*}
\begin{center} 
\includegraphics[width=0.28\hsize]{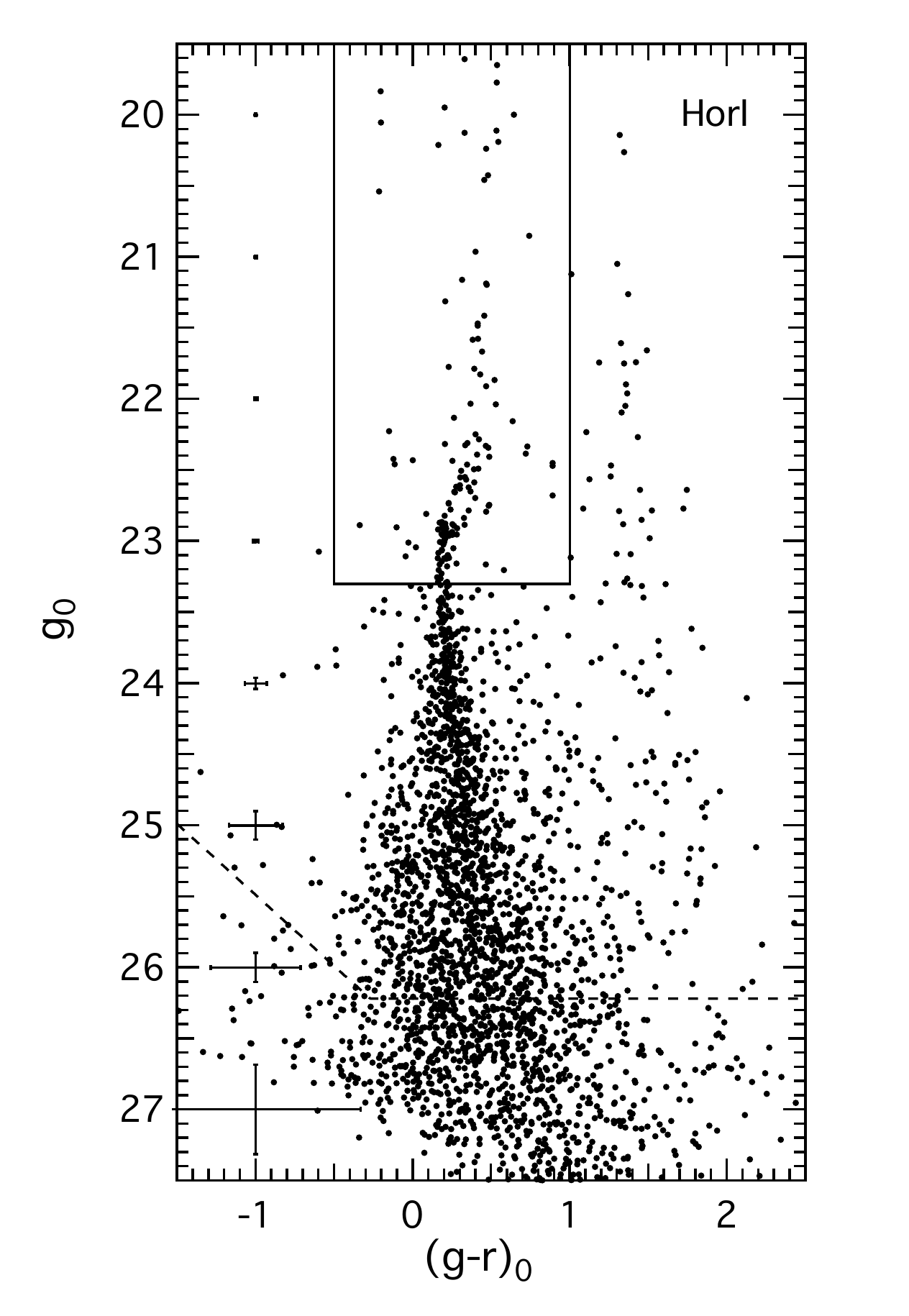}\hspace{-0.8cm}
\includegraphics[width=0.28\hsize]{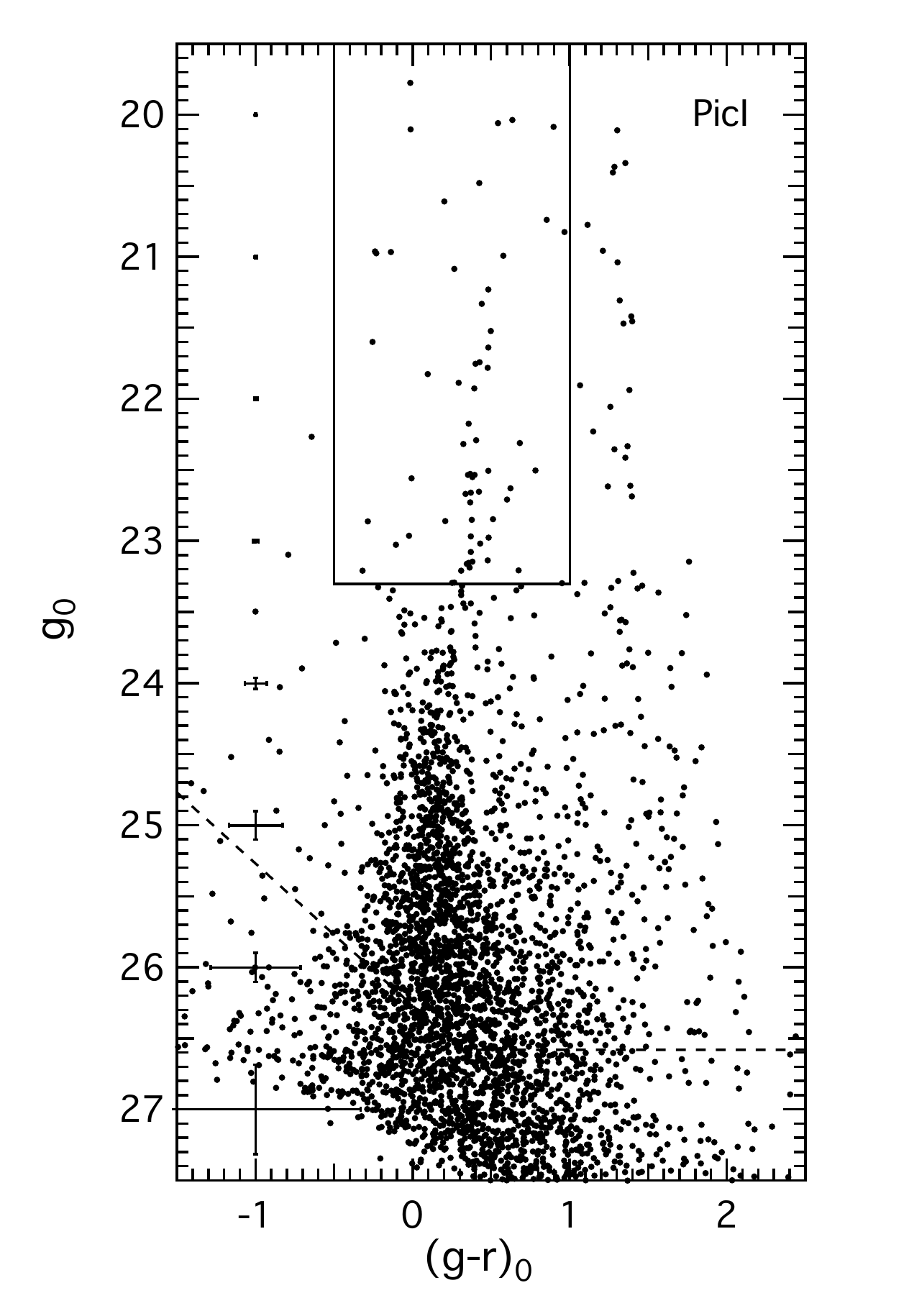}\hspace{-0.8cm}
\includegraphics[width=0.28\hsize]{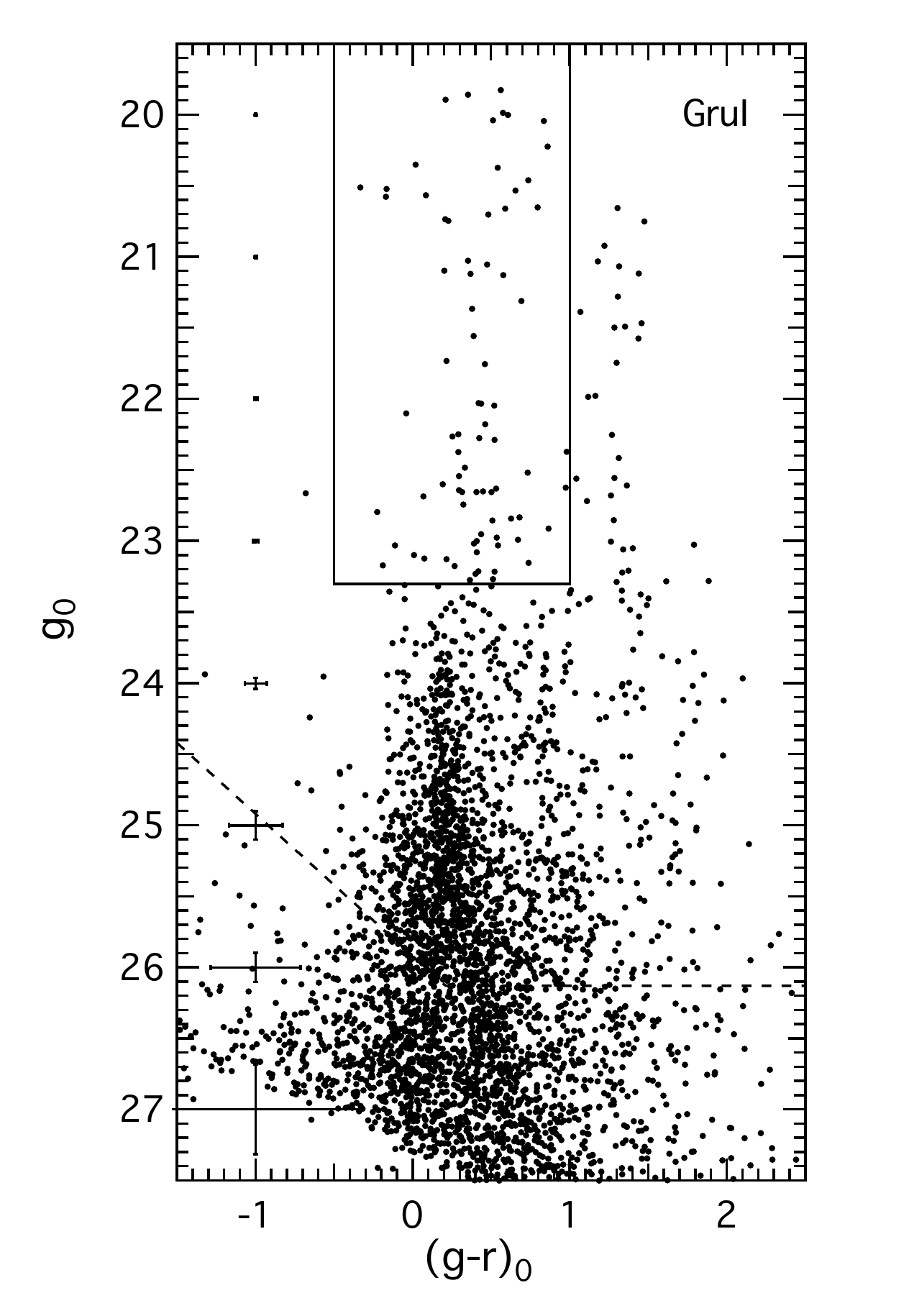}\hspace{-0.8cm}
\includegraphics[width=0.28\hsize]{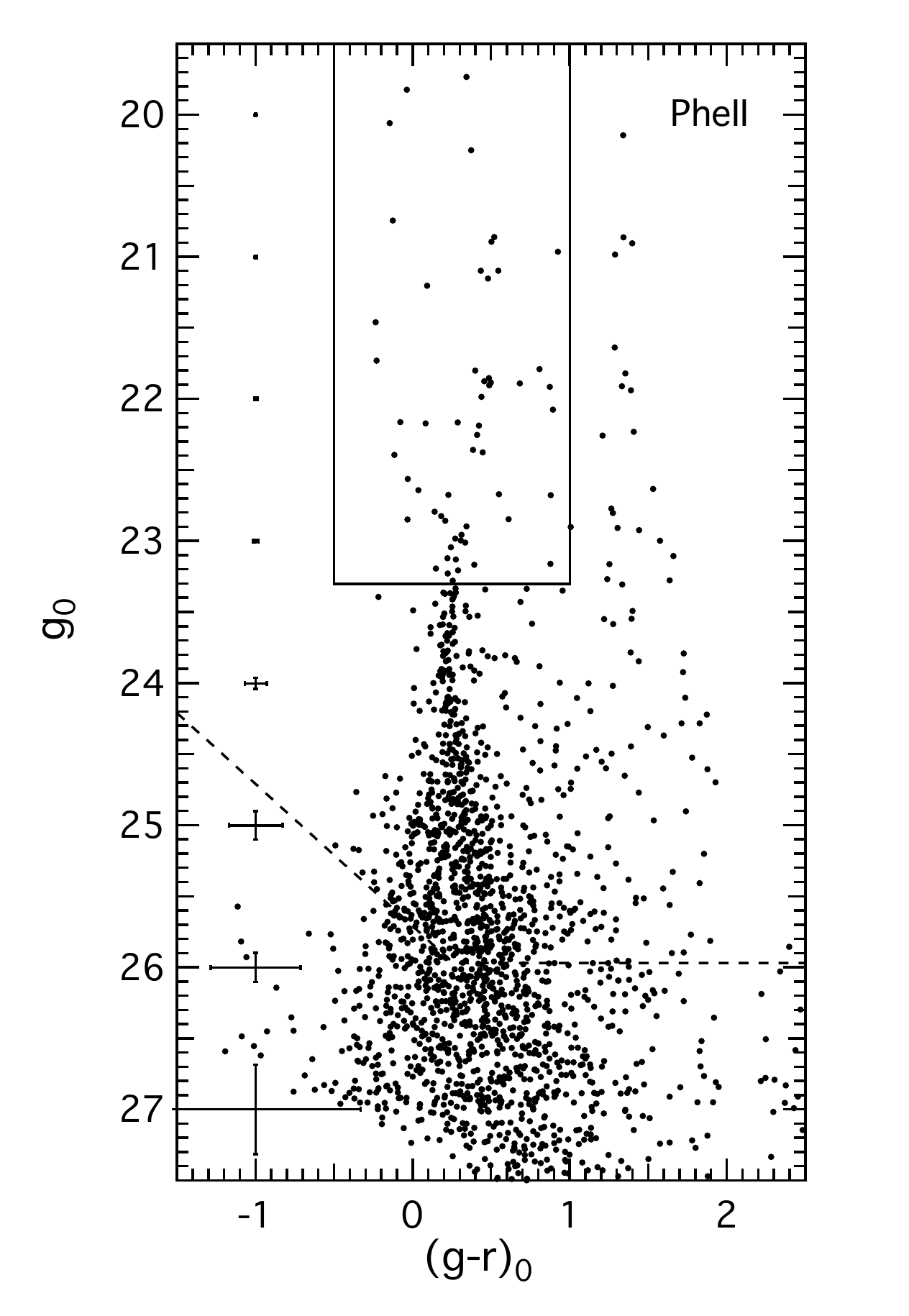}\hspace{-0.5cm}
\caption{The $g_\circ$ vs. $(g-r)_\circ$ colour-magnitude diagrams of all stars detected in the $5\farcm5\times 5\farcm5$ GMOS-S field centred on Hor I, Pic I, Gru I, and  Phe II (from left to right). The rectangular areas correspond to the colour-magnitude windows presented in the discovery papers \citep{Koposov2015, Bechtol2015}. The CMDs reveal distinct main sequence populations extending over $\sim 3-4$ magnitudes down to $g\sim 26.0-26.6$. The error bars running vertically along the colour axis in 1\,mag intervals represent the typical photometric uncertainties. The 50\% completeness levels are given as a dashed line.}
\end{center}
\label{fig:cmd_field}
\end{figure*}

The panels in Figure~\ref{fig:cmd_field} show the extinction-corrected $(g-r)_\circ$ vs. $g_\circ$ colour-magnitude diagrams (CMDs) of the entire GMOS-S fields using all detections classified as stars from our photometric analysis (see $\S$\ref{sec:photcalib}) that were found in the vicinity of each ultra-faint stellar system. The calibrated photometry was corrected for Galactic extinction based on the reddening map by \citet{SFD1998} and the correction coefficients from \citet{Schlafly2011}. The CMDs reveal stars $\sim4$ magnitudes fainter than the main-sequence turn-off (MSTO) and down to the 50\% completeness level $g_{lim}\sim 26$. The rectangular windows correspond to the colour-magnitude range presented in the discovery papers by \citet{Bechtol2015} and \citet{Koposov2015}.

\section{Parameter Analysis}\label{sec:param_analysis}
For determining the fundamental properties of each ultra-faint stellar system, ie. mean age, mean metallicity $\langle$[Fe/H]$\rangle$, the [$\alpha$/Fe]$_{avg}$ ratio, heliocentric distance ($D_\odot$), central coordinates ($\alpha_0,\delta_0$), position angle from north to east ($\theta$), ellipticity ($\epsilon=1-b/a$) and half-light radius ($r_h$) we employed an iterative process. 
First, we established the Dartmouth model isochrone \citep{Dartmouth} that best fits the CMD of the entire GMOS-S field (Figure~\ref{fig:cmd_field}) using the maximum likelihood (ML) method introduced in \cite{Frayn2002}. This method was used in our previous studies \citep{KimJerjen2015a, Kim2, Kim2016}. We calculated the maximum-likelihood values $\mathcal{L}_i$ over a grid of Dartmouth isochrones as defined by Equations\,1 and 2 in \citet{Fadely2011}. The grid points in the multi-dimensional parameter space cover ages from 7.0--13.5\,Gyr, a broad range of chemical composition $-2.5\leq$ [Fe/H] $\leq-0.5$\,dex, $-0.2\leq$ [$\alpha$/Fe] $\leq +0.6$\,dex, and a distance interval $(m-M)\pm 0.5$, where $(m-M)$ is the mean value of the distance modulus for the object reported in the discovery papers.  Grid steps were 0.5\,Gyr, 0.1\,dex, 0.2\,dex, and 0.05\,mag, respectively. For each object, we present the matrix of likelihood values after interpolation and smoothing over two grid points in Figures \ref{fig:HorI_age_metal},\ref{fig:PicI_age_metal},\ref{fig:GruI_age_metal}, and \ref{fig:PheII_age_metal}.

The best fitting model isochrone was then used to identify stars that are sufficiently close to the stellar population of the object in colour-magnitude space. These stars were defined to have a $g_*$-band magnitude in the interval $19.5<g_\circ<27.0$ and a colour ($g_*- r_*$) that fulfil the requirement:
\begin{eqnarray}\label{eqn:iso}
\frac{1}{\sqrt{2\pi \sigma^2_{tot}}} \exp(-((g_*-r_*)-(g-r)_{iso})^2/2\sigma_{tot}^2) >0.5,
\end{eqnarray}
where $(g-r)_{iso}$ is the colour of the model isochrone at $g_*$ and $\sigma^2_{tot}=\sigma_{int}^2+\sigma^2_{g_*}+\sigma^2_{r_*}$. The quantity $\sigma_{int}=0.1$\,mag was chosen as the intrinsic colour width for the isochrone mask and $\sigma^2_{g_*} $, $\sigma^2_{r_*}$ are the photometric uncertainties of a star. 
Restricting the measurement of the parameters $\alpha_0,\delta_0, \theta, \epsilon, r_h$ on this sub-sample reduces the level of contamination in the R.A.-DEC distribution and thus significantly increases the number ratio between member stars of the stellar system and foreground. 

\begin{figure}
\begin{center} 
\includegraphics[width=0.95\hsize]{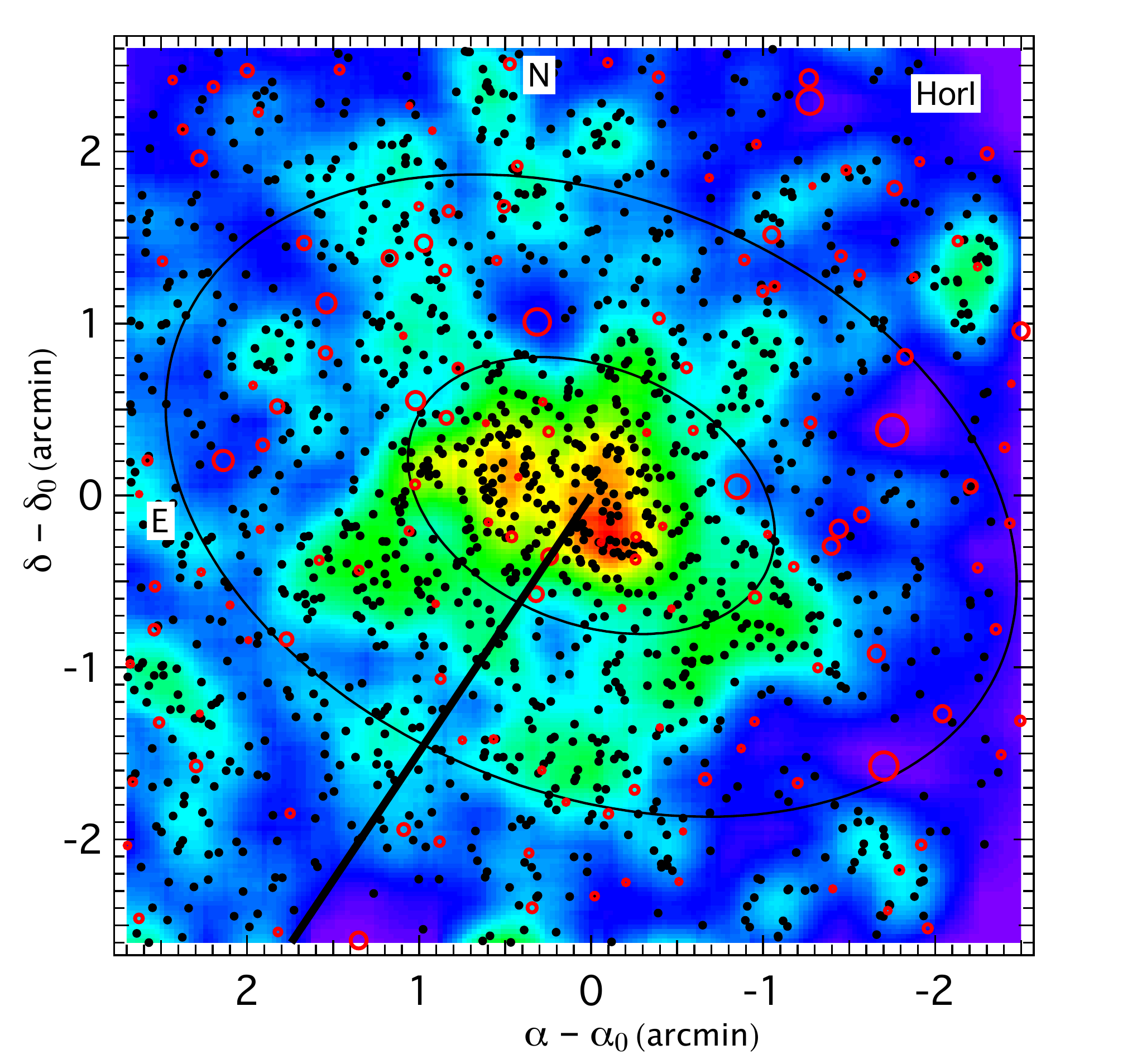}
\caption{
Sky distribution of all stars within the mask defined by the best-fitting isochrone at the Hor I CMD (for more details see $\S$\ref{sec:param_analysis}). The concentration of Hor I stars are clearly visible in the centre of the field. The two ellipses have a position angle of $68^\circ$ and a semi-major axis length of $1r_h$ and $2.32 r_h$, respectively. They cover the areas that contains 50 and 90 percent of the Hor I stellar population, assuming an exponential radial profile. The open red circles are {\sc AllWISE} catalogue objects \citep{Wright2010, Cutri2013}, scaled by their W1 magnitudes. They highlight positions of bright foreground stars and background galaxies. The solid line points in the direction of the LMC.}
\label{fig:HorIstellar_distribution}
\end{center}
\end{figure}

To determine the centre coordinates ($\alpha_0$, $\delta_0$) and structural parameters of the stellar system we employed the ML routine from \cite{Martin2008}, which was previously used by us e.g.~in \citet{Kim2016} based on the likely member stars, i.e.~stars that are within the isochrone mask. We used a 2-dimensional elliptical exponential profile plus foreground:
\begin{eqnarray}\label{eqn:elliptexp}
E(r,r_h,\Sigma_\circ,\Sigma_f)=\Sigma_\circ\exp(-1.68r/r_h)+\Sigma_f
\end{eqnarray}
to model the member star distribution on the sky where 
\begin{eqnarray}
 r = \left\{\left[\frac{1}{1-\epsilon}(x\cos\theta-y\sin\theta)\right]^2 + (x\sin\theta+y\cos\theta)^2\right\}^{1/2} \nonumber
\end{eqnarray}

\noindent is the elliptical (semi-major axis) radius and $(x,y)$ the spatial position of a star, $\epsilon$ the ellipticity of the distribution, $\theta$ the positional angle of the major axis, $r_h$ is the half-light radius, $\Sigma_\circ$ the central star density, and $\Sigma_f$ the foreground star density.
Based on the first estimates for these quantities we constructed a new CMD from stars that are within 
an ellipse with a semi-major axis length $a=2.32r_h$, semi-minor axis length $b=a(1-\epsilon)$ and position angle $\theta$ of the nominal centre of the stellar overdensity. Assuming an underlying exponential profile, this area contains 90\,percent of the total number of member stars and we refer to it as the 90\% ellipse in the following sections. We then re-calculated refined values for age, $\langle$[Fe/H]$\rangle$, [$\alpha$/Fe] and $D_\odot$, and generated the associated isochrone mask to re-calculate $\alpha_0$, $\delta_0$, $\theta$, $\epsilon$ and $r_h$. This process of measuring the two sets of parameters typically converged to the final values after 2-3 iterations. We finally estimate the number of stars $N_*$ that belong to the overdensity with Equation\,5 from~\cite{Martin2008}.
All parameters derived in this section are summarised in Tables~\ref{tab:HorI_parameters},  \ref{tab:PicI_parameters}, \ref{tab:GruI_parameters}, and \ref{tab:PheII_parameters}. We will discuss the results for Hor I in $\S$\ref{sec:HorIprop}, Pic I in $\S$\ref{sec:PicIprop}, Gru I in $\S$\ref{sec:GruIprop}, Phe II in $\S$\ref{sec:PheIIProp}, respectively.

\section{Properties of Horologium I (DES J0255.4-5406)}\label{sec:HorIprop}
\begin{figure}
\begin{center} 
\includegraphics[width=0.95\hsize]{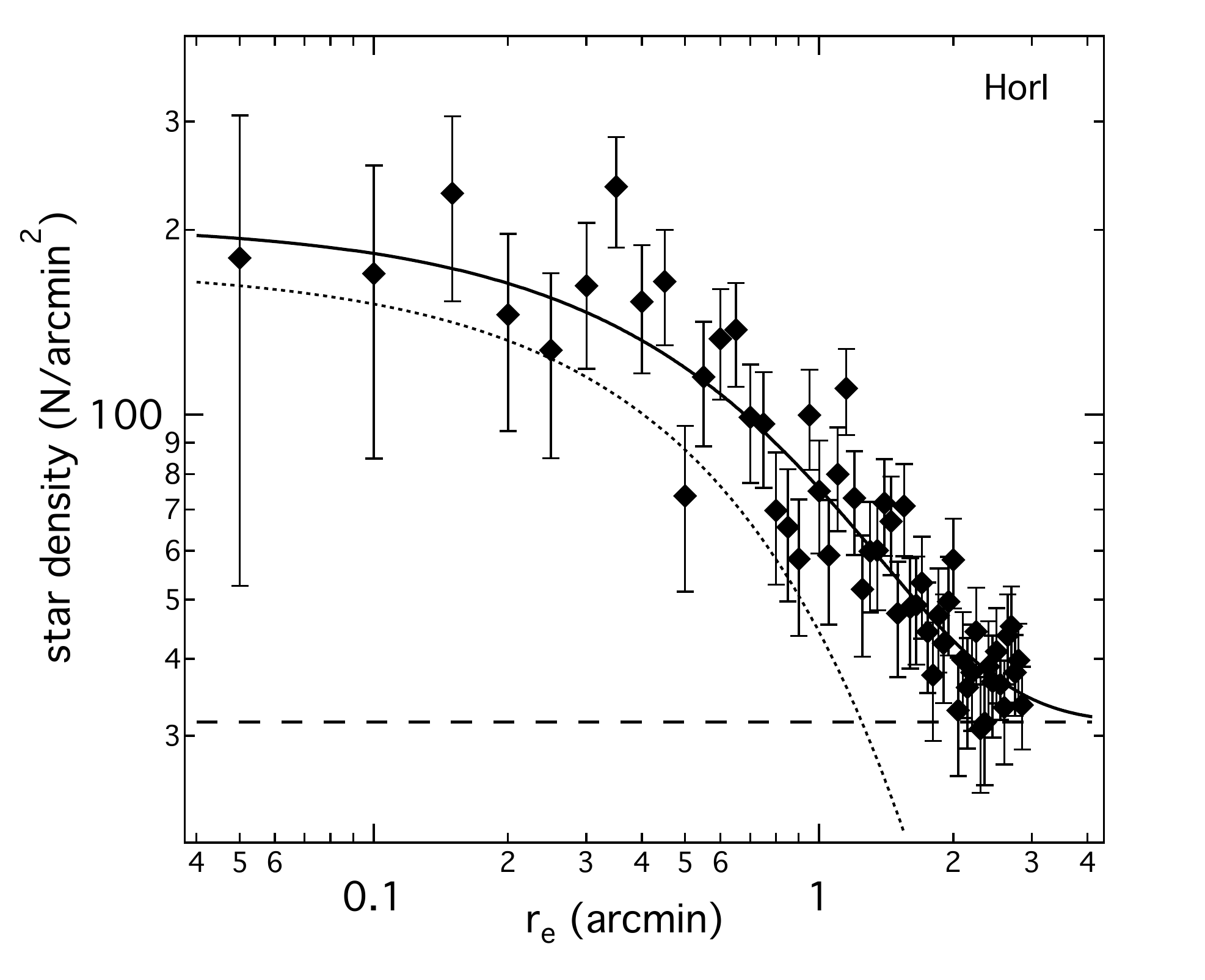}
\caption{Radial density profile of Hor I. The best-fitting Exponential profile (black dotted line) is superimposed on the data points. The horizontal dashed line is the density of the foreground stars. The solid black line represents the profiles + foreground. The error bars were derived from Poisson statistics.}
\end{center}
\label{fig:HorIrad_profile}
\end{figure}

\begin{figure*}
\begin{center} 
\includegraphics[width=0.3\hsize]{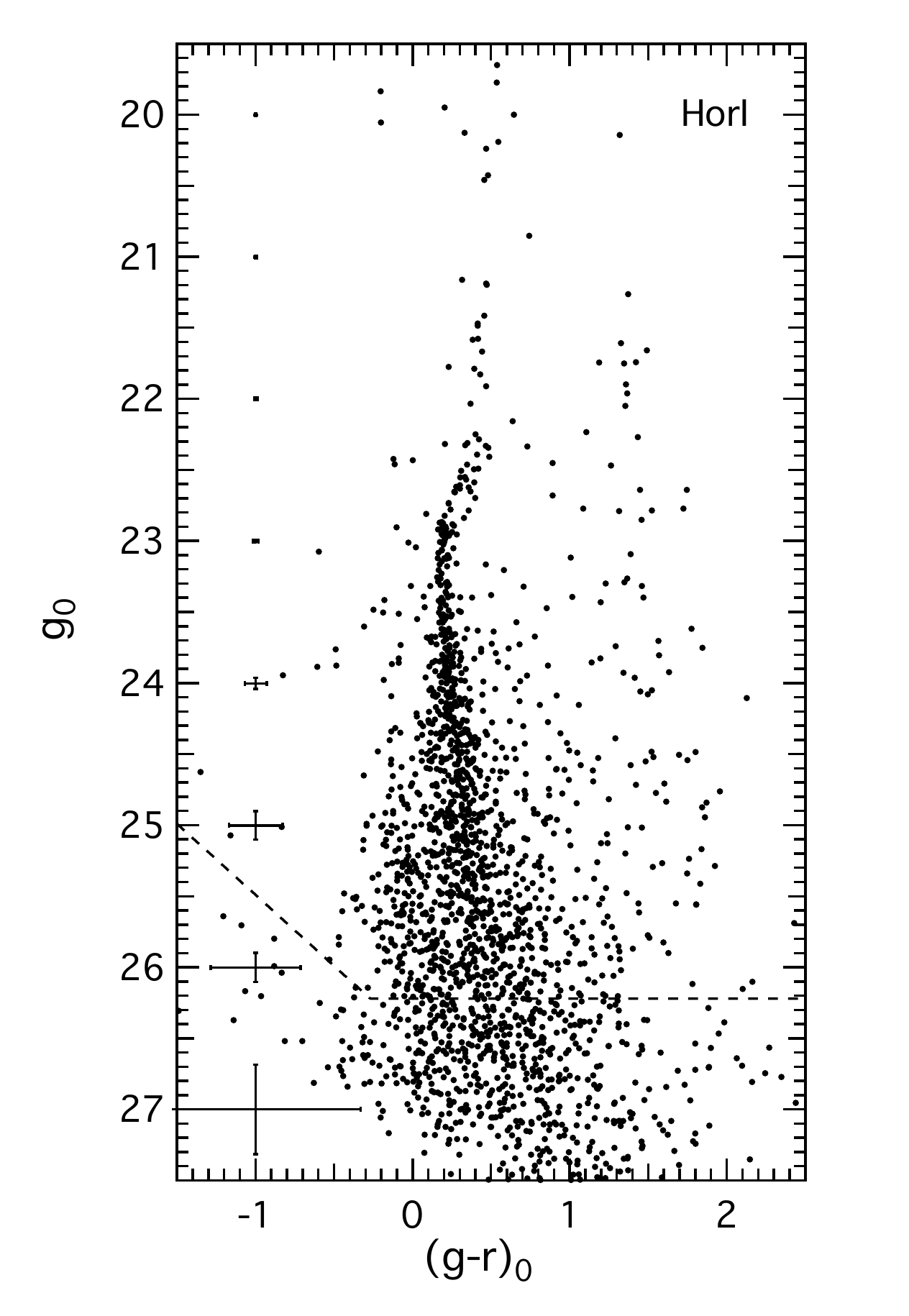}\hspace{-0.2cm}
\includegraphics[width=0.3\hsize]{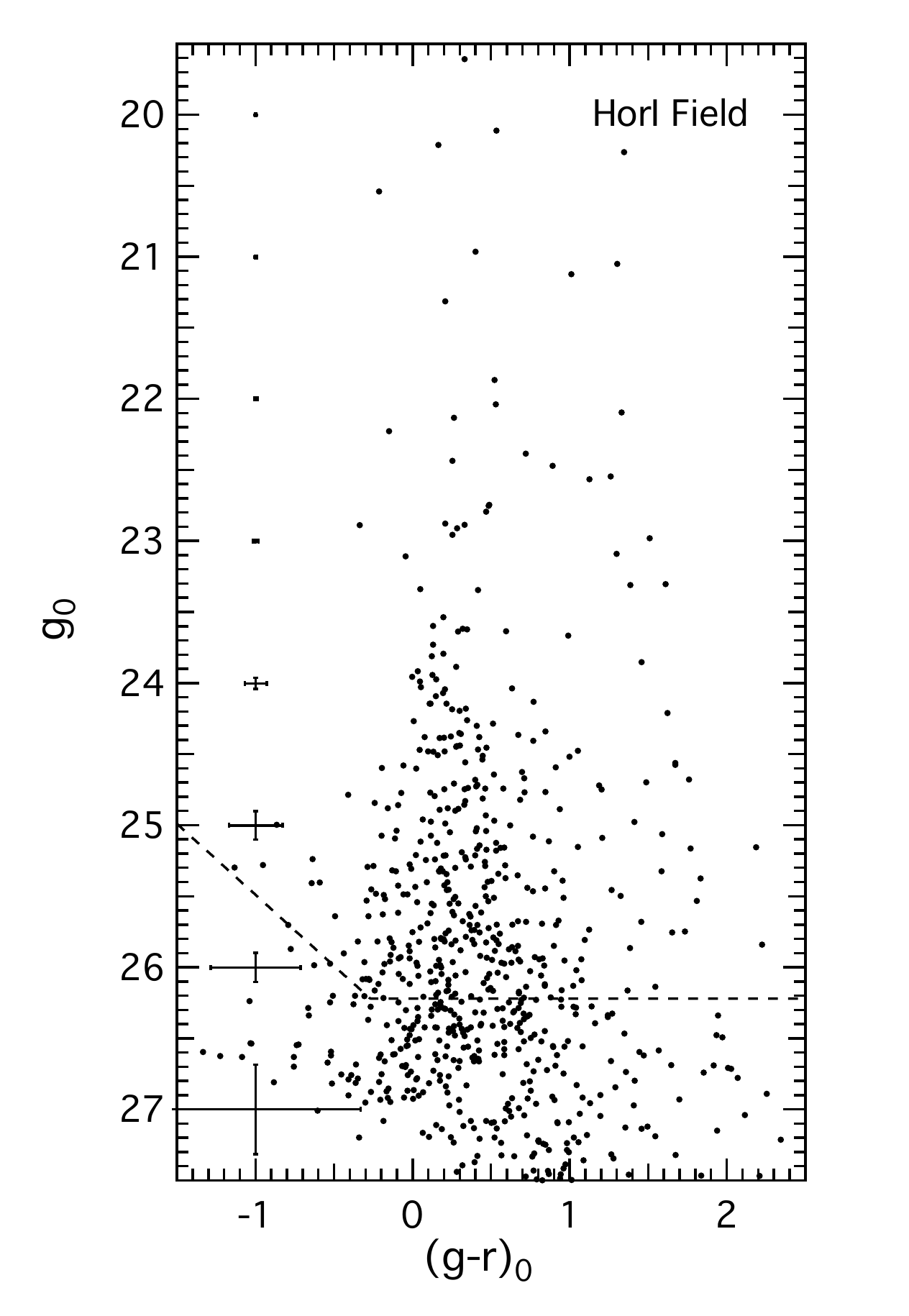}\hspace{-0.2cm}
\includegraphics[width=0.3\hsize]{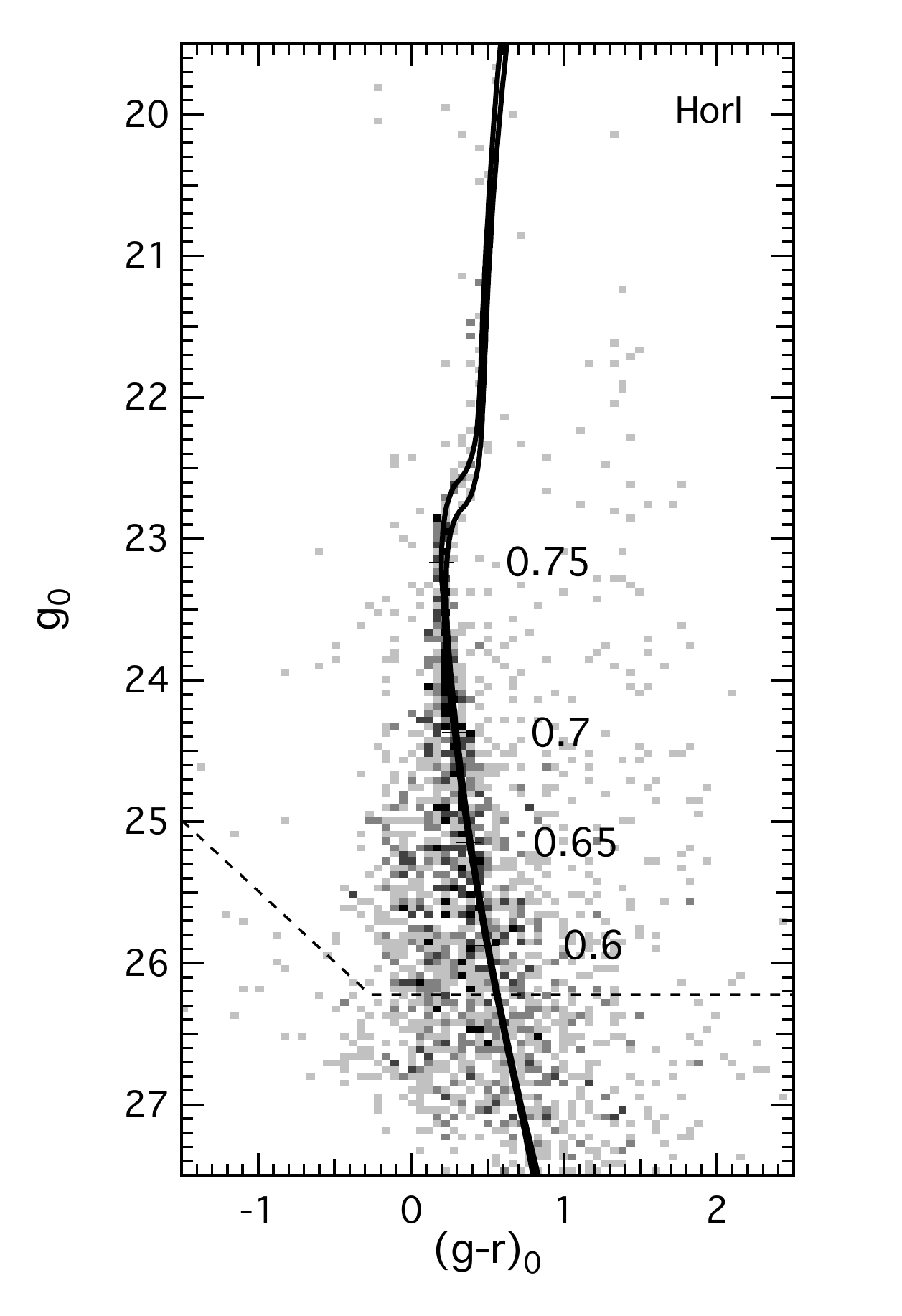}\hspace{-0.2cm}
\caption{{\it Left panel:} The colour-magnitude diagram of all stars within the 2.32$r_h$ ellipse 
centred on the nominal celestial coordinates of Hor I.
{\it Middle panel:} Comparison CMD of the stars within the FoV and outside of the 2.32$r_h$ ellipse covering the same area, showing the distribution of the foreground stars in colour-magnitude space. {\it Right panel:} Hess diagram of the foreground-subtracted CMD. The Dartmouth model isochrone that best describes the stellar population in the magnitude range $21<g_o<25.5$ has a metallicity of [Fe/H]$=-2.40$\,dex, $[\alpha$/Fe]$=+0.2$\,dex
and an age of $13.70$\,Gyr. However, the subgiant branch region of Hor I appears bimodal (see also left panel) and a  model isochrone with the same distance but with a different metallicity and slightly different $[\alpha$/Fe] and age is required to describe the redder stars in the SGB region: [Fe/H]$=-2.00$\,dex, $[\alpha$/Fe]$=+0.4$\,dex, age$=14.00$\,Gyr (isochrone with the fainter SGB). 
The masses of main-sequence stars in solar mass units are marked to show the covered mass range. In all three panels the dashed line represents the 50-percent completeness limit as determined with artificial star test and the MCMC method.}
\label{fig:cmdiso_HorI}
\end{center}
\end{figure*}

\begin{figure}
\begin{center} 
\includegraphics[width=0.95\hsize]{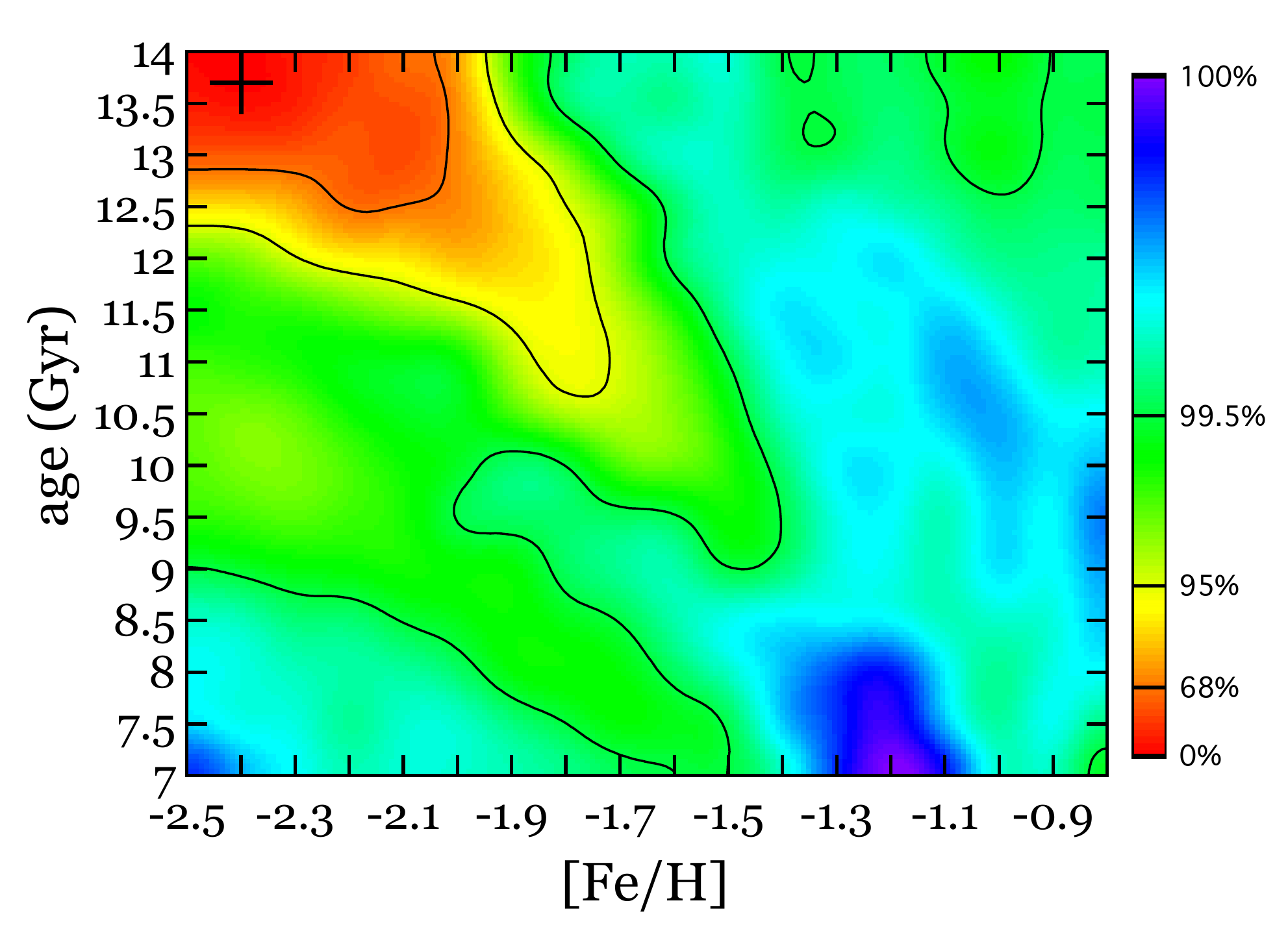}
\caption{Smoothed maximum likelihood density map in age-metallicity space for all stars within the 90\% ellipse around Hor I. Contour lines show the 68\%, 95\%, and 99\% confidence levels. The diagonal flow of the contour lines reflects the
age-metallicity degeneracy inherent to such an isochrone fitting procedure. The 1D marginalized parameters 
around the best fit with uncertainties are listed in Table~\ref{tab:HorI_parameters}.}\label{fig:HorI_age_metal}
\end{center}
\end{figure}

Hor I is located at the edge of the HI Magellanic Stream (Figure~\ref{fig:MCs}), equiangular from both the Large and Small Magellanic Clouds ($\sim$ 23 degrees). While the preliminary distance estimates from \citet{Bechtol2015} and \citet{Koposov2015} are consistent ($\sim$79 vs 87 kpc), \cite{Bechtol2015} predict a much larger physical size of the system than that reported by \citet{Koposov2015} ($r_h \sim$30 vs 60 pc). Despite this, both teams derive a similar total absolute magnitude (M$_V \sim$ $-$3.5 vs $-$3.4\,mag). To resolve these discrepancies and to obtain additional age and metallicity information for the underlying stellar population, this section presents the analysis of our deeper photometry using the procedures as outlined in Section~\ref{sec:param_analysis}. The on-sky distribution of the Hor I stars and their corresponding radial density profile are shown in Figures~\ref{fig:HorIstellar_distribution} and \ref{fig:HorIrad_profile}. The stellar population inside the 90\% ellipse is used to generate a colour-magnitude diagram for isochrone fitting (Figure~\ref{fig:cmdiso_HorI}), infer the luminosity function (Figure~\ref{fig:HorI_LF}) and estimate the system's absolute magnitude. Figures~\ref{fig:cmdzoom_HorI} and \ref{fig:HorI_MSTO_CDF} explore the possibility of two stellar populations in Hor I, while Figure~\ref{fig:HorI_age_metal} shows the best-fit solution in age-metallicity space for the entire stellar population. The parameters for Hor I are listed in Table~\ref{tab:HorI_parameters}.

\subsection{Structural Parameters}\label{sec:HorIstruct}
Figure \ref{fig:HorIstellar_distribution} shows the distribution of all objects classified as stars in the GMOS-S field centred on Hor I that are consistent with the best-fitting isochrone (for more details see $\S$\ref{sec:param_analysis}). A well-defined concentration of Hor I stars is visible at $\alpha_0=02^\mathrm{h}55^\mathrm{m}28\fs3$, $\delta_0=-54^\circ 07'\, 17''$. The increased number statistics from this study also reveals the  existence  of  faint  features  such  as  low density substructure beyond the half-light ellipse to the south of the centre, which could not be detected in previous shallower photometric  surveys.  
We derive a position angle of $\theta=68^\circ\pm9^\circ$ and an ellipticity $\epsilon=0.33\pm 0.08$. These structure parameters remained previously undetermined. The two ellipses have a semi-major axis length of $1r_h$ and $2.32 r_h$, respectively. They border the regions that contain 50 and 90 percent of the Hor I stellar population, assuming an exponential radial profile. The open red circles mark objects from the {\sc AllWISE} catalogue \citep{Wright2010, Cutri2013}, scaled with their W1 magnitudes. They highlight the positions of bright foreground stars and background galaxies (see also top left panel in Figure \ref{fig:cmd_field}). 

The best-fitting isochrone puts Hor I at a distance of 68$\pm$3 kpc from the Sun, slightly closer than the initial estimates from \citet{Bechtol2015} and \citet{Koposov2015} which, in turn, places it within proximity to the Magellanic Clouds. It is only $\sim$29 kpc from the LMC and $\sim$26 kpc from the SMC. 
Figure~\ref{fig:HorIrad_profile} shows the star number density in elliptical annuli around Hor I, where $r_e$ is the elliptical radius. Overplotted is the best-fitting Exponential (black dotted) profile using the modal values from the ML analysis. The error bars were derived from Poisson statistics. The associated elliptical half-light radius is $r_h=22.8^{+4}_{-3}$\,pc, confirming the relative compactness of Horologium I as reported by \citet{Koposov2015}.
The values for the structure parameters are listed in Table\,\ref{tab:HorI_parameters}.

\subsection{Stellar Population of Hor I}\label{sec:HorIpop}  
The left panel of  Figure~\ref{fig:cmdiso_HorI}  shows the colour-magnitude diagram of all stars within $2.32 r_h$ of the centre of Hor I. The middle panel is the CMD of field stars outside the 90\,percent ellipse, covering the same area. 
The right panel shows the Hess diagram for the foreground-corrected Hor\,I CMD with two Dartmouth isochrones superimposed, which we will discuss in the following. We note that the statistical foreground decontamination was performed the same way as in Paper I \& II. The isochrone which best represents the bulk of the Hor I population, the one with the brighter subgiant branch (SGB), has an age of 13.7\,Gyr, a metallicity of [Fe/H]$=-2.40$\,dex and [$\alpha$/Fe]$=+0.2$\,dex, shifted to a distance modulus of $m-M=19.18$\,mag ($D_\odot=68$\,kpc). 

Figure~\ref{fig:HorI_age_metal} shows the smoothed maximum likelihood density map of the age-metallicity space and the location of the best-fit is highlighted with a cross. The stars used to generate this map were selected from inside the 90\% ellipse as seen in Figure~\ref{fig:HorIstellar_distribution}.

\begin{figure}
\begin{center} 
\includegraphics[width=0.95\hsize]{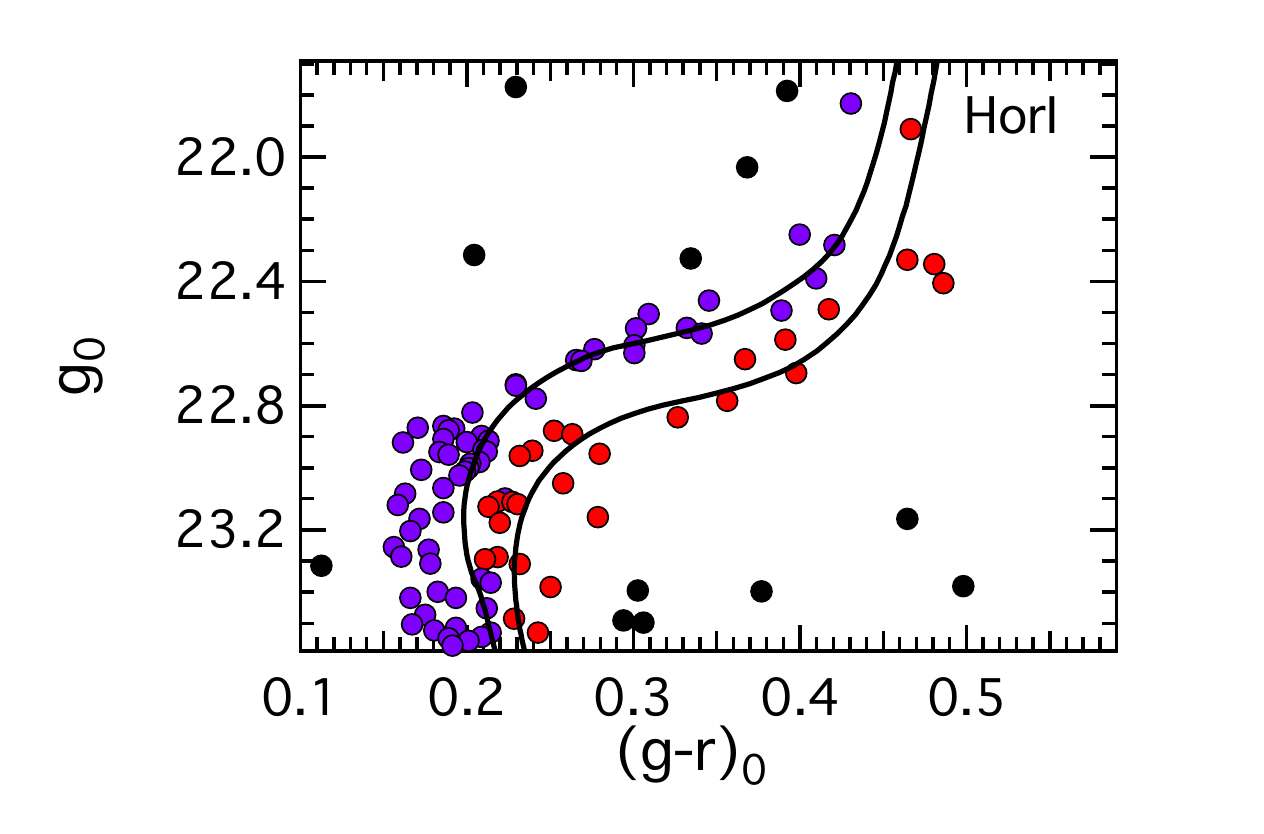}
\includegraphics[width=0.95\hsize]{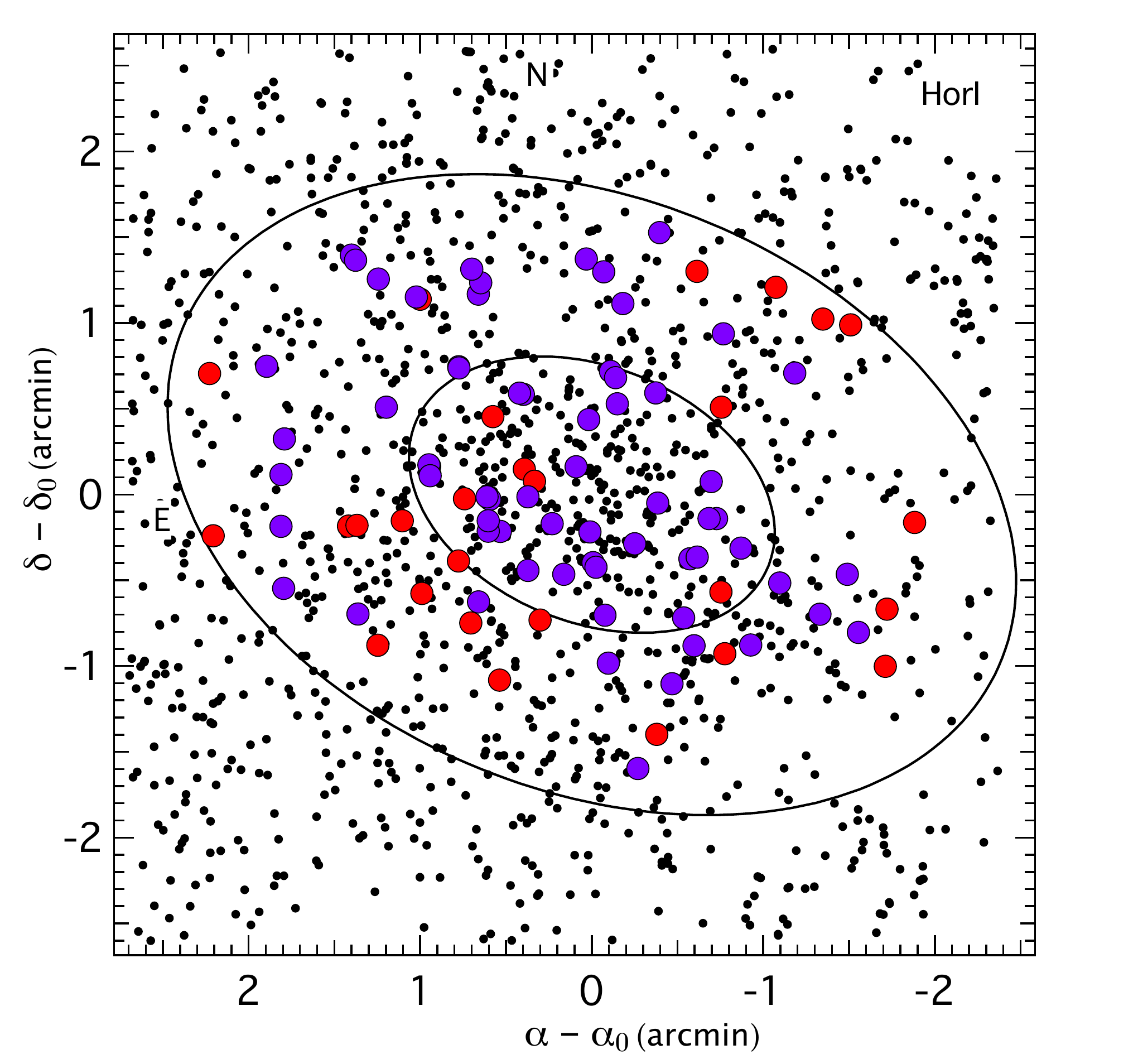}
\caption{(Top) The Hor I CMD around the main-sequence turn-off region shows a distinct bifurcation suggesting the presence of two sub-populations of stars present in Hor I. 
A change in metallicity and a small change in $\alpha$ abundance is able to describe the difference. The smaller population of 28 stars colour coded in red follows a [Fe/H]$=-2.00$\,dex, [a/Fe]$=+0.4$\,dex, age$=14.00$\,Gyr model isochrone, while the 64 stars in purple are consistent with the median solution that best described the entire population: [Fe/H]$=-2.40$\,dex, [a/Fe]$=+0.2$\,dex, age$=14.00$\,Gyr. (Bottom) The R.A.-DEC distribution of these Hor I MSTO/subgiant stars with the same colour code. The two ellipses have radii of $1r_h$ and $2.32r_h$, respectively.}
\label{fig:cmdzoom_HorI}
\end{center}
\end{figure}

\begin{figure}
\begin{center} 
\includegraphics[width=0.95\hsize]{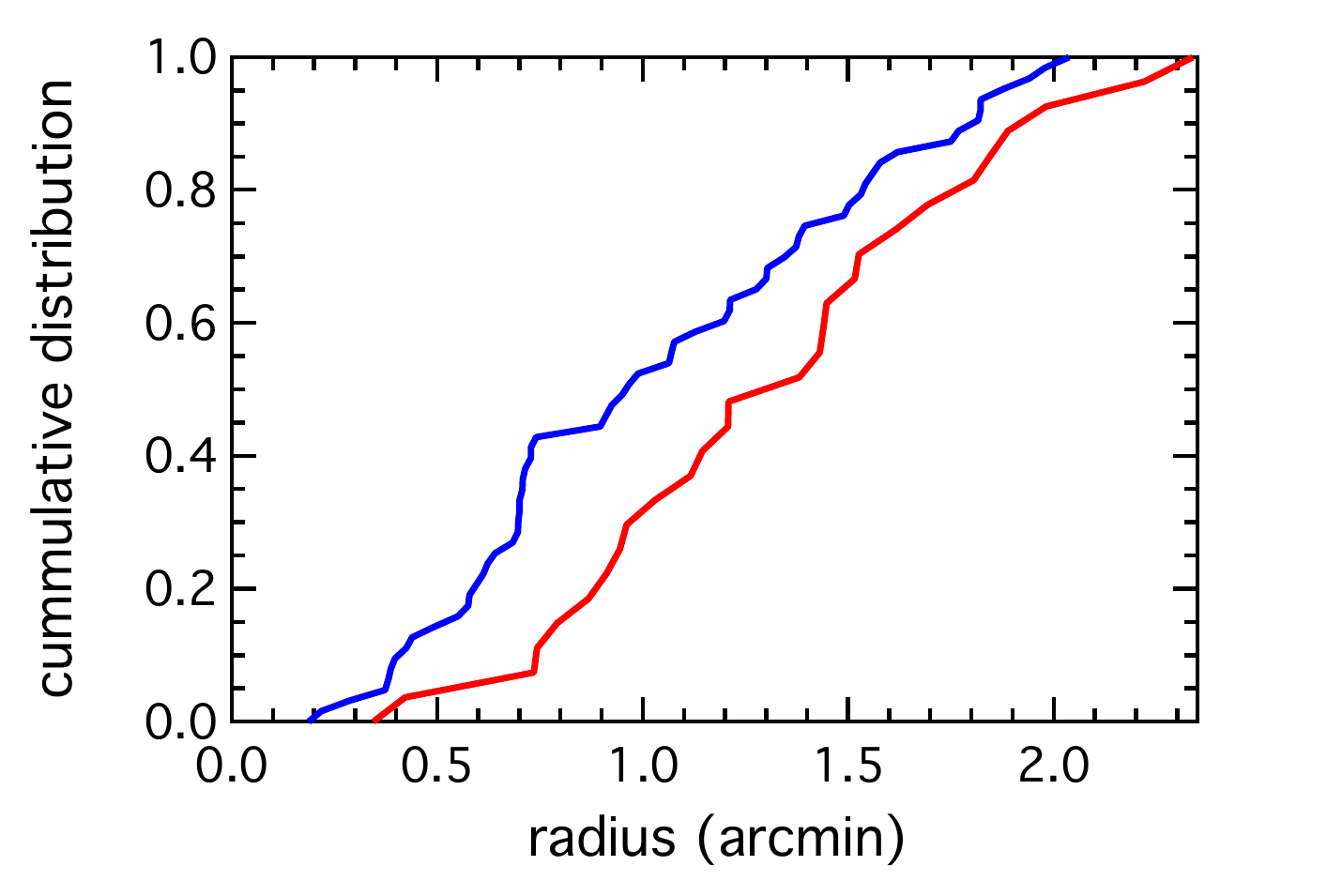}
\caption{The cumulative distribution functions for the two subsamples of Hor I stars. The MSTO/subgiant stars that follow the more metal-poor ($-2.40$\,dex) isochrone (blue line) are found systematically closer to the galaxy centre than the stars following the isochrone with a metallicity of $-2.00$\,dex. The maximum vertical offset is $\Delta=0.3504$ and the corresponding $p$-value for a Kolmogorov-Smirnov comparison test is 0.012. The null hypothesis that the two subsamples were drawn from the same population is rejected at the 98.8\% confidence level.\label{fig:HorI_MSTO_CDF}}
\end{center}
\end{figure}

Looking at the Hor I CMD more closely we notice the subgiant branch region exhibits some level of bimodality. The top panel of Figure \ref{fig:cmdzoom_HorI} shows a zoomed version of the CMD around the MSTO and subgiant branch. To further investigate this feature we divided these stars into two groups colour-coded in purple and red. The choice of the two groups of stars is done by-eye, and it is intended to demonstrate that these two groups correspond to two distinct sub-populations. The black dots are considered foreground stars. We
determined the best-fitting Dartmouth isochrone over the entire Hor I CMD, while restricting the fit to the purple or red stars in the MSTO/subgiant branch region ($21.7<g_o<23.6$), respectively. A change in metallicity and in $\alpha$-abundance is able to account for the difference. The smaller subpopulation of 28 MSTO/subgiant branch stars colour coded in red follows a [Fe/H]$=-2.00$\,dex, [$\alpha$/Fe]$=+0.4$\,dex, age$=14.00$\,Gyr isochrone, whereas the 64 stars in purple are consistent with the median solution that best described the entire population: [Fe/H]$=-2.40$\,dex, [$\alpha$/Fe]$=+0.2$\,dex, age$=14.00$\,Gyr. The bottom panel of Figure \ref{fig:cmdzoom_HorI} shows the R.A.-DEC distribution of these Hor I stars using the same colour code. Like in Figure \ref{fig:HorIstellar_distribution}, the two ellipses have radii of $1r_h$ and $2.32r_h$, respectively. The purple stars appear to be more concentrated towards the centre of Hor I. To test the significance of this impression we plot in Figure \ref{fig:HorI_MSTO_CDF} the cumulative distribution functions for the two groups of Hor I MSTO stars. The  stars that follow the more metal-poor isochrone (blue line) are systematically closer to the galaxy centre than the stars following the isochrone with a metallicity of $-2.00$\,dex. The maximum offset between the two cumulative distributions is $\Delta=0.3504$. The corresponding $p$-value computed from a Kolmogorov-Smirnov comparison test is 0.012. Hence, the null hypothesis that the two subsamples are drawn from the same population can be rejected at the 98.8\% confidence level.

Hor I is the first ultra-faint dwarf galaxy candidate to our knowledge to show the presence of two distinct stellar populations. They are both centred on the Hor I overdensity and yet it is unclear if these correspond to two bursts of star formation leading to different enrichment levels or a possible merger event early in the life of Hor I between two stellar populations with differing chemical compositions. 
Evidence of subgiant branch splitting/broadening in Milky Way globular clusters has been presented by \citet{Cassisi2008, Milone2010} and \citet{Piotto2012}. For instance, the subgiant branch region of globular clusters NGC6388 and NGC6715 in Figure 3 of \citet{Piotto2012} show a similar morphology. The authors suggest that the difference between the two stellar populations is either a 1-2 Gyr age difference or highly contrasting C+N+O content levels. In these studies they find that the population ratio varies from an almost 50-50 split, to another case of 97\% in the dominant population. In the case of Hor I, we observe an approximate 2.3:1 ratio and the errors on the age estimates could account for a 1-2 Gyr age difference. Additionally, our best-fit isochrones are consistent with both an [Fe/H] and an [$\alpha$/Fe] abundance variation between the two subgiant branches. All these pieces of information favour the possibility that Hor I has undergone at least two star formation events, the second triggered by galactic outflow and stellar feedback, rather than a ultra-faint to ultra-faint merger event. 
In this case, it raises questions on how this occurred given how small the Hor I system is today. 
What conditions would be necessary to facilitate such an event without destroying the object in the process? 

\subsection{Luminosity Function and Total Luminosity}\label{sec:HorILum}
\begin{figure}
\begin{center} 
\includegraphics[width=0.95\hsize]{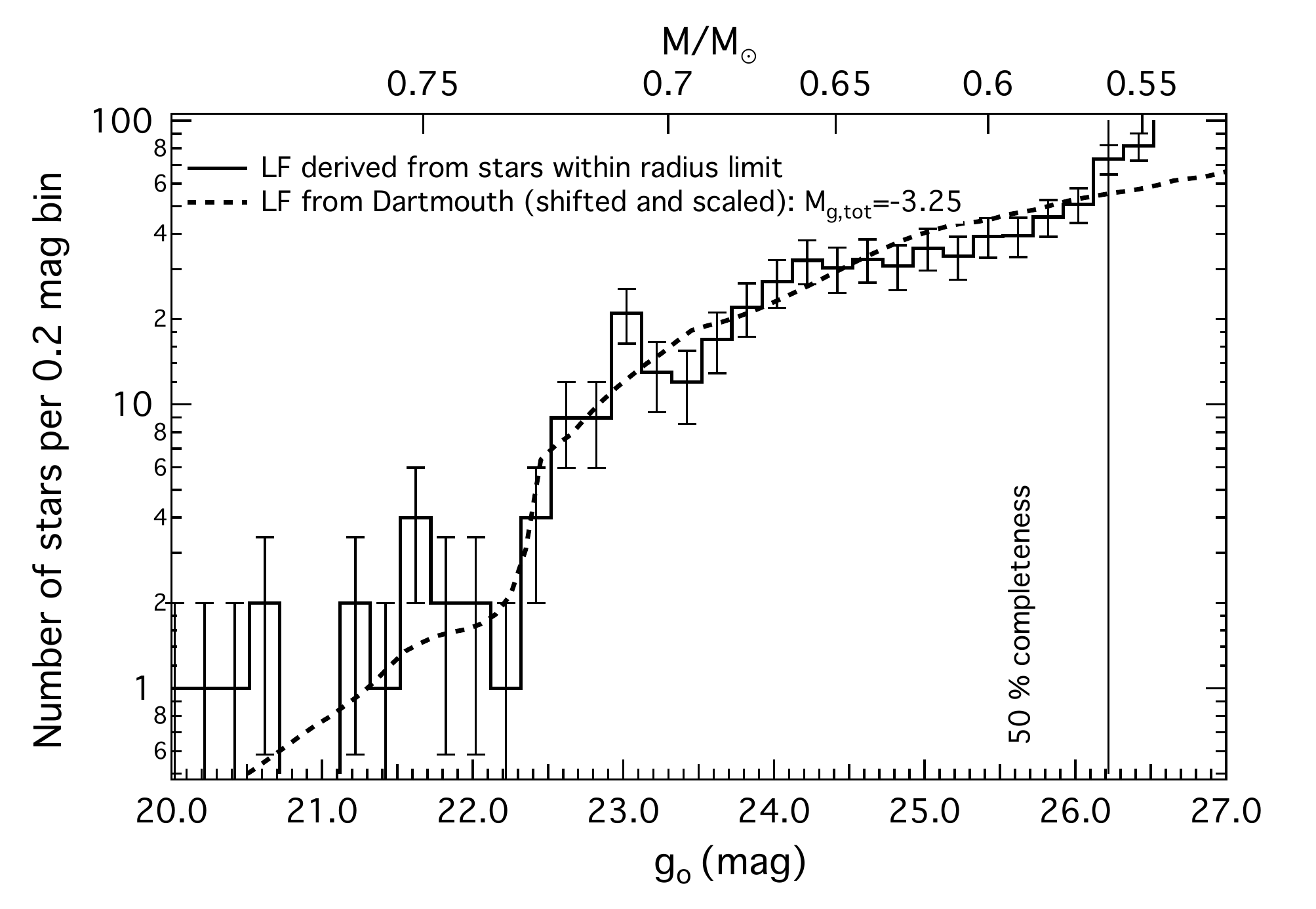}
\caption{Completeness-corrected Hor\,I luminosity function of all stars that are within the isochrone mask (histogram). The best-fitting Dartmouth model luminosity function shifted by the distance modulus $19.18$\,mag and scaled to a total luminosity of $M_g=-3.25$\,mag is overplotted (dashed line).\label{fig:HorI_LF}}
\end{center}
\end{figure}

The stellar luminosity function of Hor I has been derived and is presented in Figure~\ref{fig:HorI_LF}. We calculated the integrated light by comparing the completeness-corrected observed LF with the Dartmouth model LF that corresponds to the best-fitting isochrone of 13.7\,Gyr, [Fe/H]$=-2.40$, and [$\alpha$/Fe]$=+0.2$. We measured a total $g$-band luminosity of $M_{g}=-3.25\pm0.1$. The two integrated Dartmouth model LFs in $g$ and $V$ have a colour of $g-V=0.33$, which convert the $M_{g}$ magnitude into $M_{V}=-3.58$. 
Since this method relies on the overall shape of the LF instead of individual flux, the inclusion or exclusion of a single bright star as observed in the luminosity function carries a uncertainty typically up to $\sim0.3$\,mag. 
All derived parameters presented in this section are summarised in Table\,\ref{tab:HorI_parameters}.

\begin{table}
\caption{Derived properties and structural parameters of Horologium I \label{tab:HorI_parameters} }
{
\begin{center}
\begin{tabular}{lcc}
\hline
 & \bf{Horologium I} \\ \hline
$\alpha_0$\,(J2000) & 
$02^\mathrm{h}55^\mathrm{m}28\fs3\pm0\fs4$  \\
$\cdots$ & $43.8679\pm0.0017$\,deg \\
$\delta_0$\,(J2000) & 
$-54^\circ 07'\, 17'' \pm 5''$ \\
$\cdots$ & $-54.1214\pm0.0014$\,deg \\
$\theta$ (deg) & $68^{\circ}\pm 9^{\circ}$ \\
$\epsilon$ & $0.33^{+0.08}_{-0.08}$  \\
$r_h$ (arcmin) & $1.11^{+0.16}_{-0.14}$ \\
$N_*$  & $257\pm 38$ \\
$E(B-V)$ (mag) & 0.013 \\       
$A_g$ & 0.048 \\
$A_r$ & 0.033\\
$(m-M)$ & $19.18\pm 0.09$  \\
$D_{\odot}$ (kpc) & $68\pm 3$ \\
$r_h$ (pc) & $22.8^{+4}_{-3}$   \\
age (Gyr) &   $13.7^{+0.3}_{-0.8}$  \\
$\langle [$Fe/H$] \rangle$ (dex) &  $-2.40^{+0.10}_{-0.35}$  \\
$[\alpha$/Fe$]_{\rm avg}$ (dex)&  $+0.2^{+0.1}_{-0.1}$ \\
$M_V$ (mag) & $-3.58\pm0.30$   \\
\hline
\end{tabular}
\end{center}
}
\end{table}

\section{Properties of Pictor I (DES J0443.8-5017)}\label{sec:PicIprop}
Pic I is an outer MW halo system and can be found $\sim$20 degrees from the Large Magellanic Cloud and $\sim$33 degrees from the Small Magellanic Cloud (Figure~\ref{fig:MCs}). Interestingly, its reported heliocentric distance of $\sim$120 kpc places it equidistant from the Magellanic Clouds ($\sim$83 kpc). Unlike Hor I, Pic I has no line-of-sight overlap with the HI Magellanic Stream and is well isolated on the sky. 
It is roughly the same size and has a similar luminosity as the other satellites presented here ($r_{h} \sim $ 30 pc, $M_{V} = -3.1$). 

\begin{figure}
\begin{center} 
\includegraphics[width=0.95\hsize]
{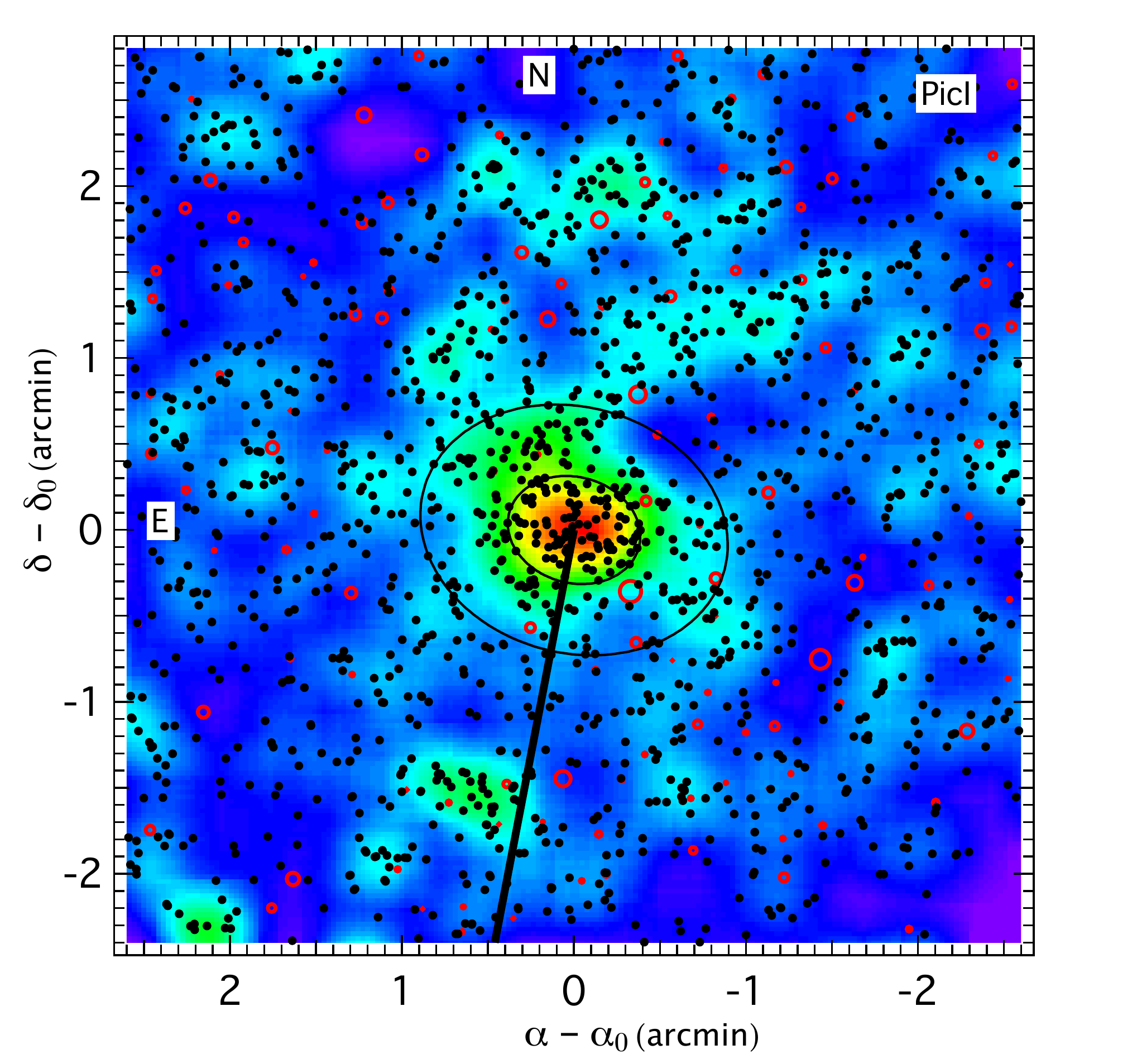}
\caption{Stars that are sufficiently close to the best-fitting isochrone (for more details see $\S$\ref{sec:param_analysis}) significantly increases the contrast between stars associated to the Pic I overdensity and Galactic foreground. The concentration of Pic I stars are clearly visible in the centre of the GMOS field. The two ellipses have a position angle of $75^\circ$ and a semi-major axis length of $1 r_h$ and $2.32 r_h$, respectively. The outer ellipse borders the region that contains 90 percent of the Pic I stellar population, assuming an exponential radial profile. The open red circles are objects from the {\sc AllWISE} catalogue \citep{Wright2010,Cutri2013}, scaled to reflect their magnitudes. These objects highlight the position of the bright objects in the field both foreground stars and background galaxies.The solid line points in the direction of the LMC.
\label{fig:PicIstellar_distribution}}
\end{center}
\end{figure}

\begin{figure}
\begin{center} 
\includegraphics[width=0.95\hsize]{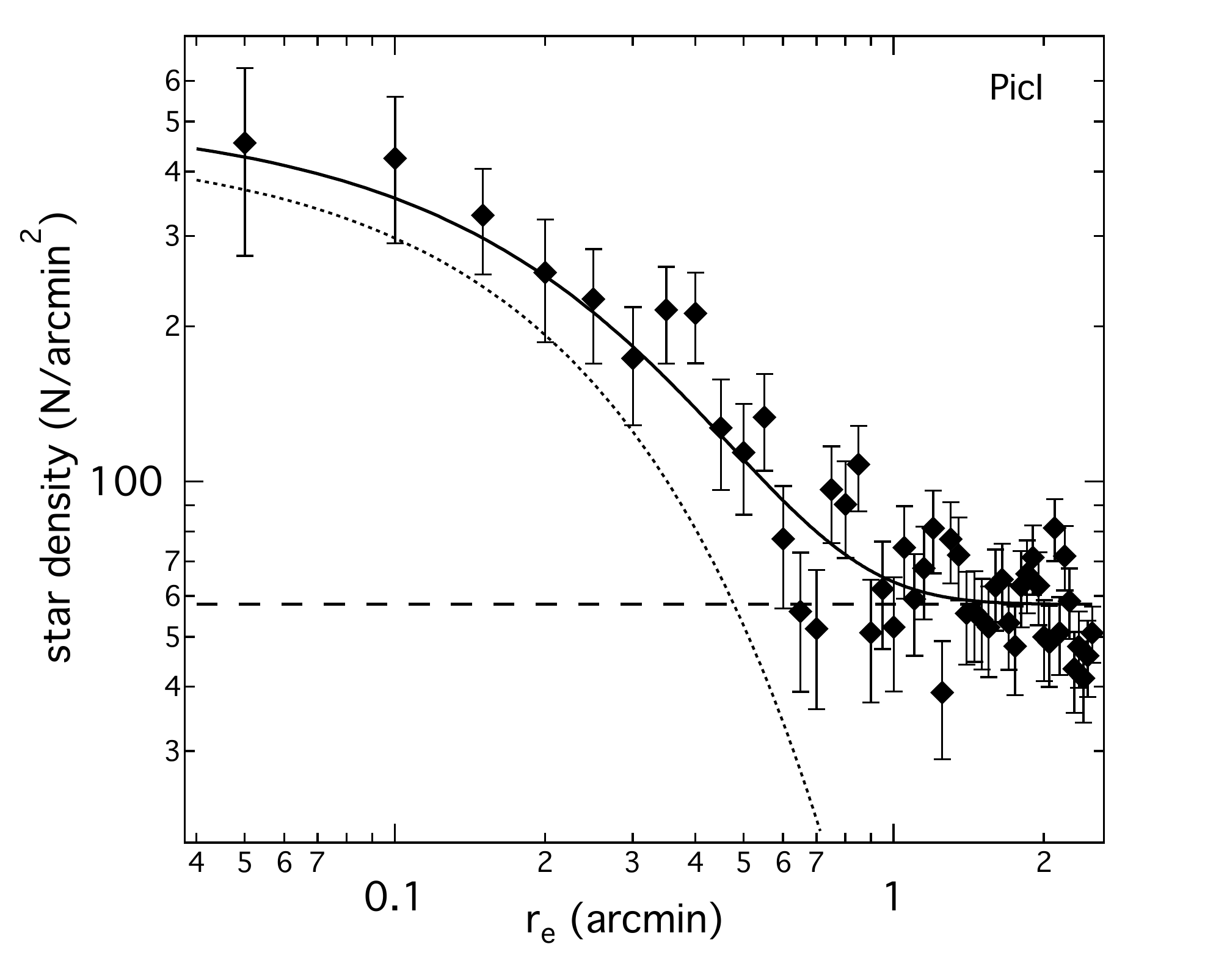}
\caption{Radial density profile of Pic I stars. The best-fitting Exponential  profile (dotted line) is superimposed on the data points. The horizontal dashed line is the 
density of the foreground stars. The solid black line represent the profile + foreground. 
The error bars were derived from Poisson statistics.\label{fig:PicIrad_profile}}
\end{center}
\end{figure}

\begin{figure*}
\begin{center} 
\includegraphics[width=0.3\hsize]{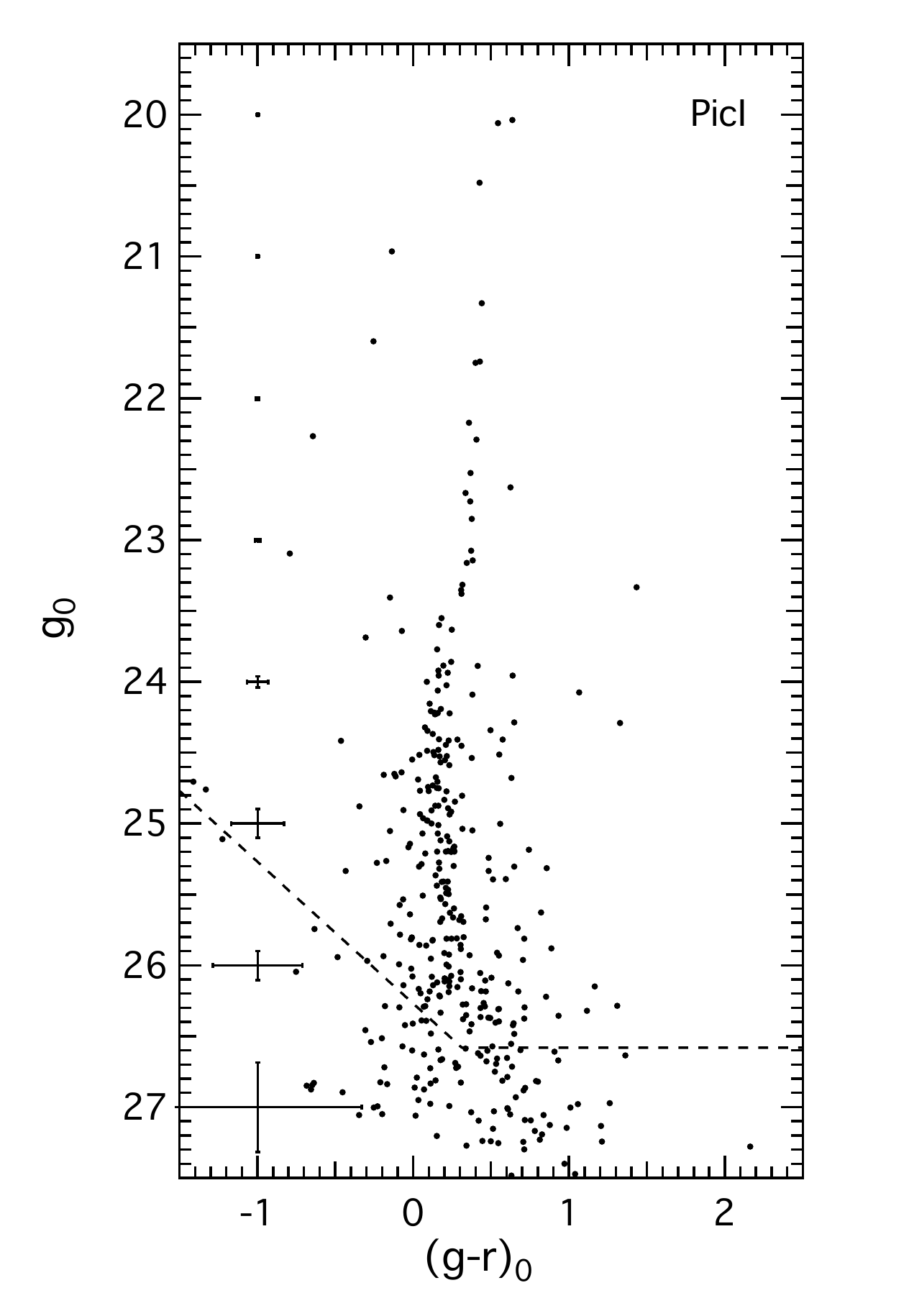}\hspace{-0.2cm}
\includegraphics[width=0.3\hsize]{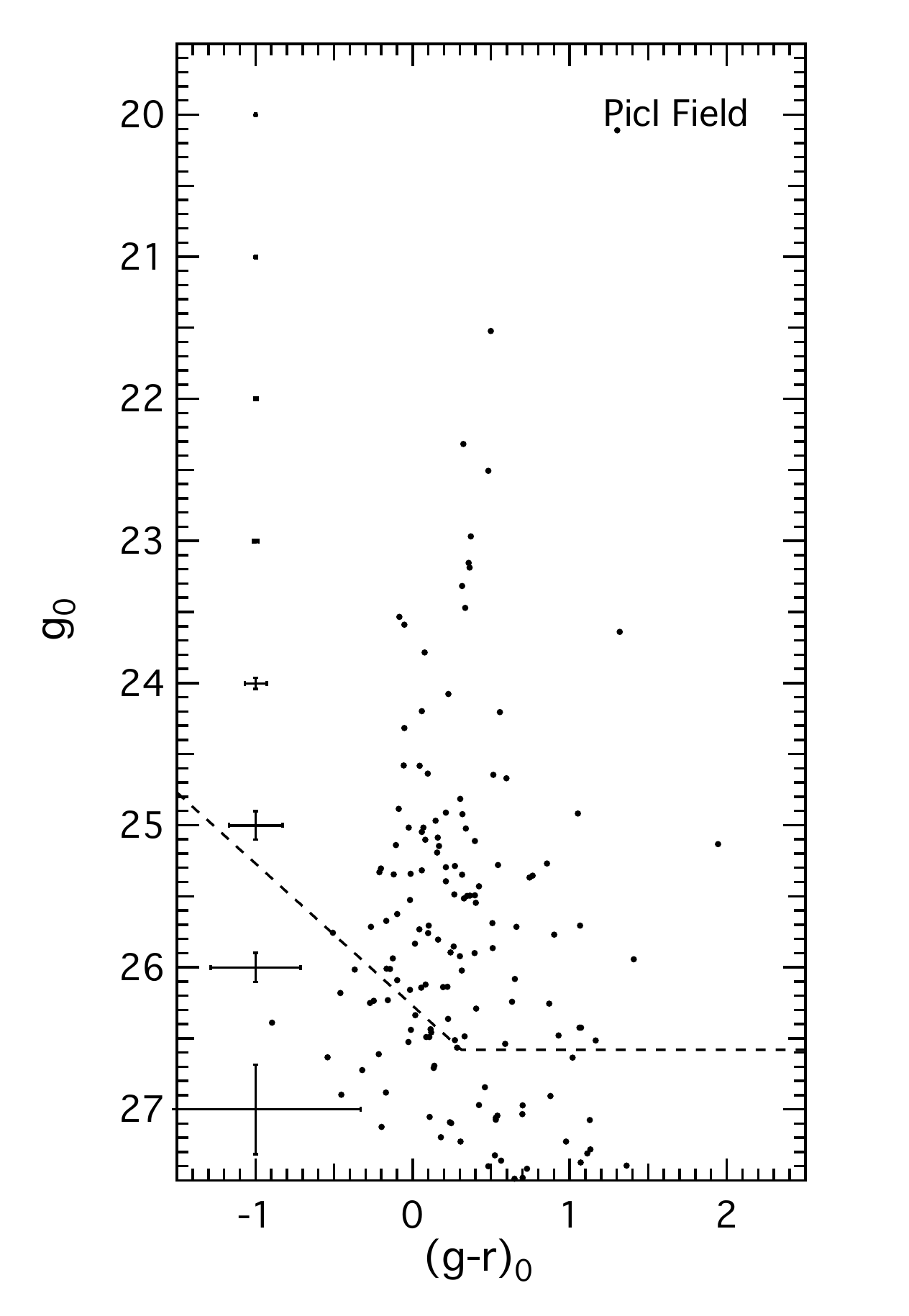}\hspace{-0.2cm}
\includegraphics[width=0.3\hsize]{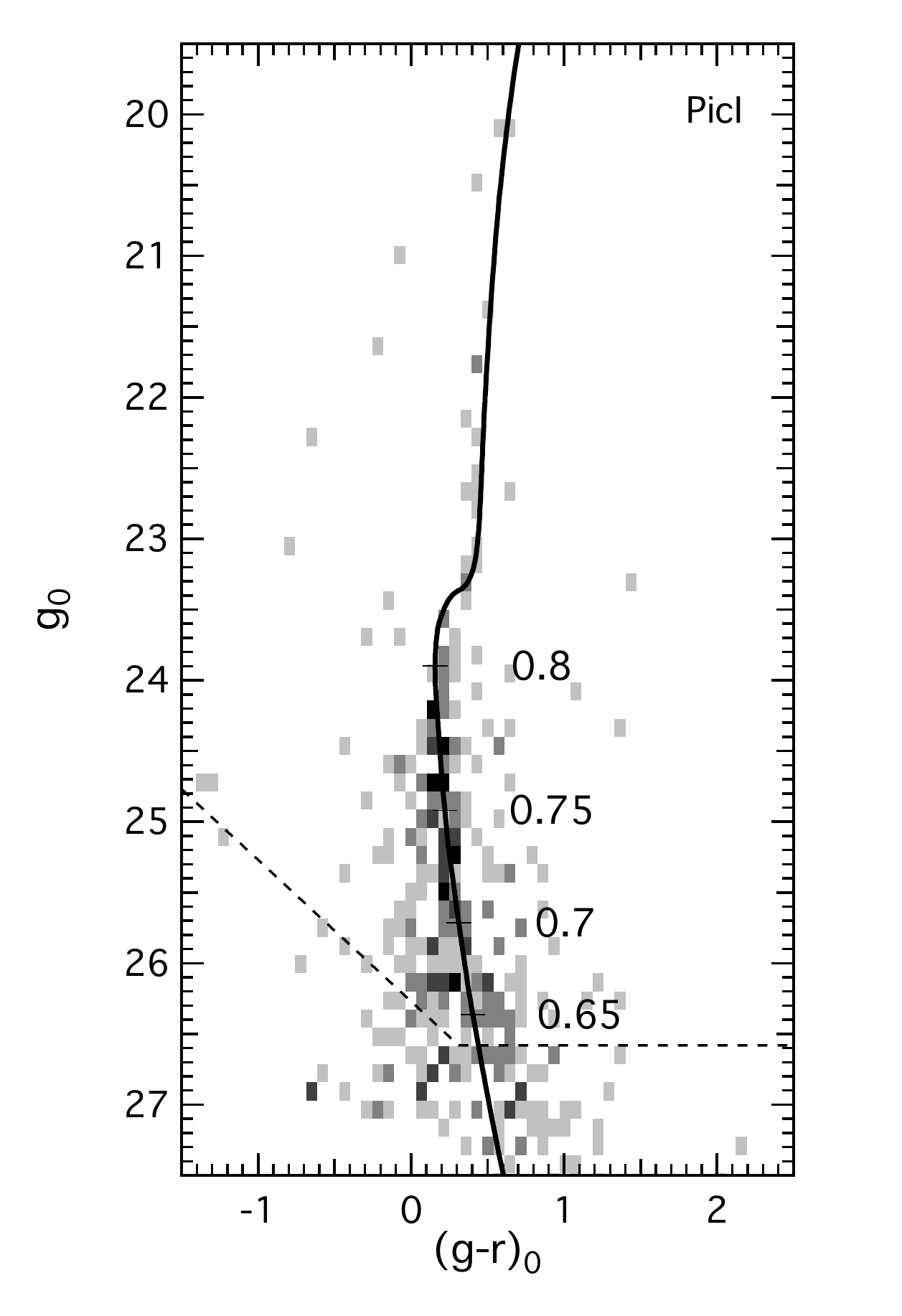}\hspace{-0.2cm}
\caption{{\it Left panel:} The colour-magnitude diagram of all stars within the ellipse centred on the nominal celestial coordinates of Pic I (Figure\,\ref{fig:PicIstellar_distribution}). 
{\it Middle panel:} Comparison CMD of stars outside the $2.32\,r_h$ ellipse, covering the same area, showing the distribution of Galactic foreground stars in colour-magnitude space.
{\it Right panel:} Hess diagram of the foreground-subtracted CMD superimposed with the best-fitting Dartmouth isochrone. The isochrone has a metallicity of [Fe/H]$=-2.28$\,dex, $[\alpha$/Fe]=+0.2\,dex and an age of 11.8\,Gyr. The distance modulus is $m-M=20.20$\,mag. The masses of main-sequence stars in solar mass units are marked to show the covered mass range. In all three panels the dashed line represents the 50-percent completeness limit.
}\label{fig:cmdiso_PicI}
\end{center}
\end{figure*}

\begin{figure}
\begin{center} 
\includegraphics[width=1\hsize]{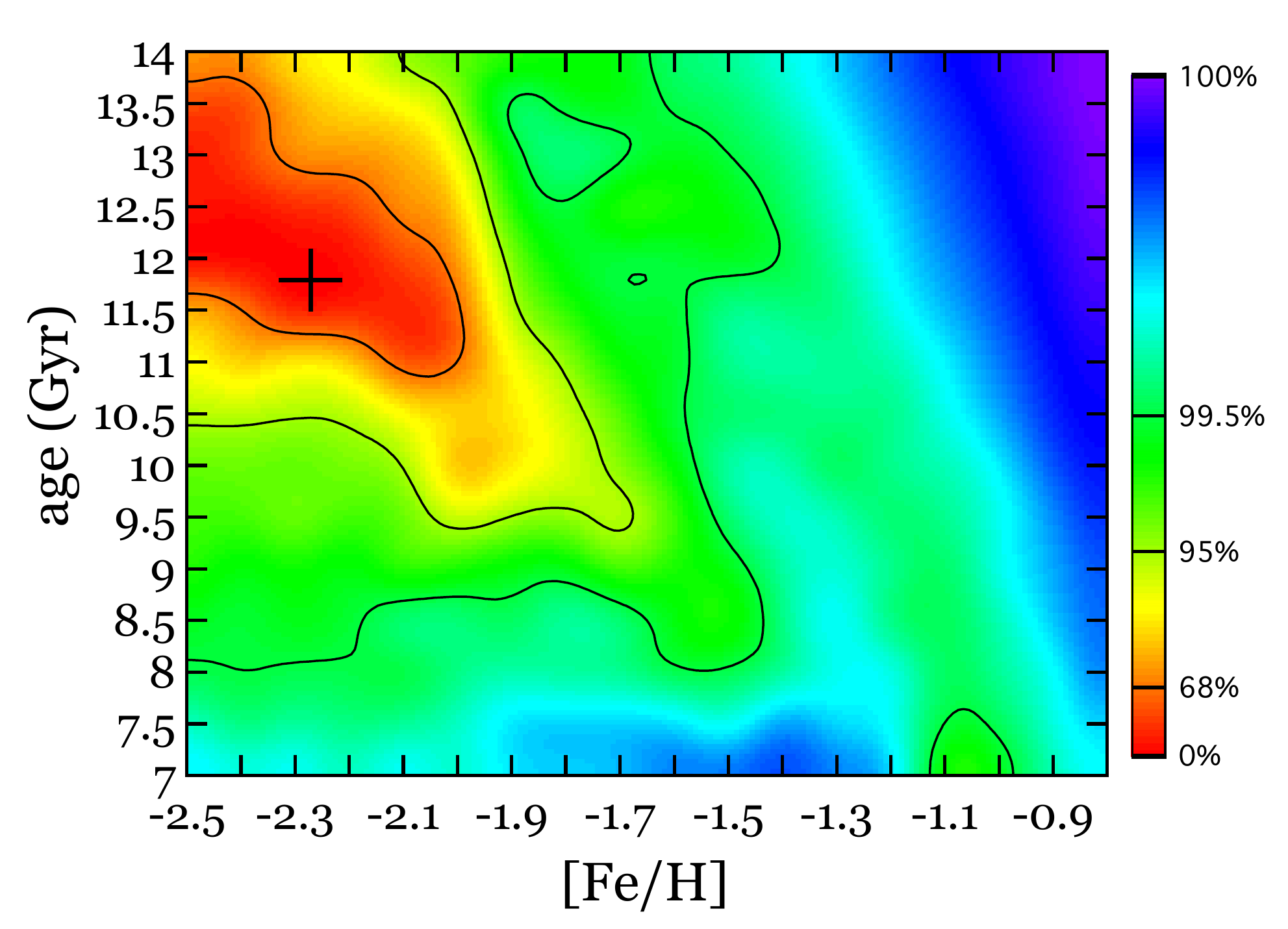}
\caption{Smoothed maximum likelihood density map in age-metallicity space for all stars within the 90\% ellipse around Pic I. Contour lines show the 68\%, 95\%, and 99\% confidence levels. The diagonal flow of the contour lines reflects the
age-metallicity degeneracy inherent to such an isochrone fitting procedure. The 1D marginalized parameters 
around the best fit with uncertainties are listed in Table~\ref{tab:PicI_parameters}.\label{fig:PicI_age_metal}}
\end{center}
\end{figure}

\subsection{Structural Parameters}\label{sec:PicIstruct}
Figure \ref{fig:PicIstellar_distribution} presents the on-sky distribution and density map of the Pic I stars selected based on their proximity to the best-fitting isochrone. Overall, it is a well defined system with a mildly elliptical distribution. The inner ellipse of the best-fitting 2D-model has a semi-major axis length of 1$r_h$ and contains 50\% of the Pic I stellar population, while the outer ellipse has a semi-major axis length of 2.32$r_h$ and contains 90\% of the Pic I stars. To highlight bright foreground stars and background galaxies, sources from the {\sc AllWISE} survey are overplotted on the field (open red circles) where the size of the circle is proportional to the apparent magnitude of the object.

Figures~\ref{fig:PicIrad_profile} shows the star number density profile of Pic I, where $r_e$ is the elliptical radius. Overplotted are the best-fitting Exponential profile (black dotted) using the modal values from the ML analysis. The error bars were derived from Poisson statistics. We derived a position angle $\theta=75^\circ\pm3^\circ$, an ellipticity $\epsilon=0.21^{+0.04}_{-0.06}$ and half-light radius of $r_h=22.8^{+4}_{-3}$\,pc. 
The values for the structure parameters are listed in Table\,\ref{tab:PicI_parameters}.

\subsection{Stellar Population}\label{sec:PicIpop}
Figure~\ref{fig:cmdiso_PicI} (left panel) shows the colour-magnitude diagram of all stars within $2.32r_h$ of the centre of Pic I. The middle panel, as per Figure~\ref{fig:cmdiso_HorI}, is the CMD of field stars outside the 90\,percent ellipse, covering the same area. The right panel shows the Hess diagram for the foreground-corrected Pic I CMD with the best-fitting Dartmouth isochrone superimposed. We note that the statistical decontamination was performed the same way as for Hor I.

Figure~\ref{fig:PicI_age_metal} shows the smoothed maximum likelihood density map of the age-metallicity space and the location of the best-fit is highlighted with a cross. The stars used to generate this map were selected from inside the 90\% ellipse as seen in Figure~\ref{fig:PicIstellar_distribution}. Similar to Hor I, Pic I consists of an old (11.8\,Gyr), metal-poor ([Fe/H]$=-2.28$) stellar population with [$\alpha$/Fe$]_{\rm avg}=+0.2$. Pic I is at a heliocentric distance of 110\,kpc ($m-M=20.20$\,mag). 
Based on its CMD, Pic I appears to be a less luminous system than Hor I. 

\subsection{Luminosity Function and Total Luminosity}\label{sec:PicILum}

\begin{figure}
\begin{center}
\includegraphics[width=0.9\hsize]{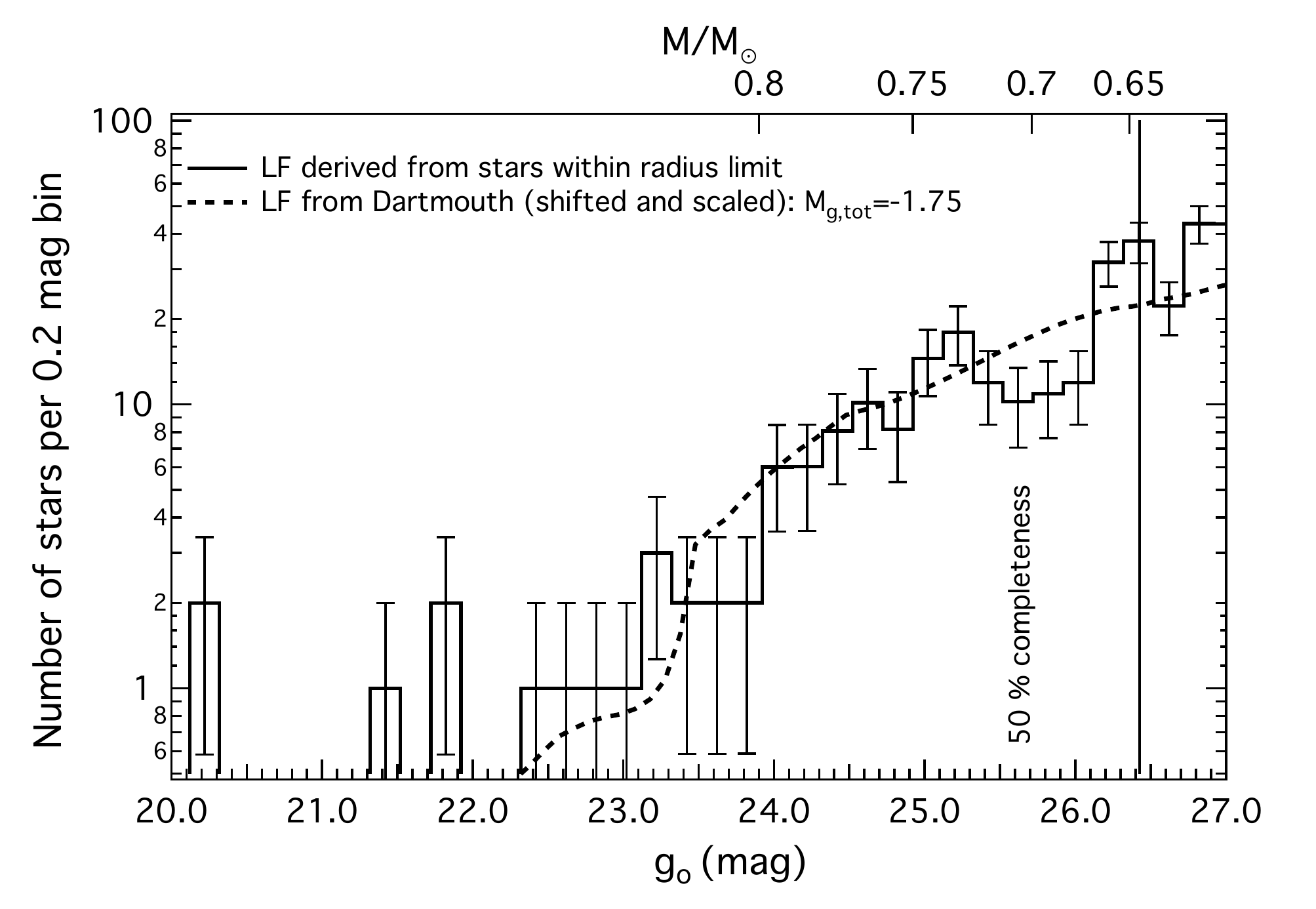}\hspace{-0.2cm}
\caption{Completeness-corrected Pic I luminosity function of all stars that are within the isochrone mask and within the 90\% ellipse (histogram). The best-fitting Dartmouth model luminosity function shifted by the distance modulus $20.20$\,mag and scaled to a total luminosity of $M_g=-1.75$\,mag is overplotted (dashed line).}.
 \label{fig:LF_PicI} 
\end{center}
\end{figure}

The total luminosity of Pic I has been derived in the same manner as Hor I ($\S$\ref{sec:HorILum}) and presented in Figure~\ref{fig:LF_PicI}. We calculated the integrated light by comparing the completeness-corrected observed LF with the Dartmouth model LF that corresponds to the best-fitting isochrone of 11.8\,Gyr, [Fe/H]$=-2.28$, and [$\alpha$/Fe]$=+0.2$. We measured a total $g$-band luminosity of $M_{g}=-1.75\pm0.2$. The integrated Dartmouth model LFs in $g$ and $V$ have a colour of $g-V=0.30$, which convert the $M_{g}$ magnitude into $M_{V}=-2.05$. 
For the same reasons as outlined in $\S$\ref{sec:HorIprop} a more realistic estimate for the uncertainty of the total luminosity of Pic I is $\sigma_{M_{V}}=0.50$. We also note that adding the extra flux of a single bright RGB candidate (see the LF in Figure 16) would increase the total absolute magnitude to $M_V=-2.40$\,mag, well within the quoted uncertainty. All derived parameters presented in this section are summarized in Table\,\ref{tab:PicI_parameters}.

\begin{table}
\caption{Derived properties and structural parameters of Pictor I, see $\S$\ref{sec:param_analysis} for details on the listed parameters.
\label{tab:PicI_parameters} }
{
\begin{center}
\begin{tabular}{l|c}
\hline
 & \bf{Pictor I} \\ \hline
$\alpha_0$\,(J2000) & $04^\mathrm{h}43^\mathrm{m}47\fs5\pm0\fs3$ \\
$\cdots$ & $70.9478\pm0.0012$\,deg\\
$\delta_0$\,(J2000) & $-50^\circ 17'\, 11'' \pm 5''$ \\
$\cdots$ & $-50.2864\pm0.0014$\,deg \\
$\theta$ (deg) & $75^{\circ}\pm 3^{\circ}$ \\
$\epsilon$ & $0.21^{+0.04}_{-0.06}$  \\
$r_h$ (arcmin) & $0.39^{+0.08}_{-0.06}$ \\
$N_*$  & $106\pm 16$ \\
$E(B-V)$ (mag) & 0.0116 \\       
$A_g$ & 0.044 \\
$A_r$ & 0.030\\
$(m-M)$ & $20.20\pm 0.08$  \\
$D_{\odot}$ (kpc) & $110\pm 4$ \\
$r_h$ (pc) & $12.9^{+0.3}_{-0.2}$   \\
age (Gyr) &   $11.8^{+0.9}_{-0.5}$  \\
$\langle [$Fe/H$] \rangle$ (dex) &  $-2.28^{+0.30}_{-0.25}$  \\
$[\alpha$/Fe$]_{\rm avg}$ (dex)&  $+0.2^{+0.1}_{-0.1}$ \\
$M_V$ (mag) & $-2.05\pm0.50$   \\
\hline
\end{tabular}
\end{center}
}
\end{table}

\section{Properties of Grus I}\label{sec:GruIprop}
Like the grouping of the newly discovered ultra-faint satellite candidates Tuc II, Tuc III, Tuc IV, Tuc V, Gru II, and Phe II, Gru I is situated on the connection line between LMC and SMC, approximately $\sim$46 degrees away from the LMC and $\sim$26 degrees from the SMC. At our derived heliocentric distance $D_\odot=115\pm6$ kpc, Gru I is around 94 kpc from the LMC and 69 kpc from the SMC. Given these relatively large distances it would appear that Gru I is an outer MW halo system unrelated to the Magellanic Clouds in the foreground. 

\begin{figure}
\begin{center} 
\includegraphics[width=1\hsize]{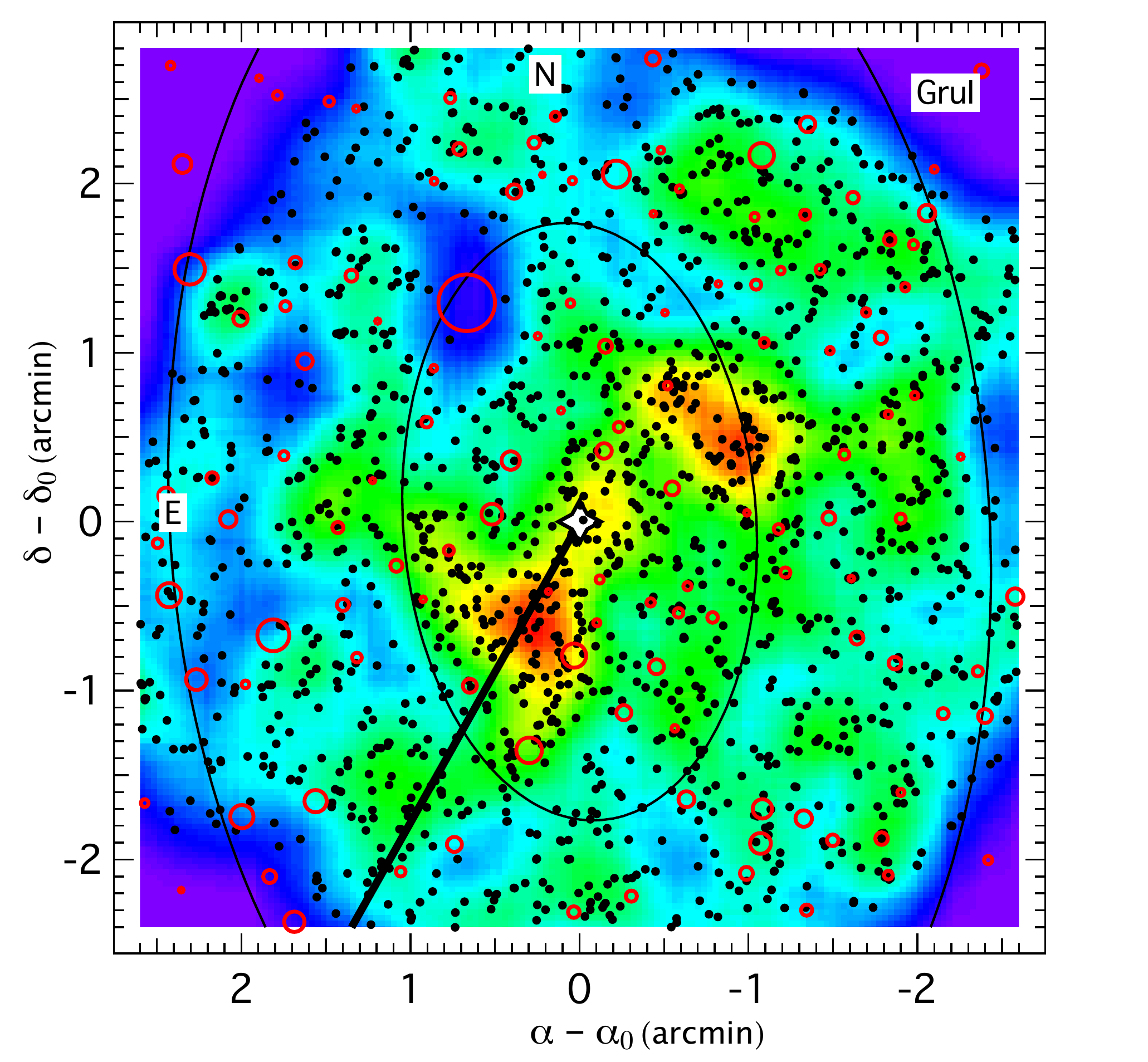}
\caption{The central region of Gru I contains two stellar over-densities on either side of the centre of mass (white star). The two ellipses have a position angle of $4^\circ$ and ellipticity of $\epsilon=0.41$ adopted from 
\citet{Koposov2015}. The semi-major axis lengths are $1r_h=1.77$\,arcmin and $2.3 r_h$, respectively, where $r_h$ was taken from \citet{Koposov2015}. The outer ellipse borders the region that contains 90 percent of the Gru I stellar population and is larger than the GMOS-S FoV. The open red circles are objects from the {\sc AllWISE} catalogue \citep{Wright2010,Cutri2013}, scaled to reflect their magnitudes. These objects highlight the position of the bright objects in the field both foreground stars and background galaxies. The solid line points in the direction of the LMC.}
\label{fig:GruIstellar_distribution}
\end{center}
\end{figure}

\begin{figure}
\begin{center} 
\includegraphics[width=0.495\hsize]{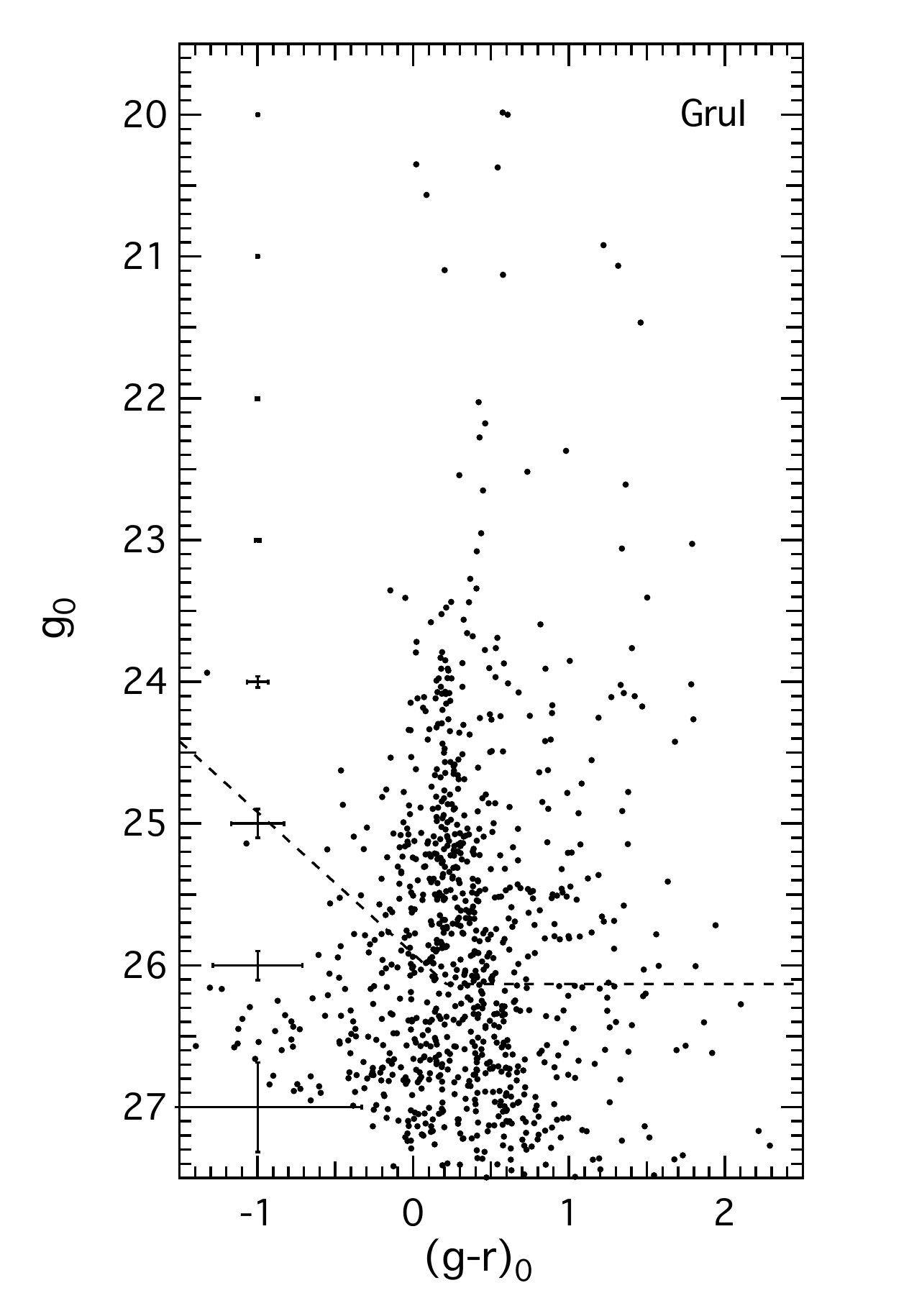}
\includegraphics[width=0.495\hsize]{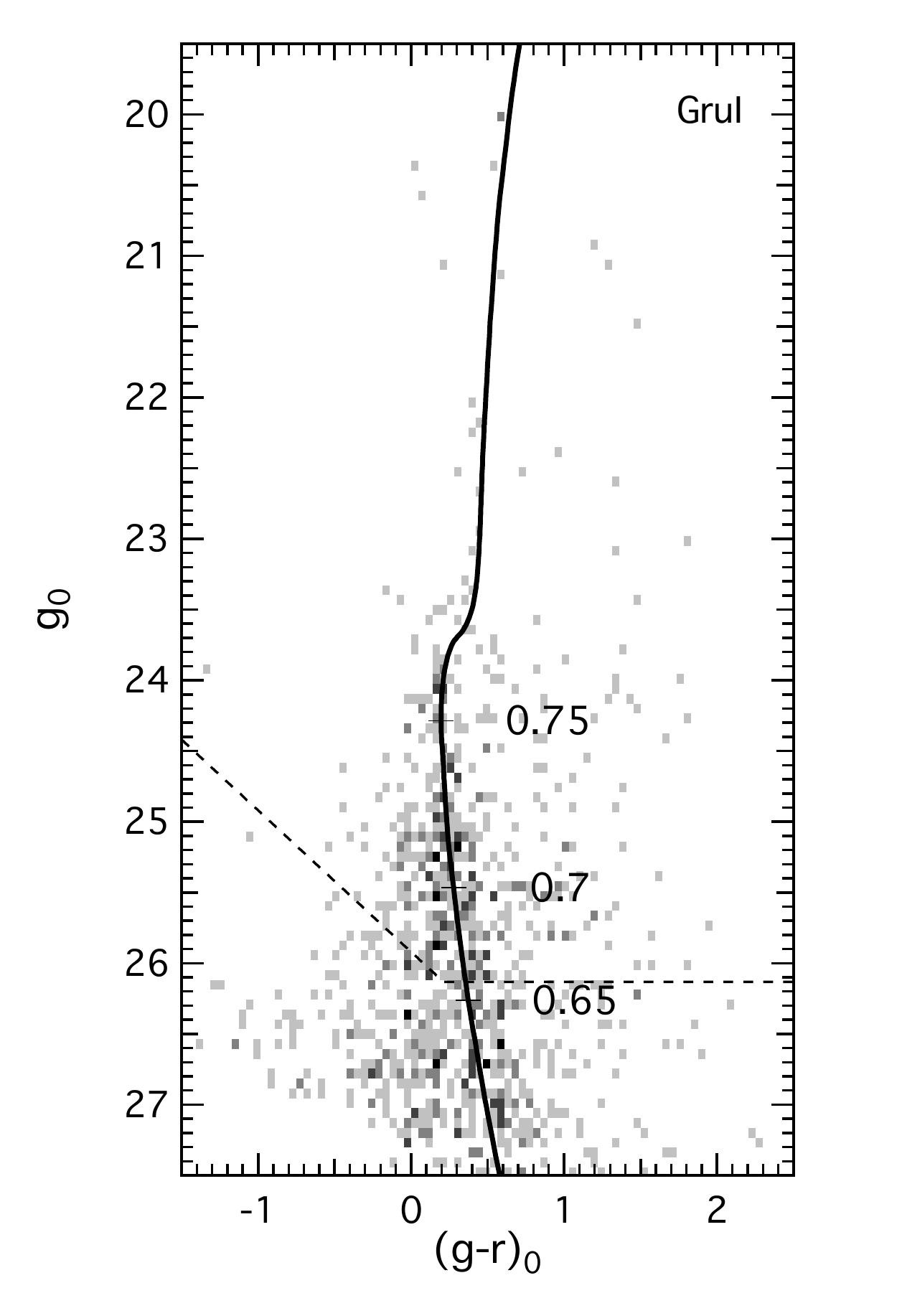}\\
\includegraphics[width=0.48\hsize]{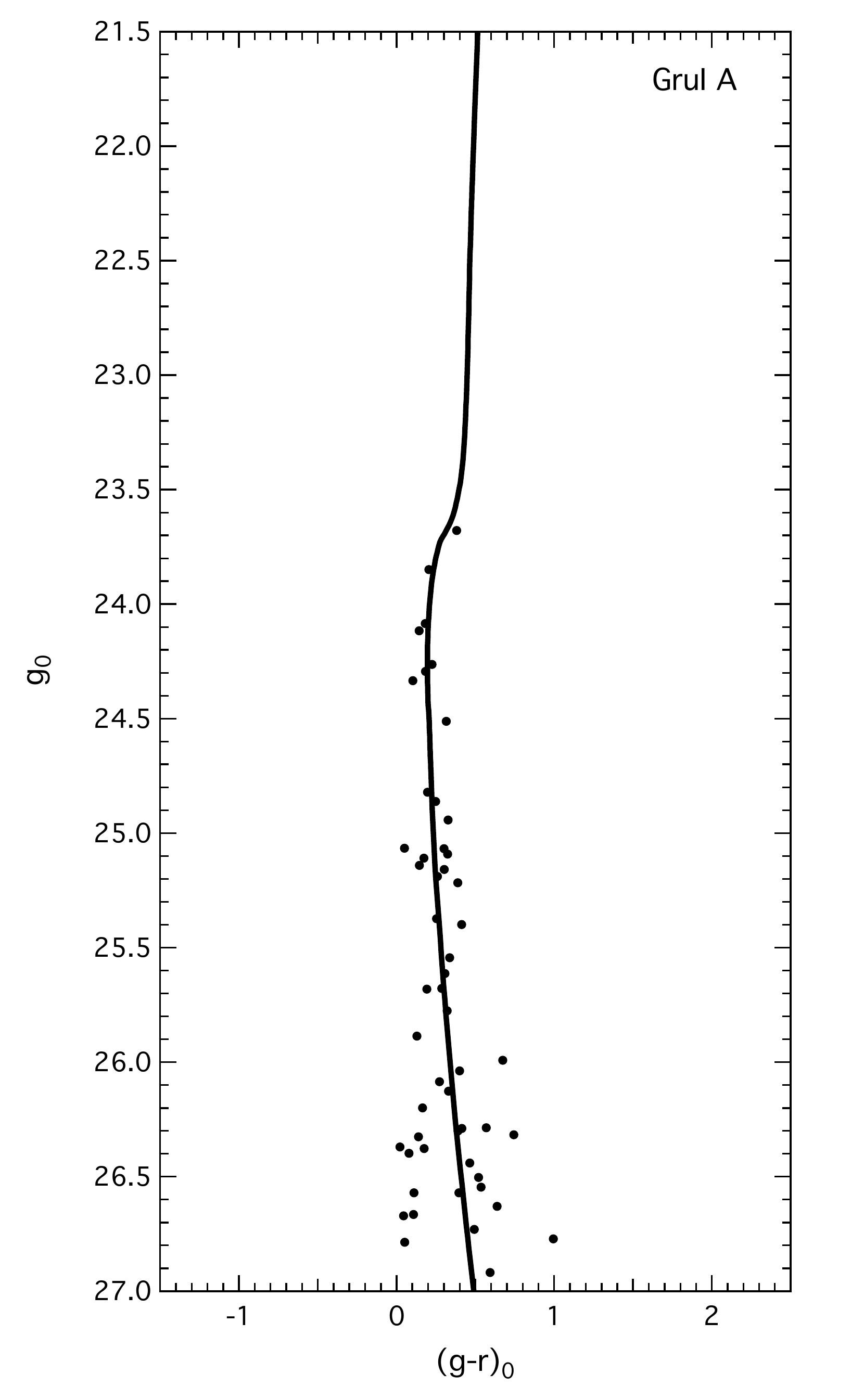}
\includegraphics[width=0.48\hsize]{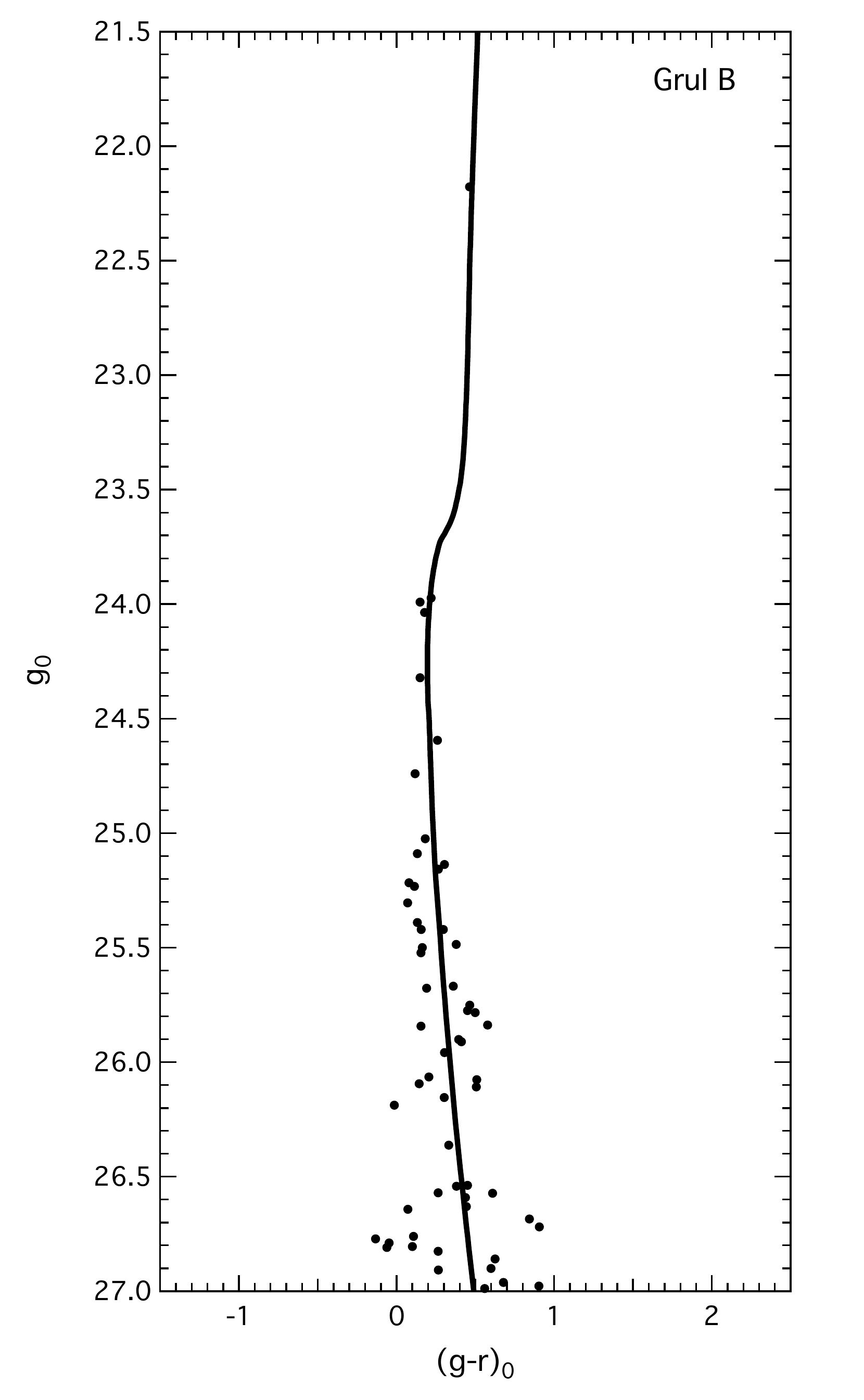}
\caption{{\it Top Left panel:} The colour-magnitude diagram of all stars within the $1r_h$ ellipse centred on the nominal celestial coordinates of Gru I shown in Figure\,\ref{fig:GruIstellar_distribution}. 
{\it Top Right panel:} Hess diagram of the CMD superimposed with the best-fitting Dartmouth isochrone.  The masses of main-sequence stars in solar mass units are marked to show the covered mass range. The dashed line represents the 50-percent completeness limit as determined with artificial star test and the MCMC method.
{\it Lower panels:} Colour-magnitude diagrams of the two stellar overdensities, as seen in Figure~\ref{fig:GruIstellar_distribution}, where Gru I{\sc a} (left) refers to overdensity at $[(\alpha - \alpha_0), (\delta -\delta_0)]\approx [+0.2,-0.5]$ and Gru I{\sc b} (right) refers to the overdensity at $[(\alpha - \alpha_0), (\delta -\delta_0)]\approx [-0.6,+0.8]$. The isochrone in both panels is the best-fitting isochrone from the top right panel.} \label{fig:cmdiso_GruI}
\end{center}
\end{figure}

\begin{figure}
\begin{center} 
\includegraphics[width=0.95\hsize]{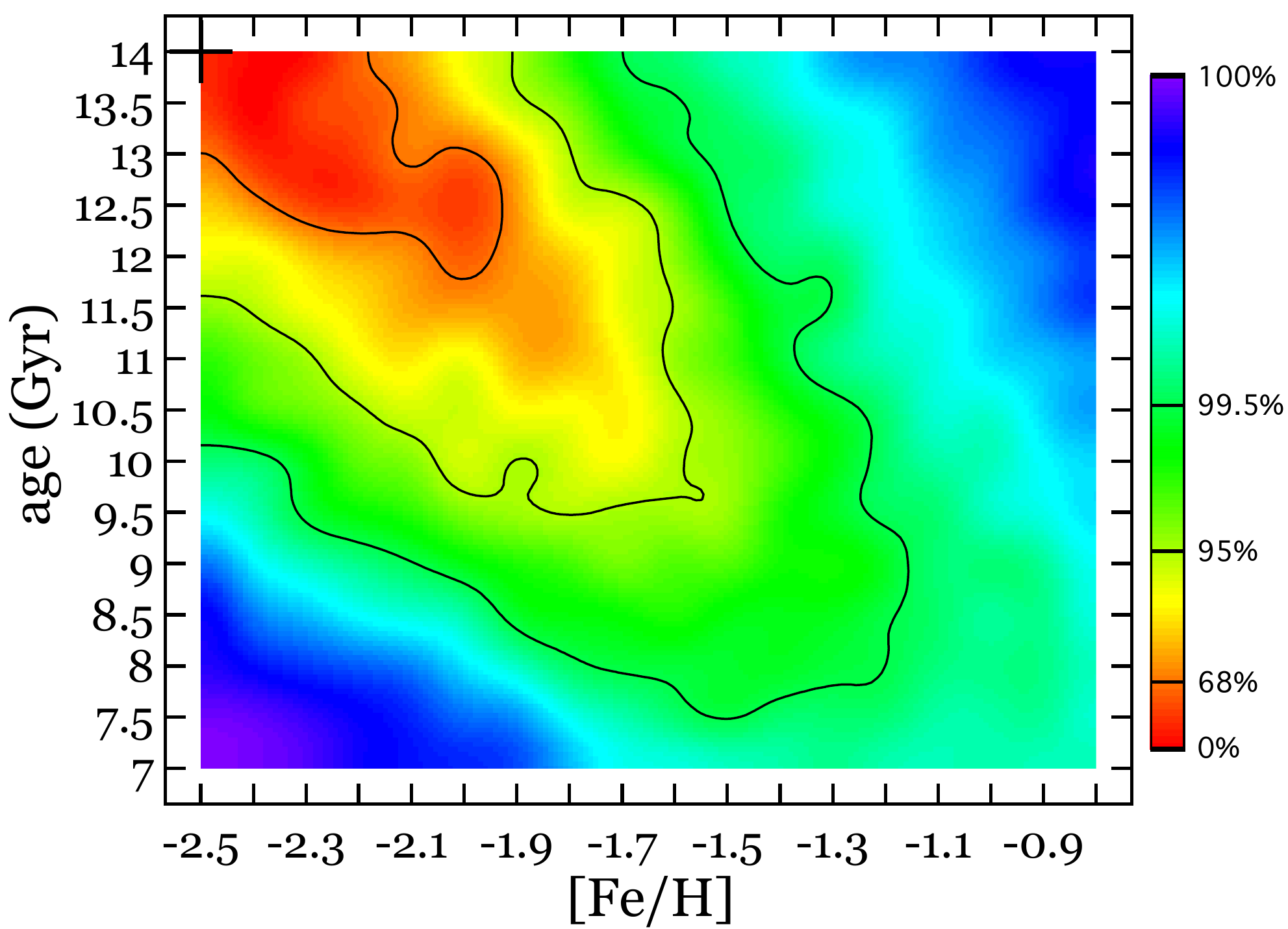}
\caption{Smoothed maximum likelihood density map in age-metallicity space for all stars within the 90\% ellipse around Gru I. Contour lines show the 68\%, 95\%, and 99\% confidence levels. The diagonal flow of the contour lines reflects the age-metallicity degeneracy inherent to such an isochrone fitting procedure. The 1D marginalized parameters around the best fit (cross) with uncertainties are listed in Table~\ref{tab:GruI_parameters}. The peak of the ML density map is located in the upper-left corner marked by a cross.}\label{fig:GruI_age_metal}
\end{center}
\end{figure}

\subsection{Structural Parameters}\label{sec:GruIstruct}
Figure \ref{fig:GruIstellar_distribution} shows the stellar distribution corresponding to the best-fit isochrone seen in the top right panel of Figure~\ref{fig:cmdiso_GruI}. The increased number statistics in the density map resolves the central region of Gru I and reveals that the system is not made up of a well defined central concentration, but rather consists of two smaller stellar overdensities that are distributed on either side of the centre of mass at $(\alpha - \alpha_0), (\delta -\delta_0)$=(0,0) in Figure \ref{fig:GruIstellar_distribution}. We calculated the centre of mass by allocating each Gru I star its stellar mass with the help of the best-fitting Dartmouth isochrone. The two substructures reside within a broader Gru I stellar population that extends beyond the GMOS-S field of view. The overdensities are spatially extended suggesting they may have been tidally distorted. The line drawn on Figure~\ref{fig:GruIstellar_distribution} points into the direction of the Large Magellanic Cloud. The values for the central coordinates $(\alpha_0,\delta_0)$, position angle ($\theta$), ellipticity ($\epsilon$), and half-light radius ($r_h$) given in Table~\ref{tab:GruI_parameters} are taken from \citet{Koposov2015}. The other parameters including the stellar properties have been determined as part of our analysis.

\subsection{Stellar population}
Grus I has a very clear stellar population visible in the GMOS-S field as shown in Figures~\ref{fig:cmd_field} and \ref{fig:cmdiso_GruI}, where there is a significant overdensity of main sequence stars below the turn-off region around $g_{\circ}>23.6$. Gru I has a sparsely populated but well-defined Red Giant Branch (RGB) and from the isochrone-fitting results (Figure~\ref{fig:GruI_age_metal}), it is an old and metal-poor stellar population. Indeed, given that the best-fitting isochrone is at the extreme limits of the age-metallicity range of the Dartmouth isochrone models, it is conceivable that Gru I is even more metal-poor than we are able to estimate here.

The central region of Gru I consists of two small stellar overdensities distributed on either side of the centre of mass.
The presence of these two overdensities could be due to Gru I having been tidally disrupted recently, or perhaps they are evidence of multiple stellar populations within this system. In the following, we will refer to the overdensity at $[(\alpha - \alpha_0), (\delta -\delta_0)]\approx [+0.2,-0.5]$ as Gru I{\sc a} and the overdensity at $\approx [-0.6,+0.8]$, as Gru I{\sc b}. The separation between the two cores is approximately 52 pc, with Gru I{\sc a} having physical dimensions of $\approx 22 \times 25$ pc and Gru I{\sc b} $\approx 13 \times 28$ pc as estimated from the yellow isodensity contours in Figure \ref{fig:GruIstellar_distribution}.  

Investigating the stellar populations of each component separately, we select all 67 Gru I{\sc a} stars inside the yellow isodensity contour and plot their CMD in Figure~\ref{fig:cmdiso_GruI} (lower left) and all 62 Gru I{\sc b} stars in Figure~\ref{fig:cmdiso_GruI} (lower right). The isochrone in each case is the best-fit isochrone for the entire Gru I population (Figure~\ref{fig:GruI_age_metal}) and it is well matched to these sub-regions. 

A possibility for the presence of these overdensities could be that  Gru I{\sc a} and Gru I{\sc b} represent two halves of a larger star formation event in Gru I that have maintained coherency through the lifetime of Gru I, perhaps through rotation support or that Gru I{\sc a} and Gru I{\sc b} were formed in response to recent external tidal forces exerted on the system forcing stars out of centre of mass and into these nascent tidal arms. The alignment of the features with the direction of the LMC (black line in Figure~\ref{fig:GruIstellar_distribution}) is worth noticing in this context and is suggestive of a tidal origin. That possibility will be considered further in the discussion ($\S$\ref{sec:discussion}).

\begin{table}
\caption{Properties and structural parameters of Gru I. The coordinates $(\alpha_0,\delta_0)$ are the centre of stellar mass. The values for the position angle ($\theta$), ellipticity ($\epsilon$), and half-light radius ($r_h$) were taken from \citet{Koposov2015}. See $\S$\ref{sec:param_analysis} for details on the listed parameters.
\label{tab:GruI_parameters}}
{
\begin{center}
\begin{tabular}{lc}
\hline
 & \bf{Gru I} \\ \hline
$\alpha_0$\,(J2000) & $22^\mathrm{h}56^\mathrm{m}40\fs8$  \\
$\cdots$ & $344.1700$\,deg  \\
$\delta_0$\,(J2000) & $-50^\circ 09'\, 51''$  \\
$\cdots$ & $-50.1641$\,deg  \\
$\theta$ (deg) & $4^{\circ}\pm60^{\circ}$ \\
$\epsilon$ & $0.41^{+0.20}_{-0.28}$  \\
$r_h$ (arcmin) & $1.77^{+0.085}_{-0.039}$ \\
$E(B-V)$ (mag) & 0.071 \\  
$A_g$ (mag) & 0.027\\
$A_r$ (mag) & 0.019\\
$(m-M)$ & $20.30\pm 0.11$  \\
$D_{\odot}$ (kpc) & $115\pm 6$ \\
age (Gyr) &   $14.0^{+1.0}_{-1.0}$  \\
$\langle [$Fe/H$] \rangle$ (dex) &  $-2.5^{+0.3}_{-0.3}$  \\
$[\alpha$/Fe$]_{\rm avg}$ (dex)&  $+0.2^{+0.1}_{-0.1}$ \\
\hline
\end{tabular}
\end{center}
}
\end{table}

\section{Properties of Phoenix II (DES J2339.9-5424)}\label{sec:PheIIProp}
Phe II is the closest of the four study objects to the SMC ($\sim$20 degrees) and is about 39 degrees from the LMC. With a reported heliocentric distance of $\sim$80 kpc, Phe II is approximately 32 kpc from the SMC and 54 kpc from the LMC. As can be seen in Figure~\ref{fig:MCs}, like Hor I, it resides in the outskirts of the HI gas component from the Magellanic Stream, raising the possibility that both of these systems might be part of the Magellanic Cloud satellite group. 

\begin{figure}
\begin{center} 
\includegraphics[width=0.95\hsize]{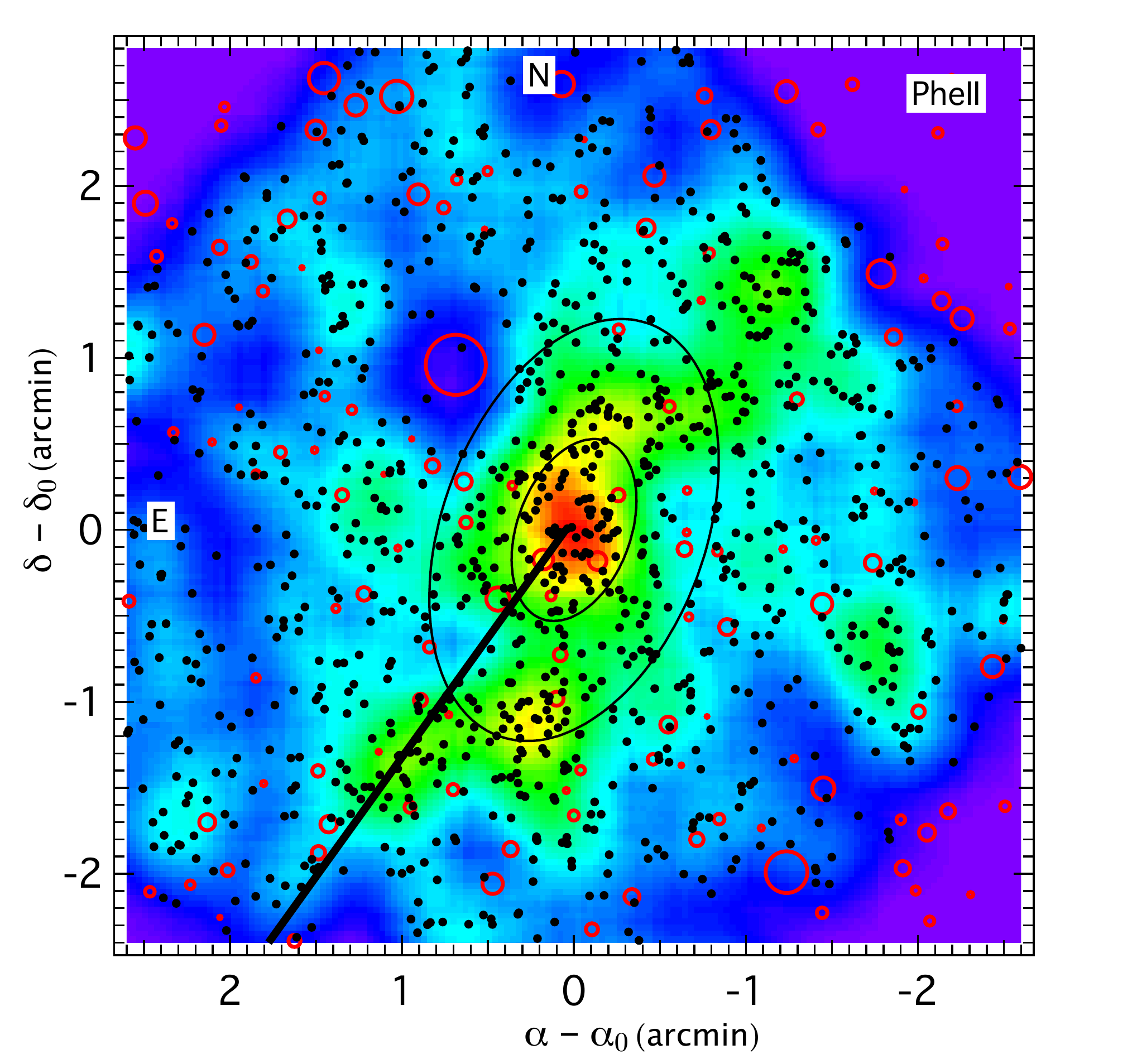}
\caption{Phe II's stellar density map shows clear evidence of tidal stripping with a nearly symmetrical and elongated, S-shaped distribution of stars extending on either side of a compact core. The two ellipses have an ellipticity of $\epsilon=0.45$, a position angle of $162^\circ$ and a semi-major axis length of $1r_h=0.52$\,arcmin and $2.32 r_h$, respectively, reflecting the properties of the core region. The outer ellipse borders the region that contains 90 percent of the Phe II stellar population, assuming an exponential radial profile. The open red circles are objects from the {\sc AllWISE} catalogue \citep{Wright2010,Cutri2013}, scaled to reflect their magnitudes. These objects highlight the position of the bright objects in the field both foreground stars and background galaxies. The solid line points in the direction of the LMC.}
\label{fig:PheIIstellar_distribution}
\end{center}
\end{figure}

\begin{figure}
\begin{center} 
\includegraphics[width=0.48\hsize]{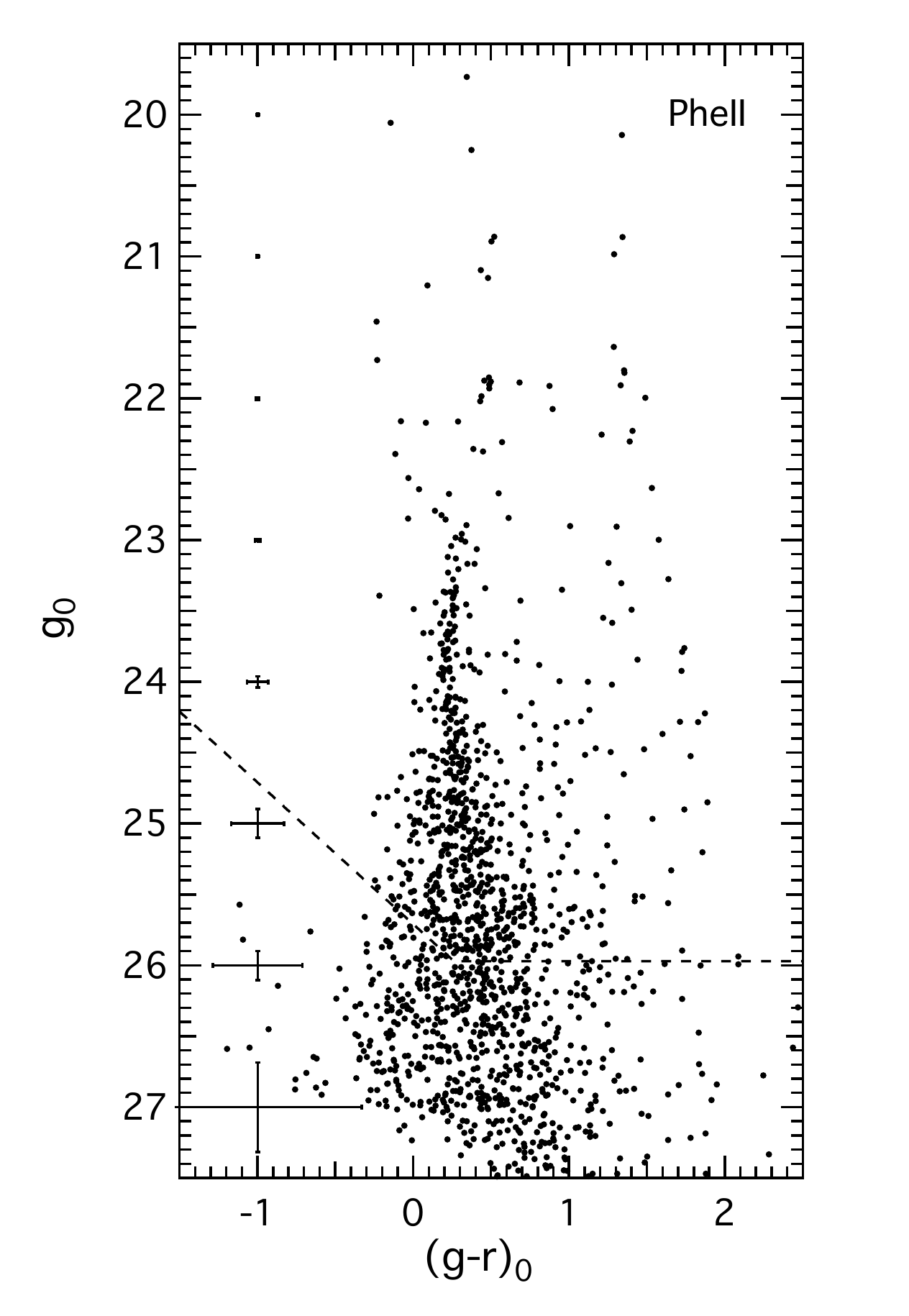}
\includegraphics[width=0.48\hsize]{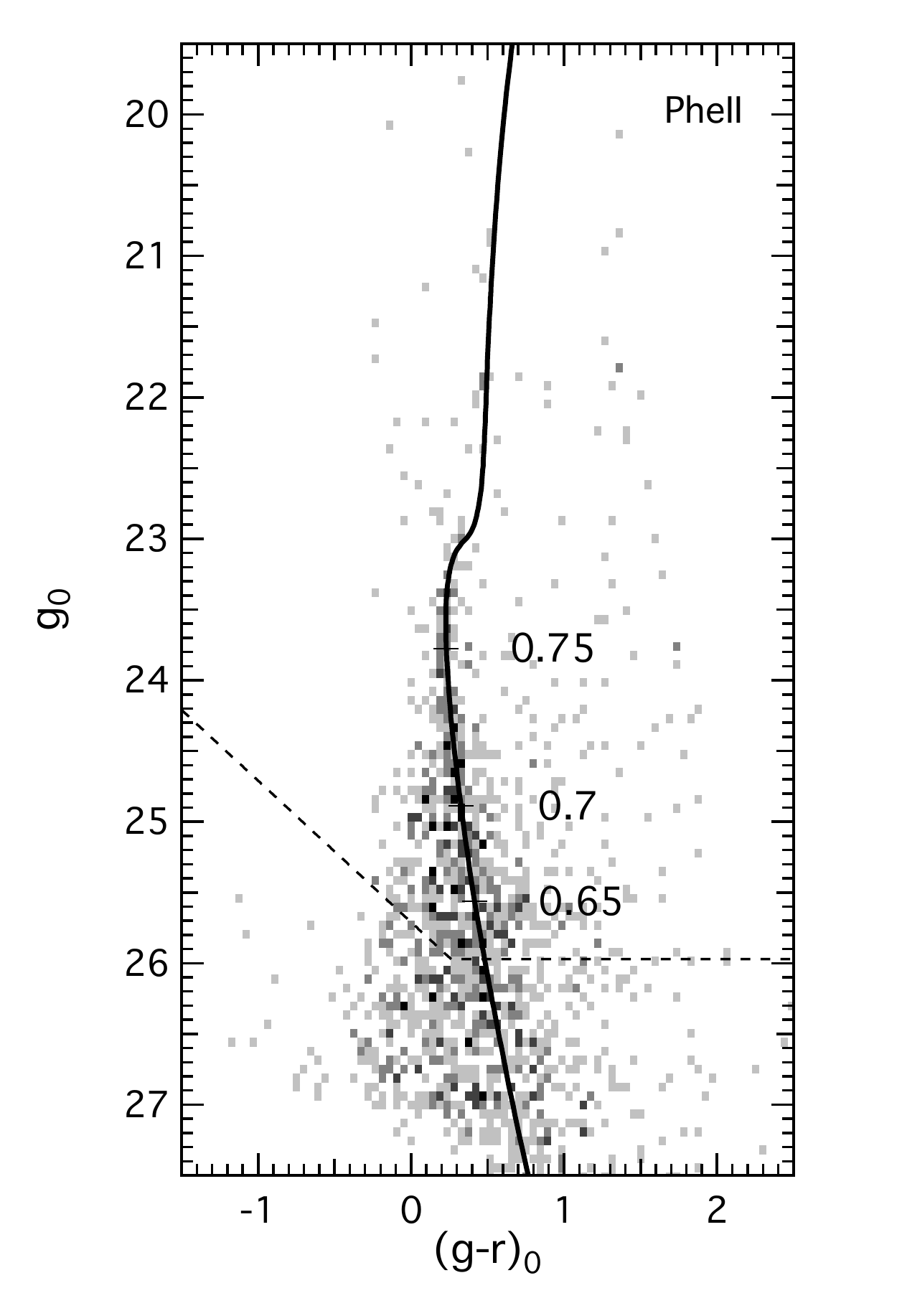}
\caption{{\it Left panel:} The colour-magnitude diagram of all stars within an elliptical radius of 2.2 arcmin centred on the nominal celestial coordinates of Phe II shown in Figure\,\ref{fig:PheIIstellar_distribution}. This ellipse contains the Phe II core region as well as its tidal arms. {\it Right panel:} Hess diagram of the CMD superimposed with the best-fitting Dartmouth isochrone.  No subtraction of foreground stars was conducted due to the complexity of the Phe II morphology. The masses of main-sequence stars in solar mass units are marked to show the covered mass range. The dashed line represents the 50-percent completeness limit as determined with artificial star test and the MCMC method.}
\label{fig:cmdiso_PheII}
\end{center}
\end{figure}

\begin{figure}
\begin{center} 
\includegraphics[width=1\hsize]{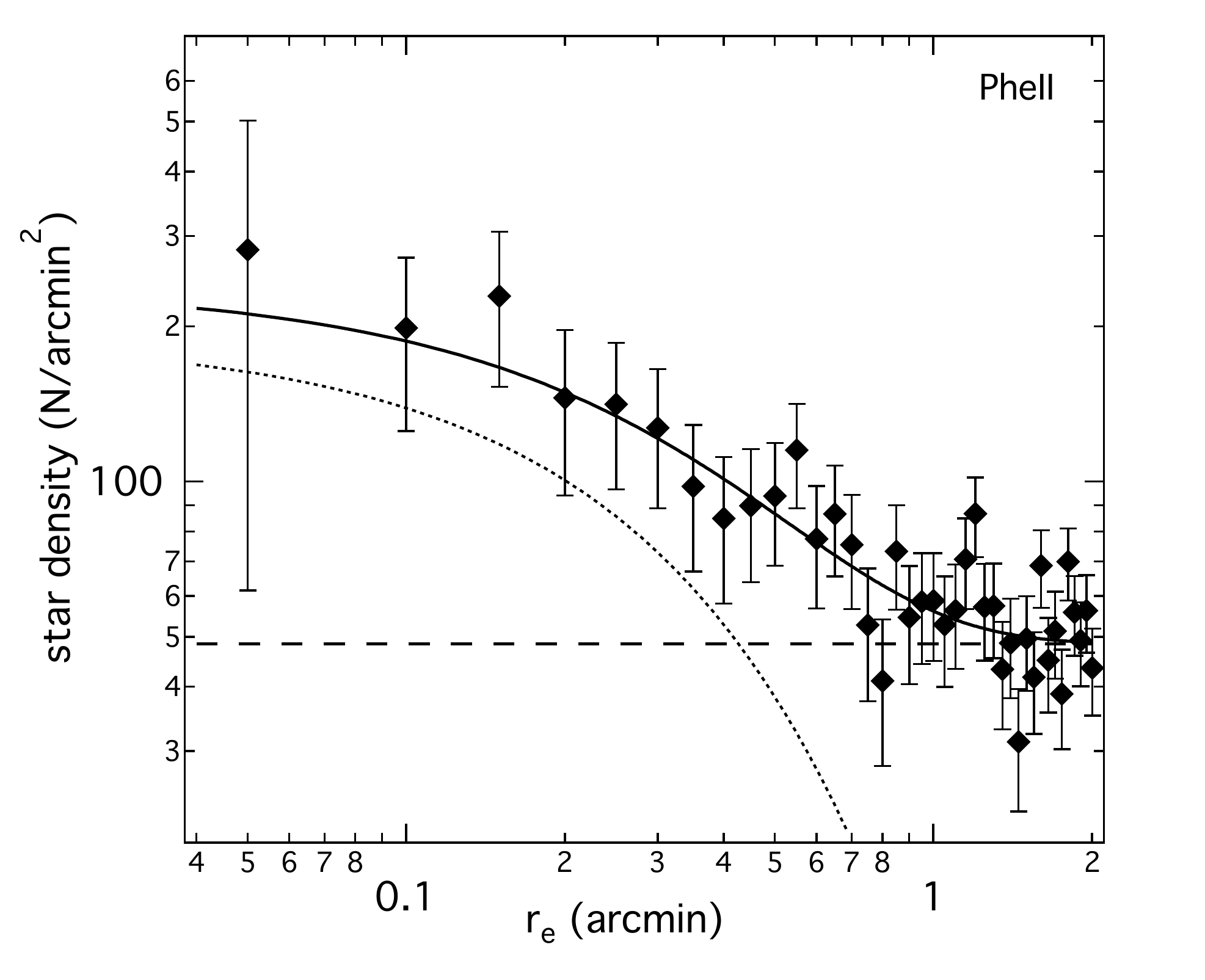}
\caption{Radial density profile of Phe II. The best-fitting Exponential profile (dotted line) is superimposed on the data points. The horizontal dashed line is the density of the foreground stars. The solid black line represents the profile + foreground. The central concentration of Phe II stars has a half-light radius of 0.52 arcmin. The error bars were derived from Poisson statistics.\label{fig:PheIIrad_profile}}
\end{center}
\end{figure}

\begin{figure}
\begin{center} 
\includegraphics[width=0.95\hsize]{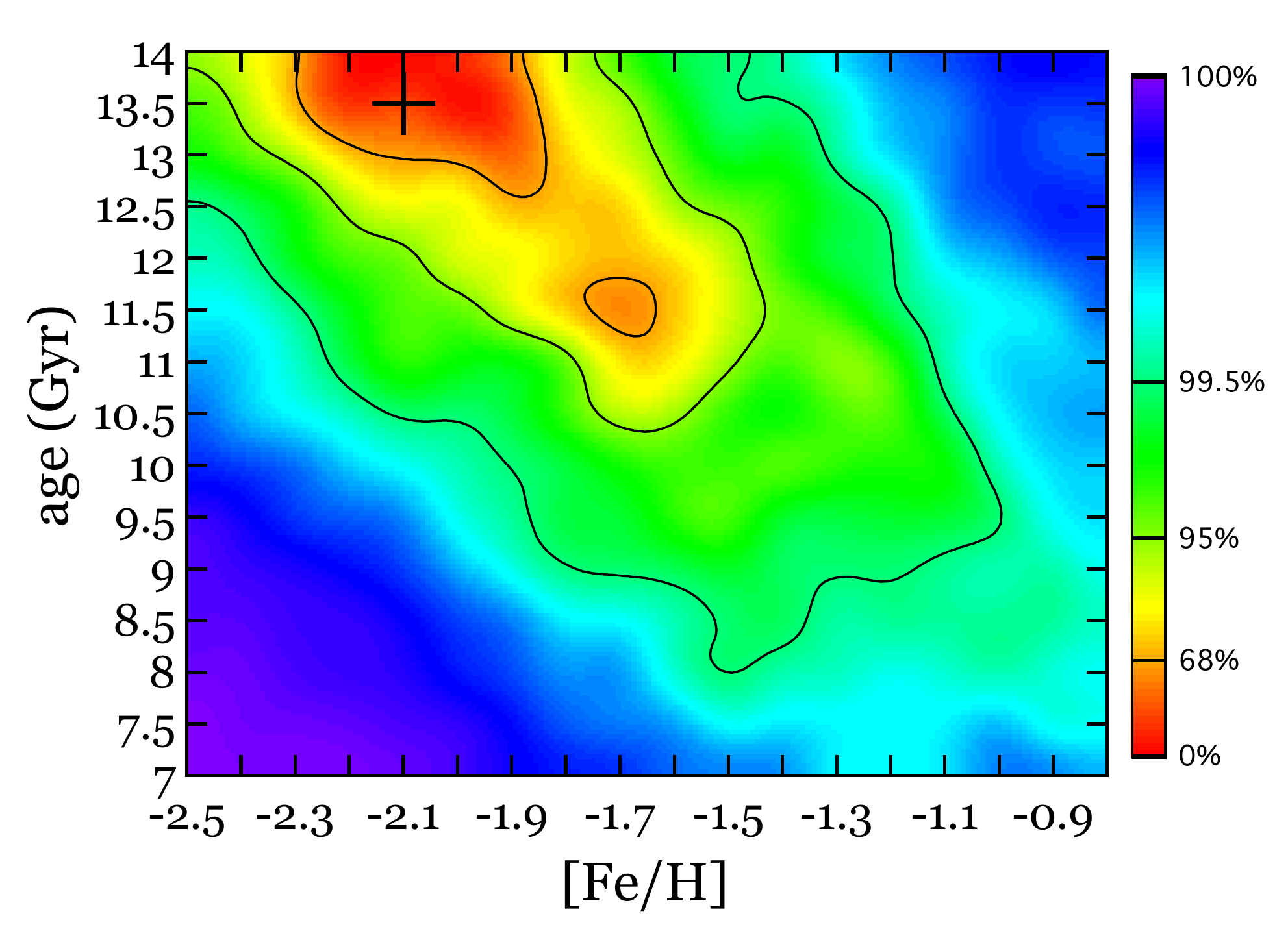}
\caption{Smoothed maximum likelihood density map in age-metallicity space for all stars within the 90\% ellipse around Phe II. Contour lines show the 68\%, 95\%, and 99\% confidence levels. The 1D marginalized parameters around the best fit (cross) with uncertainties are listed in Table~\ref{tab:PheII_parameters}.}\label{fig:PheII_age_metal}
\end{center}
\end{figure}

\subsection{Structural Parameters}\label{sec:PheIIstruct}
Phe II's stellar density map shows remarkably clear evidence of tidal stripping (Figure~\ref{fig:PheIIstellar_distribution}) with a nearly symmetrical and elongated distribution of stars  extending on either side of a compact centre reaching beyond the half-light radius and following the S-shaped characteristic of mass-loss found around tidally disturbed globular clusters like Pal 5 \citep{Rockosi2002, Odenkirchen2003, Grillmair2006}, NGC 5466 \citep{Belokurov2006}, Eridanus 3, Pal 15 \citep{Myeong2017} and NGC 7492 \citep{Navarrete2017}. As demonstrated by the black line in Figure~\ref{fig:PheIIstellar_distribution}, Phe II's tidal arms are well aligned with the direction of the Large Magellanic Cloud. Uncovering the complex morphology of Phe II's stellar structure demonstrates that the true size of Phe II is significantly smaller than the initial estimates. Its true half-light radius is $r_h = 12.6\pm2.5$ pc, less than half the size quoted in previous studies ($\approx 33$\,pc). Our distance estimate, on the other hand, is consistent with the literature where we find Phe II resides at $D_{\odot} = 81\pm5$ kpc. To avoid our ellipticity and position angle measurements being heavily influenced by the tidal arms, these estimates are derived from those stars inside the yellow isodensity contour line seen in Figure~\ref{fig:PheIIstellar_distribution}. Figure~\ref{fig:PheIIrad_profile} shows the radial profile of Phe II traced out to $r_e=2$\,arcmin. The shape of the profile demonstrates that the star counts decrease smoothly over this angular distance and reach the forground level only at $r_e\approx 1.6$\,arcmin. 

\subsection{Stellar Population}\label{sec:PheIIpop}
Figure~\ref{fig:cmdiso_PheII} presents the Phe II colour-magnitude diagrams for all stars within the best-fitting ellipse (Figure~\ref{fig:PheIIstellar_distribution})  on the left and the associated Hess diagram on the right. 
We note that no subtraction of foreground stars was conducted due to the complexity of the Phe II morphology. 
The foreground contamination is expected to be small, given the objects Galactic latitude of $b=-58^\circ.2$.
The best-fit isochrone is overlayed on the Hess diagram along with the locations of stars with stellar masses corresponding to 0.75 M$_\odot$, 0.7 M$_\odot$ and 0.65 M$_\odot$. The location of the best-fit isochrone can be seen in age-metallicity space in Figure~\ref{fig:PheII_age_metal}. Like other ultra-faint dwarf candidates, Phe II consists of an old and significantly metal-poor stellar population ([Fe/H]$= -2.10^{+0.25}_{-0.20}$) with an enhanced alpha abundance ([$\alpha$/Fe]$=+0.2\pm0.2$).

\subsection{Luminosity Function and Total Luminosity}\label{sec:PheIILum}

\begin{figure}
\begin{center}
\includegraphics[width=0.9\hsize]{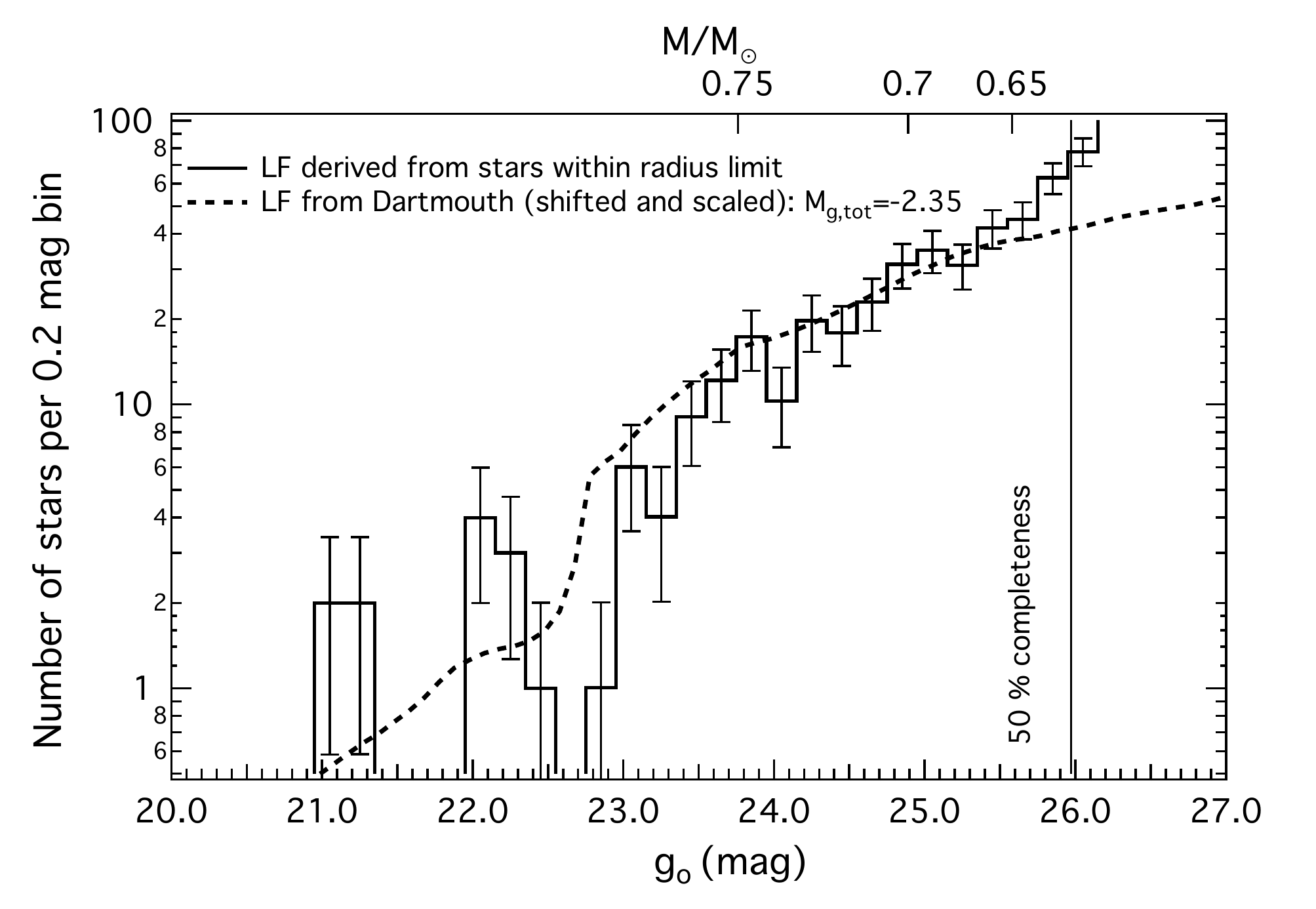}\hspace{-0.2cm}
\caption{Completeness-corrected Phe  II luminosity function of all stars that are within the isochrone mask and within the 90\% ellipse (histogram). The best-fitting Dartmouth model luminosity function shifted by the distance modulus $19.55$\,mag and scaled to a total luminosity of $M_g=-2.35$\,mag is overplotted (dashed line).}.
 \label{fig:LF_PheII} 
\end{center}
\end{figure}

The total luminosity of Phe II has been derived in the same manner as Hor I ($\S$\ref{sec:HorILum}) and presented in Figure~\ref{fig:LF_PheII}. We calculated the integrated light by comparing the completeness-corrected observed LF with the Dartmouth model LF that corresponds to the best-fitting isochrone of 13.5\,Gyr, [Fe/H]$=-2.10$, and [$\alpha$/Fe]$=+0.2$. We measured a total $g$-band luminosity of $M_{g}=-2.35\pm0.2$. The integrated Dartmouth model LFs in $g$ and $V$ have a colour of $g-V=0.29$, which convert the $M_{g}$ magnitude into $M_{V}=-2.74$. For the same reasons as outlined in $\S$\ref{sec:HorIprop} a more realistic estimate for the uncertainty of the total luminosity of Phe II is $\sigma_{M_{V}}=0.50$. All derived parameters presented in this section are summarized in Table\,\ref{tab:PheII_parameters}.

\begin{table}
\caption{Derived properties and structural parameters of Phe II, see $\S$\ref{sec:param_analysis} for details on the listed parameters. We note that the values quoted for $\theta$, $\epsilon$, and $r_h$ refer to the core region of Phe II as outlined by the yellow isodensity contours in Figure~\ref{fig:PheIIstellar_distribution}. $\dagger$ - values for central core.
\label{tab:PheII_parameters} }
{
\begin{center}
\begin{tabular}{l|c}
\hline
 & \bf{Phe II} \\ \hline\hline
$\alpha_0$\,(J2000)  &
$23^\mathrm{h}39^\mathrm{m}58\fs4$   \\
$\cdots$  & $354.9934$\,deg\\
$\delta_0$\,(J2000) & $-54^\circ 24'\, 08''$   \\
$\cdots$ & $-54.4024$\,deg \\
$\theta$ (deg)$^{\dagger}$ & $19$ \\
$\epsilon ^{\dagger}$ & $0.33$  \\
$r_h$ (arcmin)$^\dagger$ & $0.52\pm0.10$ \\
$r_h$ (pc)$^{\dagger}$ & $12.6\pm2.5$ \\
$N_*$  & $54\pm 7$ \\
$E(B-V)$ (mag) & 0.0103 \\  
$A_g$ & 0.039\\
$A_r$ & 0.027\\
$(m-M)$ & $19.55\pm 0.12$  \\
$D_{\odot}$ (kpc) & $81\pm 5$ \\
age (Gyr) &   $13.5^{+0.5}_{-0.5}$  \\
$\langle [$Fe/H$] \rangle$ (dex) &  $-2.10^{+0.25}_{-0.20}$  \\
$[\alpha$/Fe$]_{\rm avg}$ (dex)&  $+0.2^{+0.1}_{-0.1}$ \\
$M_V$ (mag) & $-2.74\pm0.50$   \\
\hline
\end{tabular}
\end{center}
}
\end{table}

\section{Discussion}\label{sec:discussion}

\subsection{Size-Luminosity and Luminosity-Metallicity relations}\label{sec:SLLZdiscuss}
\begin{figure}
\begin{center}
\includegraphics[width=\hsize]{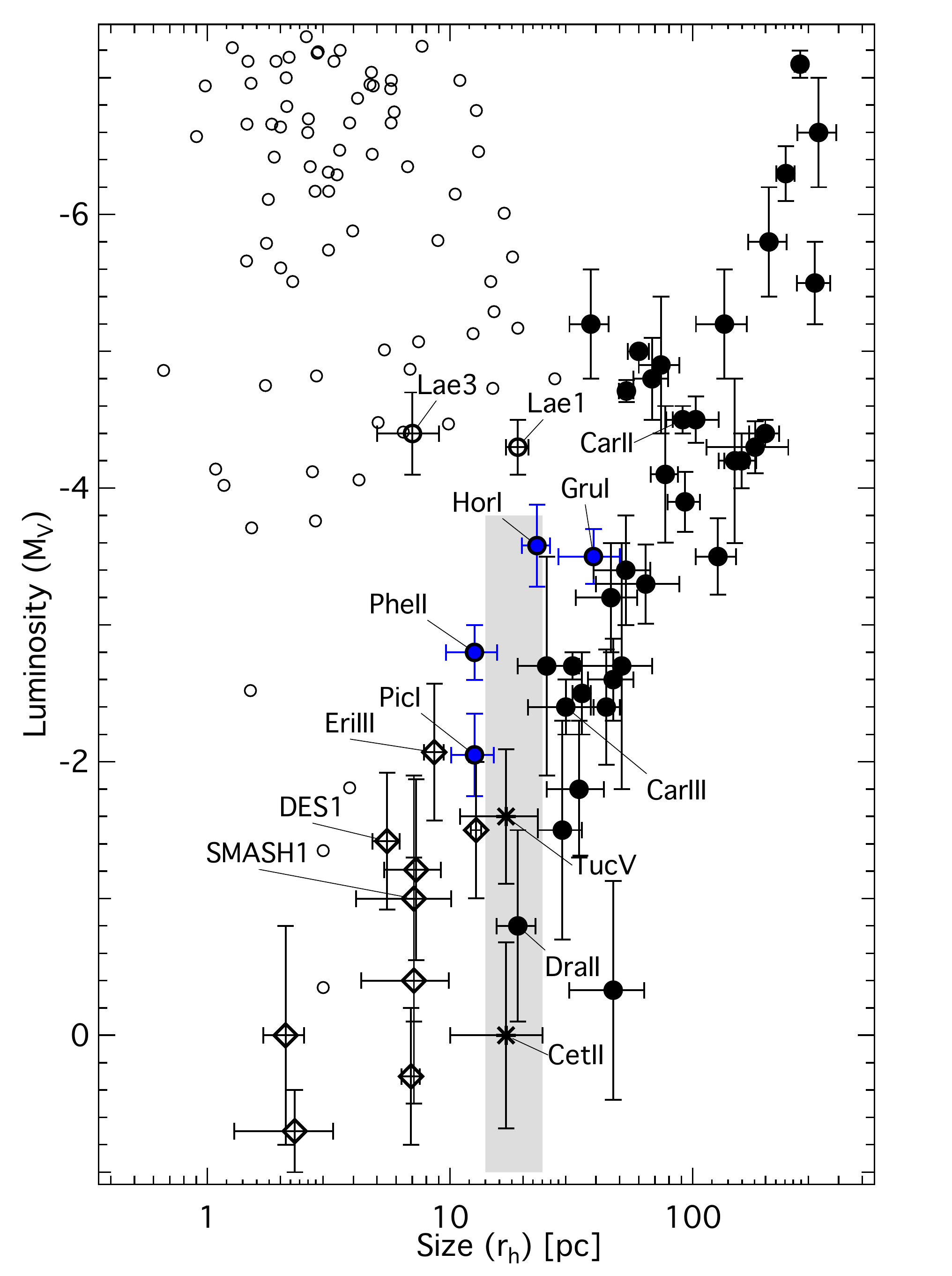}
\caption{ The size-luminosity (S-L) relation for classical Milky Way satellite galaxies (black filled circles) complemented with data for Galactic globular clusters [\citet{Harris1996}, open circles] and recently discovered Milky Way satellites. The grey bar is part of the S-L plane in which the unusual objects Tuc\,V and Cet II (marked as stars) reside, which we discuss in Paper I \& II as ``The Trough of UnCertainty" (TUC). Objects to the left of the TUC are plotted as diamonds.}
 \label{fig:SL-relation} 
\end{center}
\end{figure}

\begin{figure}
\begin{center}
\includegraphics[width=\hsize]{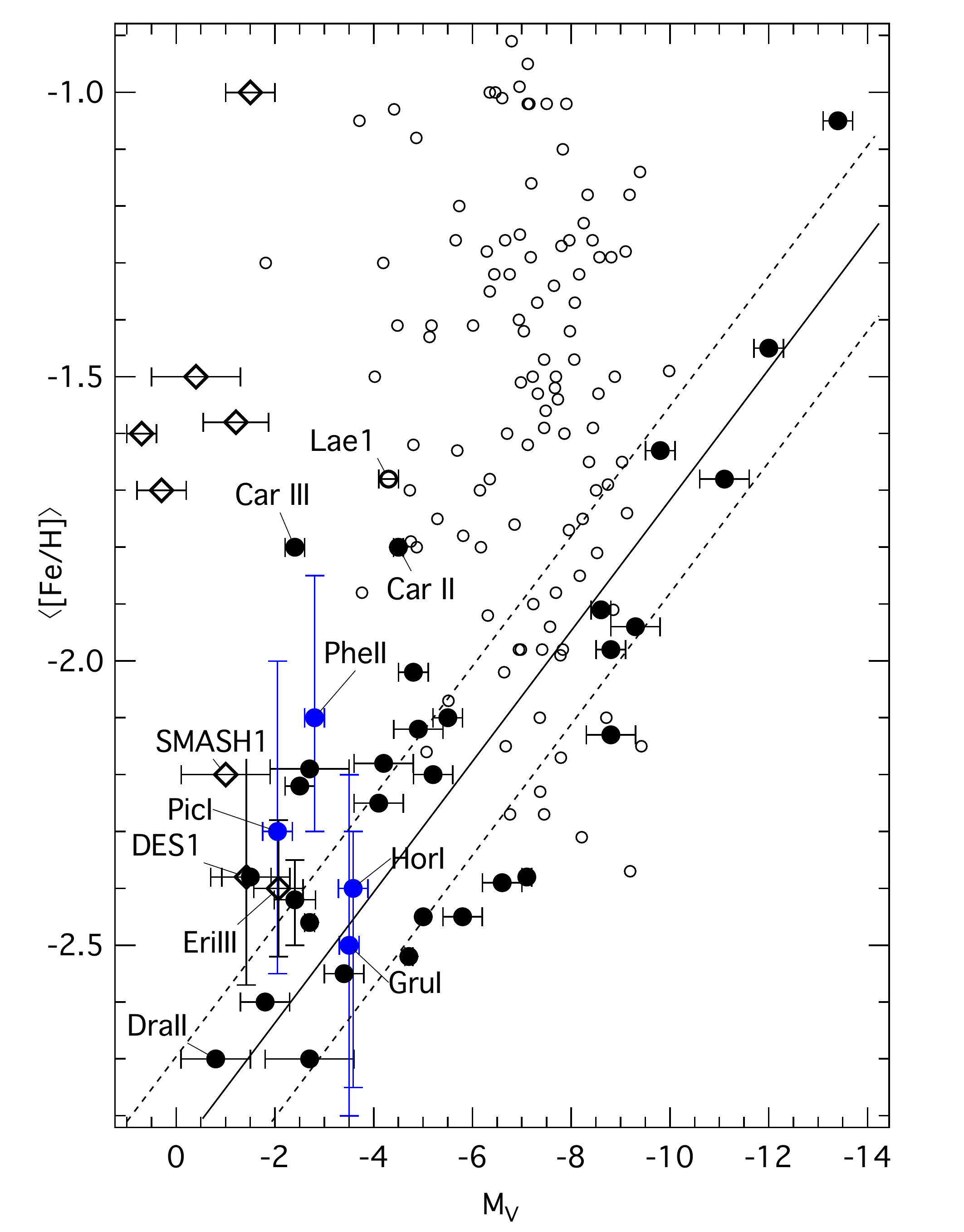}
\caption{ The luminosity-stellar metallicity plane for classical Milky Way satellite galaxies complemented with data for ultra-faint satellites from the literature and our study. Symbols are the same as in Figure~\ref{fig:SL-relation}. The black and two dotted lines represent the least-squares fit and $1\sigma$\,rms from ~\citet{Kirby2013}, based on spectroscopically studied stars in 14 dwarf galaxies (Segue\,2 was excluded). Phe II and Pic I do fall just outside of the 1$\sigma$ confidence band. Hor I and Grus I are found fully consistent with the LZ-relation for dwarfs. ADD DRACO II new parameters } 
 \label{fig:LZ-relation} 
\end{center}
\end{figure}

In Figures~\ref{fig:SL-relation} and \ref{fig:LZ-relation}, we present the updated size-luminosity and luminosity-metallicity planes for the Milky Way satellite population with the objects discussed in this paper shown in blue. Confirmed dwarf galaxies and ultra-faint candidates are shown as black filled circles and globular clusters as open circles. The "Trough of Uncertainty" (grey-shaded region in Fig.~\ref{fig:SL-relation}, TUC) from \citet{Conn2018a} highlights the locations of Tuc V and Cet II. Both objects were shown not to be ultra-faint dwarf galaxies but most likely SMC halo stars (Tuc V) or part of the Sagittarius tidal stream in the case of Cet II \citep{Conn2018a, Conn2018b}. Objects to the left of TUC, with radii typically smaller than $\sim10$\,pc, are exclusively star clusters and plotted as diamonds. Objects with half-light radii larger than $r_h\sim 20$\,pc are dwarf galaxies. Draco II remains the only object residing in the TUC region. It was classified as disrupting dwarf galaxy based on a recent study by \cite{Longeard18}.
A few ultra-faint systems are labeled with their names, including the four systems of our present study and the most recently discovered two neighbouring satellites Carina II \& III \citep{Torrealba2018}.  

In the luminosity-metallicity plane (Figure~\ref{fig:LZ-relation}) where we use the same symbols as in Figure~\ref{fig:SL-relation}, we find that the stellar systems with half-light radii to the RHS of the TUC generally follow the metallicity-luminosity relation for dwarf galaxies while those below the TUC do not. This hints at either the possibility of a size boundary limit for dwarf galaxies occurring on scales between $r_h \approx 10 - 20$ pc or this is the limit where tidal stripping and the associated mass-loss move these objects to the left of the LZ-relation.

Our derived size and luminosity of Hor I is largely consistent with \citet{Koposov2015} and \citet{Bechtol2015} keeping Hor I in the portion of the size-luminosity plane where the other ultra-faint dwarfs reside. It remains on the larger side of the gap opened up by Tuc V and Cet II when they were rejected as bona fide dwarf galaxies or star clusters in Papers I \& II. We do however measure a slightly higher value for the mean metallicity of the stellar population ($\langle [$Fe/H$] \rangle = -2.40^{+0.10}_{-0.35}$\,dex), which is statistically consistent with the spectroscopic metallicity originally reported in \cite[][$-2.76\pm0.10$\,dex]{Koposov2015b}. This refinement shifts Hor I closer onto the luminosity-metallicity relation for dwarf galaxies (Figure~\ref{fig:LZ-relation}). Given the new set of parameters for Hor I, we conclude that it is most likely an ultra-faint dwarf galaxy. 

Pic I has a half-light radius slightly larger but better constrained than that of Phe II (12.9$^{+0.3}_{-0.2}$ pc vs 12.6$\pm$2.5 pc). Compared to Gru I, it has roughly the same heliocentric distance (110$\pm$4 kpc vs 115$\pm$6 kpc) but is a significantly smaller and more compact system. It is slightly more metal-poor than Phe II ($-2.28^{+0.30}_{-0.25}$ vs $-2.10^{+0.25}_{-0.20}$) but more metal-rich than the other two satellites. This places Pic I off the dwarf galaxy luminosity-metallicity relation (Figure~\ref{fig:LZ-relation}) suggesting it is a star cluster. 

Due to its large angular size compared to the GMOS-S field, we were unable to accurately constrain a half-light radius for Gru I. However, the isochrone fitting provides a better distance estimate of $D_\odot=115\pm6$ kpc. Utilising the depth of the GMOS-S data, we find Gru I to have an extreme stellar population such that the age ($14.0\pm1.0$ Gyr) and metallicity ([Fe/H]$=-2.5\pm0.3$) are at the limits of the parameter space covered by the Dartmouth isochrone models (Figure~\ref{fig:GruI_age_metal}). As with the other objects, Gru I also has a raised alpha abundance ([$\alpha$/Fe]$=+0.2\pm0.1$). Like Hor I, we find Gru I to be consistent with the luminosity-metallicity relation for dwarf galaxies.

Phe II is perhaps the clearest example of an ultra-faint dwarf galaxy candidate losing mass through tidal stripping. It is thus not surprising that this system has a relative small half-light radius and its metallicity/luminosity are inconsistent with the LZ relation. Phe II resembles closely a tidally disturbed outer halo star cluster.

\subsection{Alignments of Tidal Features}\label{sec:Tidesdiscuss}
\begin{figure}
\begin{center} 
\includegraphics[width=1\hsize]{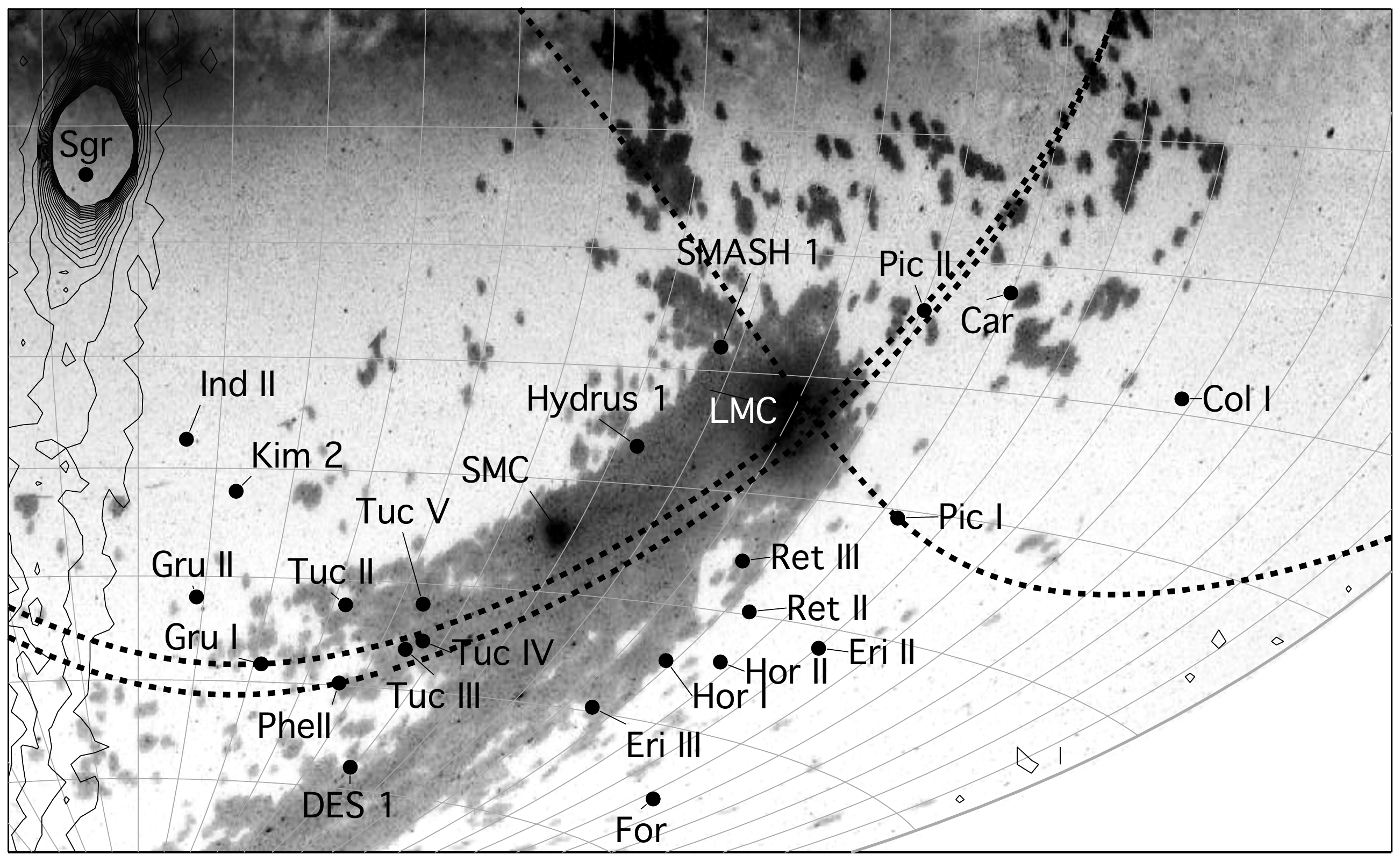}
\caption{Sky distribution of ultra-faint satellites around the Magellanic Clouds. The dotted lines are great circles aligned with the tidal features detected in Pic I, Gru I and Phe II. All three follow closely the direction of the Magellanic Stream and intersect at the LMC. This suggests that each of these objects is or was a satellite of the Large Magellanic Cloud.}
\label{fig:dwarf_orbit}
\end{center}
\end{figure}

As shown in Figures~\ref{fig:PicIstellar_distribution},~\ref{fig:GruIstellar_distribution} and \ref{fig:PheIIstellar_distribution}, there are extended substructures detected in the stellar distribution of Pic I, Gru I, and Phe II, which are roughly aligned with the direction of the Large Magellanic Cloud. In the case of Pic I, there a number of low-level overdensities which appear arranged to the north and south of the core. In Grus I, it is the alignment of the two sub-cores that reside symmetrically around the centre of mass and in Phe II, it is the S-shaped tidal arms. 
It seems these features have been formed by tidal forces from interactions with the LMC. To explore this possibility further, we take the precise alignment of these features and draw a great circle through them. For Pic I, we connect the great circle through the two symmetric overdensities located around $[(\alpha - \alpha_0), (\delta -\delta_0)]\approx [+0.6,-1.5]$ and $[(\alpha - \alpha_0), (\delta -\delta_0)]\approx [-0.3,-2.0]$. For Grus I, the great circle passes through the two sub-cores (Grus I{\sc a} and Grus I{\sc b}) and for Phe II, the great circle joins the overdense regions at the starting point of the s-shaped tidal tails (yellow isochrones), which are located at $[(\alpha - \alpha_0), (\delta -\delta_0)]\approx [+0.2,-1.2]$ and $[(\alpha - \alpha_0), (\delta -\delta_0)]\approx [-0.2,+0.6]$, respectively. 

Remarkably, despite each of these objects having heliocentric distances which far exceed that of the current location of the LMC, they all have extended stellar material which is pointing towards the LMC. This is highly suggestive that these objects are, or once were, members of the LMC satellite system and that the LMC has been actively disrupting those satellites. It seems we are witnessing the tail end of the hierarchical galaxy formation process, a satellites of satellites merger scenario whereby ultra-faint dwarf galaxy candidates are contributing material to the halo of a dwarf galaxy and now to our own Milky Way.

With the recent release of the Gaia DR2, \citet{Simon2018} finds that Hor I, along with Hydrus I \citep{Koposov2018}, Carina III \citep{Torrealba2018} and Tucana II \citep{Bechtol2015,Koposov2015} have orbits which suggest they are or were Magellanic Satellites. Curiously, of the four satellites presented in this paper, all of them appear likely to be LMC satellites, although Hor I is the only one which does not show obvious tidal features in its stellar distribution. This is somewhat surprising given that Hor I is the closest to the LMC of the four objects. \citet{Simon2018} report that Hor I is likely very close to its pericentric passage ($87^{+12}_{-15}$ kpc) which is further, but not inconsistent with our current distance estimate of $68\pm3$ kpc. \citet{Fritz2018} finds that the current velocity of Hor I does not exceed the escape velocity of the Milky Way assuming an NFW halo \citep{NFW1997} with a virial mass of $0.8\times 10^{12} M_\odot$ and with this information we can be sure that Hor I is now a bound satellite of the Milky Way. 

Interestingly, while Pic I does not have the signature tidal arms that are visible in Phe II, there are two lower significance overdensities that appear symmetrically around the core of Pic I that can be found at $[(\alpha - \alpha_0), (\delta -\delta_0)]\approx [+0.6,-1.5]$ and $[(\alpha - \alpha_0), (\delta -\delta_0)]\approx [-0.3,-2.0]$ in Figure~\ref{fig:PicIstellar_distribution}. 
Those two overdensities are aligned with the Large Magellanic Cloud suggesting that Pic I has also undergone some level of tidal disruption due to the influence of the LMC. Whether or not it is still a gravitationally bound satellite of the LMC is uncertain and exactly how much mass Pic I might have lost due to its interactions with the LMC is unclear. If we were to consider Pic I as a disrupting ultra-faint dwarf galaxy it would need to have been at least two magnitudes brighter (5 times more stellar mass) for it to align with the other dwarf galaxies on the luminosity-metallicity relation. At this stage, given Pic I is located on the size-luminosity relation with the other known star clusters it is more likely that Pic I is or was a star cluster of the LMC.

Gru I differs from most other known ultra-faint dwarf galaxy candidates in that it is structurally reasonably well defined but does not have a single central overdensity. As can be seen in Figure~\ref{fig:GruIstellar_distribution} two sub-cores reside symmetrically around the putative centre-of-mass of the system. Furthermore, the two sub-cores strongly align with the direction of the Large Magellanic Cloud suggesting that these features may have been tidally induced due to a close encounter with the LMC in the past. It is clear that the presence of these two sub-cores is not only responsible for the large error on the previous position angle estimates but they will also influence any estimate of the half-light radius.

As demonstrated in Figure~\ref{fig:dwarf_orbit}, Phe II joins Pic I and Gru I in having features which strongly align with the Large Magellanic Cloud. In this case, it is the line connecting the starting points of the tidal arms through the centre of the object. We propose that Phe II is or was a satellite of the LMC and that, like the other objects, it has contributed stellar material to the LMC halo. This picture based on morphological properties is consistent with the conclusion drawn by \citet{Fritz2018b} from Gaia DR2 astrometry and FLAMES/GIRAFFE spectroscopy that Phe II is the only object for which the firm case of a former association to the LMC can be made.

\section{Summary}\label{sec:conclusion}
We have presented the results from deep follow-up stellar photometry of the four ultra-faint dwarf galaxy candidates Horologium I, Grus I, Pictor I, and Phoenix II. Based on structural and stellar population parameters we found Pic I \& Phe II to be star clusters, while Hor I \& Gru I have properties similar to dwarf galaxies. All have very old, metal-poor stellar populations with elevated alpha abundances. We have refined all the distance estimates of these objects whereby Hor I is the closest with a heliocentric distance of $68\pm3$ kpc, followed by Phe II ($81\pm5$kpc), then Pic I is $110\pm4$ kpc, while Gru I is $115\pm6$ kpc.

Exploring the stellar populations further we find that Hor I consists of two stellar populations, as revealed by the presence of a split sub-giant branch. Interestingly, the two components, with a 70:30 percent ratio do not have the same on-sky distribution. A KS-test revealed  that the more populated and slightly more metal poor component is more concentrated than its weaker counterpart. Gru I despite having two sub-cores and no strong overdensity at the location of the putative centre-of-mass, does not appear to have two distinct stellar populations. The substructures present in this object are most likely the results of tidal interactions with the LMC. Pic I and Phe II, have well defined stellar populations that are consistent with a single star formation event.

With the exception of Hor I, each of the other objects show signs of tidal disruption. Phe II has prominant tidal arms, Gru I has a dual-core and Pic I has small overdensities outside of the half-light radius. The alignment of these substructures is not random, but all consistently point towards the LMC, strongly suggesting that each of them has, in the past, been bound and interacted with it. Hor I contains stars bright enough to be visible in the Gaia DR2 and has been found to be or have been a satellite of the LMC which makes all four of these objects members of the a larger LMC satellite system. 

As with other deep follow-up observations of newly discovered ultra-faint dwarf galaxy candidates, higher quality photometric and spectroscopic data reveal that these objects have hidden properties that were not visible in the discovery data. Each object has a rather rich and complex history that requires our understanding. As the only representatives of tail-end of the galaxy formation scenario, it is crucial that we have a complete picture of their true nature so that inferences made in terms of near field cosmology are supported by the best information. 

\section{Acknowledgements}
HJ and BCC acknowledge the support of the Australian Research Council through Discovery project DP150100862. The authors would like to thank Jennifer Miller of Gemini Observatory who generated the false-colour images of each object for us from the processed and stacked frames. 
This paper is based on observations obtained at the Gemini Observatory (GS-2016B-Q-7), which is operated by the Association of Universities for Research in Astronomy, Inc., under a cooperative agreement with the NSF on behalf of the Gemini partnership: the National Science Foundation (United States), the National Research Council (Canada), CONICYT (Chile), Ministerio de Ciencia, Tecnolog\'{i}a e Innovaci\'{o}n Productiva (Argentina), and Minist\'{e}rio da Ci\^{e}ncia, Tecnologia e Inova\c{c}\~{a}o (Brazil). 

This research has made use of: "Aladin sky atlas" developed at CDS, Strasbourg Observatory, France. \citep{2000A&AS..143...33B, 2014ASPC..485..277B}; the AAVSO Photometric All-Sky Survey (APASS), funded by the Robert Martin Ayers Sciences Fund; {\sc Topcat} in exploring and understanding this dataset \citep{2005ASPC..347...29T}; Astropy, a community-developed core Python package for Astronomy (Astropy Collaboration, 2013); SIMBAD database, operated at CDS, Strasbourg, France.

This project used public archival data from the Dark Energy Survey (DES). Funding for the DES Projects has been provided by the U.S. Department of Energy, the U.S. National Science Foundation, the Ministry of Science and Education of Spain, the Science and Technology Facilities Council of the United Kingdom, the Higher Education Funding Council for England, the National Center for Supercomputing Applications at the University of Illinois at Urbana-Champaign, the Kavli Institute of Cosmological Physics at the University of Chicago, the Center for Cosmology and Astro-Particle Physics at the Ohio State University, the Mitchell Institute for Fundamental Physics and Astronomy at Texas A\&M University, Financiadora de Estudos e Projetos, Funda\c{c}\~{a}o Carlos Chagas Filho de Amparo \`{a} Pesquisa do Estado do Rio de Janeiro, Conselho Nacional de Desenvolvimento Cient\'{i}fico e Tecnol\'{o}gico and the Minist\'{e}rio da Ci\^{e}ncia, Tecnologia e Inova\c{c}\~{a}o, the Deutsche Forschungsgemeinschaft and the Collaborating Institutions in the Dark Energy Survey. The Collaborating Institutions are Argonne National Laboratory, the University of California at Santa Cruz, the University of Cambridge, Centro de Investigaciones En\'{e}rgeticas, Medioambientales y Tecnol\'{o}gicas-Madrid, the University of Chicago, University College London, the DES-Brazil Consortium, the University of Edinburgh, the Eidgen\"{o}ssische Technische Hochschule (ETH) Z\"{u}rich, Fermi National Accelerator Laboratory, the University of Illinois at Urbana-Champaign, the Institut de Ci\`{e}ncies de l'Espai (IEEC/CSIC), the Institut de F\'{i}sica d'Altes Energies, Lawrence Berkeley National Laboratory, the Ludwig-Maximilians Universit\"{a}t M\"{u}nchen and the associated Excellence Cluster Universe, the University of Michigan, the National Optical Astronomy Observatory, the University of Nottingham, the Ohio State University, the University of Pennsylvania, the University of Portsmouth, SLAC National Accelerator Laboratory, Stanford University, the University of Sussex, and Texas A\&M University. 

This publication makes use of data products from the Wide-field Infrared Survey Explorer, which is a joint project of the University of California, Los Angeles, and the Jet Propulsion Laboratory/California Institute of Technology, funded by the National Aeronautics and Space Administration.

\appendix
\section{Unrecovered Artificial Stars} \label{App:AppendixA}
In this appendix, we present more of our findings from the artificial star experiments in each field. For a set of six magnitude bins, we show where in the frame we are unable to recover artificial stars. This allows the reader to see which parts of the field are more heavily affected by photometric completeness considerations. In many frames, the main sources of contamination in the field are the On Instrument Wave Front Sensor (OIWFS), chip gaps, hot pixel columns, bright stars and their halos.

\begin{figure*}
\begin{center} 
\includegraphics[width=0.99\hsize]{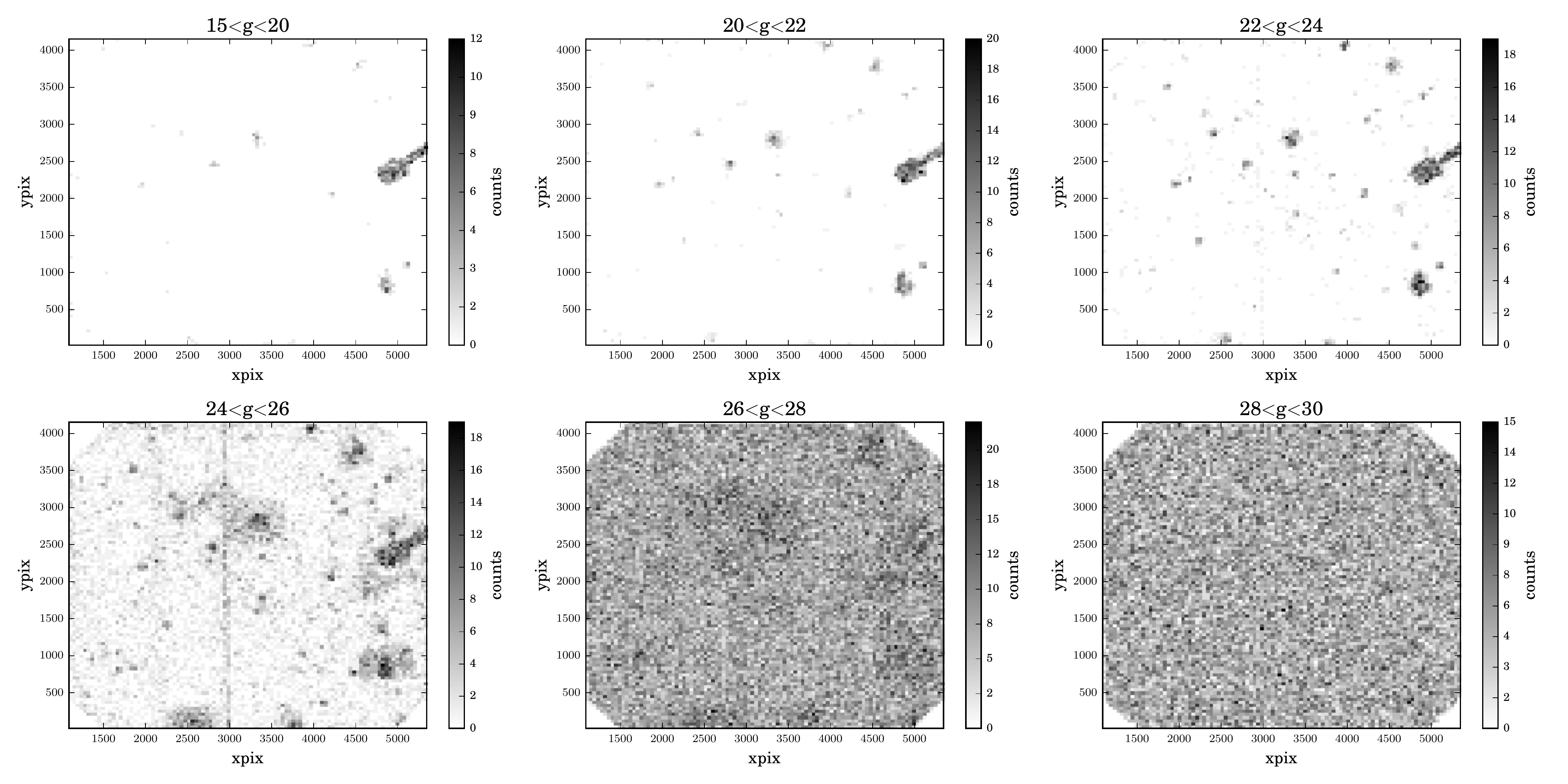}
\caption{Density map of unrecovered artificial stars in the Horologium I field. The magnitude binning is shown at the top of each panel. Initially, only the guide probe and the very bright stars are causing the artificial stars to be lost. As the artifical stars become fainter, they begin being affected by the halos of the bright stars and hot, dead or cold pixels. Eventually the magnitude of the stars begin to approach the background level and we see the field is essentially flat at the faint end, as expected. \label{fig:HorIUnRecoveredStars}}
\end{center}
\end{figure*}

\begin{figure*}
\begin{center} 
\includegraphics[width=0.99\hsize]{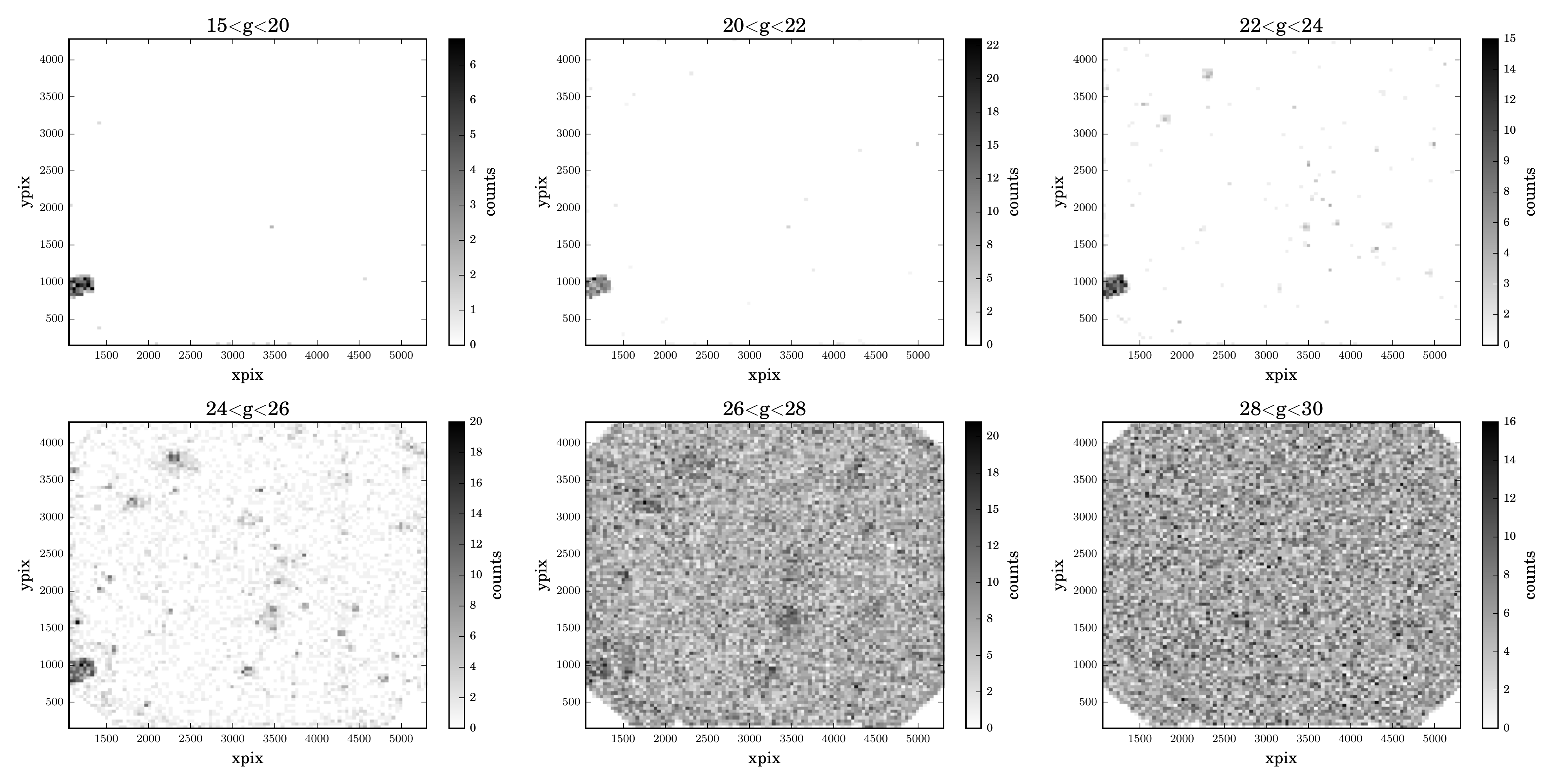}
\caption{Pic I Unrecovered Stars\label{fig:PicIUnRecoveredStars}}
\end{center}
\end{figure*}

\begin{figure*}
\begin{center} 
\includegraphics[width=0.99\hsize]{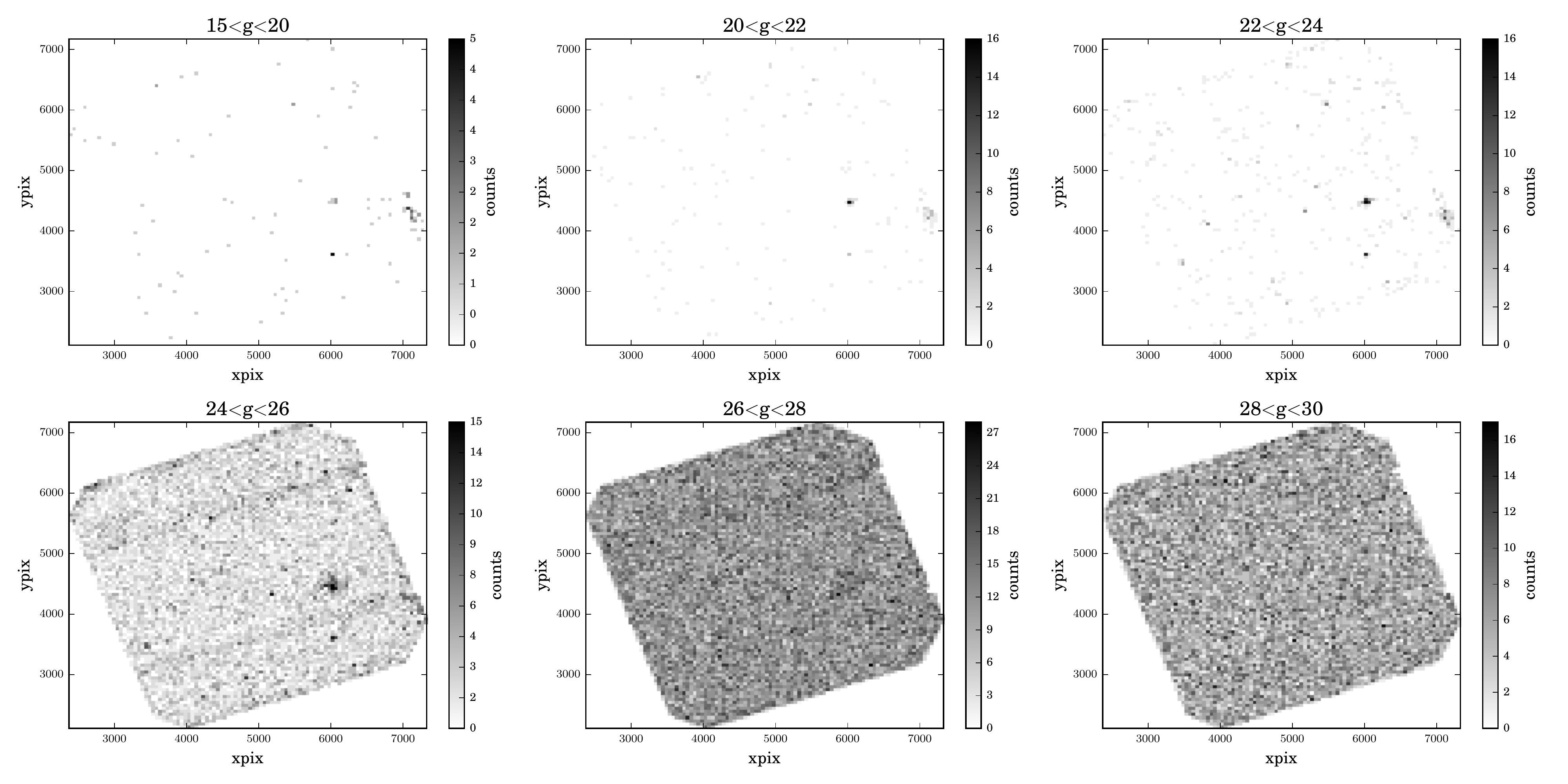}
\caption{Grus I Unrecovered Stars\label{fig:GrusIUnRecoveredStars}}
\end{center}
\end{figure*}

\begin{figure*}
\begin{center} 
\includegraphics[width=0.99\hsize]{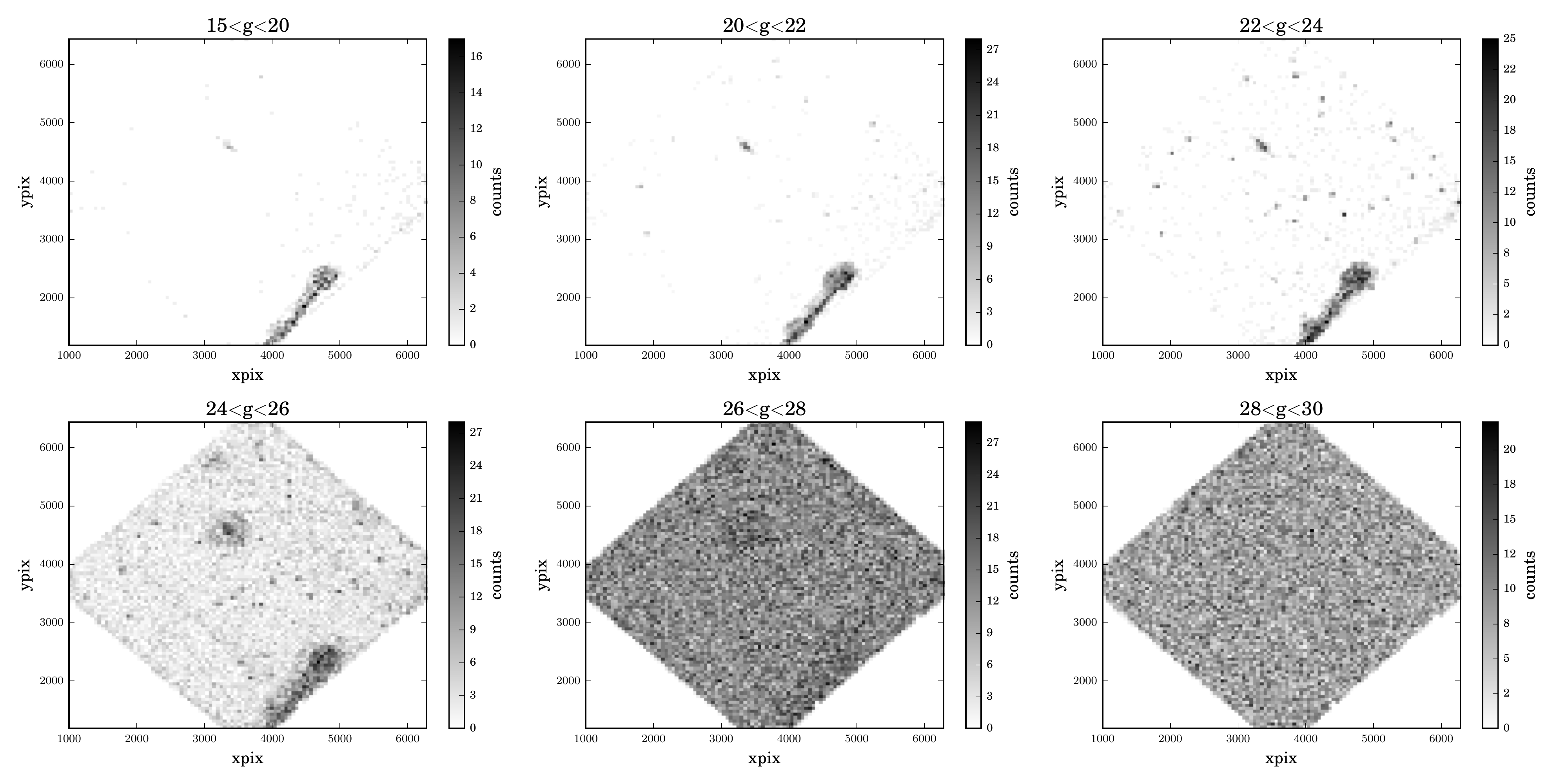}
\caption{Phe II Unrecovered Stars\label{fig:PheIIUnRecoveredStars}}
\end{center}
\end{figure*}


%



\label{lastpage}

\begin{thebibliography}{DUM}
\bibitem[Astropy Collaboration et al.(2013)]{Astropy2013} Astropy Collaboration, Robitaille, T.~P., Tollerud, E.~J., et al.\ 2013, \aap, 558, A33 
\bibitem[Baca{\"e}r (2011)]{Bacaer2011} Baca{\"e}r, N. 2011, Springer London, 35-39
\bibitem[Belokurov et al.(2006)]{Belokurov2006} Belokurov, V., Evans, N.~W., Irwin, M.~J., Hewett, P.~C., \& Wilkinson, M.~I.\ 2006, \apjl, 637, L29 
\bibitem[Balbinot et al.(2013)]{Balbinot2013} Balbinot, E., Santiago, B.~X., da Costa, L., et al.\ 2013, \apj, 767, 101 
\bibitem[Bechtol et al.(2015)]{Bechtol2015} Bechtol, K., Drlica-Wagner, A., Balbinot, E., et al. 2015, \apj, 807, 50
\bibitem[Belokurov et al.(2007)]{Belokurov2007} Belokurov, V., Zucker, D.~B., Evans, N.~W., et al.\ 2007, \apj, 654, 897
\bibitem[Belokurov et al.(2009)]{Belokurov2009} Belokurov, V., Walker, M.~G., Evans, N.~W., et al.\ 2009, \mnras, 397, 1748
\bibitem[Belokurov et al.(2014)]{Belokurov2014} Belokurov, V., Irwin, M.~J., Koposov, S.~E., et al. 2014, \mnras, 441, 2124
\bibitem[Boch \& Fernique(2014)]{2014ASPC..485..277B} Boch, T., \& Fernique, P.\ 2014, Astronomical Data Analysis Software and Systems XXIII, 485, 277 
\bibitem[Bonnarel et al.(2000)]{2000A&AS..143...33B} Bonnarel, F., Fernique, P., Bienaym{\'e}, O., et al.\ 2000, \aaps, 143, 33 
\bibitem[Bressan et al.(2012)]{Bressan2012} Bressan, A., Marigo, P., Girardi, L., et al. 2012, \mnras, 427, 127
\bibitem[Carraro(2009)]{Carraro2009} Carraro, G.\ 2009, \aj, 137, 3809
\bibitem[Cassisi et al.(2008)]{Cassisi2008} Cassisi, S., Salaris, M., Pietrinferni, A., et al.\ 2008, \apjl, 672, L115 
\bibitem[Chabrier(2001)]{Chabrier2001} Chabrier, G.\ 2001, \apj, 554, 1274
\bibitem[Chambers et al.(2016)]{2016arXiv161205560C} Chambers, K.~C., Magnier, E.~A., Metcalfe, N., et al.\ 2016, arXiv:1612.05560 
\bibitem[\protect\citeauthoryear{Conn et al.}{2018a}]{Conn2018a} Conn B.~C., Jerjen H., Kim D., Schirmer M., 2018, ApJ, 852, 68 
\bibitem[\protect\citeauthoryear{Conn et al.}{2018b}]{Conn2018b} Conn B.~C., Jerjen H., Kim D., Schirmer M., 2018, ApJ, 857, 70 
\bibitem[Crnojevi{\'c} et al.(2016)]{2016ApJ...824L..14C} Crnojevi{\'c}, D., Sand, D.~J., Zaritsky, D., et al.\ 2016, \apjl, 824, L14
\bibitem[\protect\citeauthoryear{Cutri et al.}{2013}]{Cutri2013} Cutri R.~M., et al., 2013, yCat, 2328,  
\bibitem[de Grijs \& Bono(2015)]{deGrijs2015} de Grijs, R., \& Bono, G.\ 2015, \aj, 149, 179
\bibitem[Deason et al.(2015)]{2015MNRAS.453.3568D} Deason, A.~J., Wetzel, A.~R., Garrison-Kimmel, S., \& Belokurov, V. 2015, \mnras, 453, 3568 
\bibitem[Dotter et al.(2008)]{2008ApJS..178...89D} Dotter, A., Chaboyer, B., Jevremovi{\'c}, D., et al. 2008, \apjs, 178, 89-101
\bibitem[Dolphin(2000)]{2000PASP..112.1383D} Dolphin, A.~E. 2000, \pasp, 112, 1383
\bibitem[Dotter et al.(2008)]{Dartmouth} Dotter, A., Chaboyer, B., Jevremovi{\'c}, D., et al. 2008, \apjs, 178, 89 
\bibitem[Drlica-Wagner et al.(2015)]{Drlica-Wagner2015} Drlica-Wagner, A., Bechtol, K., Rykoff, E.~S., et al. 2015, \apj, 813, 109
\bibitem[Erben et al.(2005)]{Erben2005} Erben, T., Schirmer, M., Dietrich, J.~P., et al.\ 2005, Astronomische Nachrichten, 326, 432 
\bibitem[Fadely et al.(2011)]{Fadely2011} Fadely, R., Willman, B., Geha, M., et al.\ 2011, \aj, 142, 88 
\bibitem[Foreman-Mackey et al.(2013)]{2013PASP..125..306F} Foreman-Mackey, D., Hogg, D.~W., Lang, D., \& Goodman, J. 2013, \pasp, 125, 306 
\bibitem[Fraternali et al.(2009)]{Fraternali2009} Fraternali, F., Tolstoy, E., Irwin, M.~J., \& Cole, A.~A.\ 2009, \aap, 499, 121
\bibitem[Frayn \& Gilmore(2002)]{Frayn2002} Frayn, C.~M., \& Gilmore, G.~F.\ 2002, \mnras, 337, 445
\bibitem[Frebel, Simon \& Kirby(2014)]{Frebel2014} Frebel, A., Simon, J.~D., \& Kirby, E.~N.\ 2014, \apj, 786, 74
\bibitem[Fritz et al.(2018a)]{Fritz2018} Fritz, T.~K., Battaglia, G., Pawlowski, M.~S., et al.\ 2018, arXiv:1805.00908 
\bibitem[Fritz et al.(2018b)]{Fritz2018b} Fritz, T.~K., Carrera, R., Battaglia, G.,\ 2018, arXiv:1805.07350 
\bibitem[Geha et al.(2009)]{Geha2009} Geha, M., Willman, B., Simon, J.~D., et al.\ 2009, \apj, 692, 1464 
\bibitem[Grillmair \& Dionatos(2006)]{Grillmair2006} Grillmair, C.~J., \& Dionatos, O.\ 2006, \apjl, 641, L37 
\bibitem[Harris 1996 (2010 version)]{Harris1996} Harris, W.~E.\ 1996, \aj, 112, 1487 
\bibitem[Henden et al.(2015)]{2015AAS...22533616H} Henden, A.~A., Levine, S., Terrell, D., \& Welch, D.~L. 2015, American Astronomical Society Meeting Abstracts, 225, 336.16 
\bibitem[Jethwa et al.(2016)]{2016MNRAS.461.2212J} Jethwa, P., Erkal, D., \& Belokurov, V. 2016, \mnras, 461, 2212
\bibitem[Kim \& Jerjen(2015b)]{KimJerjen2015b} Kim, D., \& Jerjen, H. 2015, \apjl, 808, L39
\bibitem[Kim \& Jerjen(2015a)]{KimJerjen2015a} Kim, D., \& Jerjen, H.\ 2015, \apj, 799, 73 
\bibitem[Kim et al.(2015a)]{Kim2} Kim, D., Jerjen, H., Milone, A.~P., Mackey, D., \& Da Costa, G.~S. 2015, \apj, 803, 63 
\bibitem[Kim et al.(2015b)]{Kim2015b} Kim, D., Jerjen, H., Mackey, D., Da Costa, G.~S., \& Milone, A.~P.\ 2015, \apjl, 804, L44
\bibitem[Kim et al.(2016a)]{Kim2016} Kim, D., Jerjen, H., Mackey, D., Da Costa, G.~S., \& Milone, A.~P.\ 2016, \apj, 820, 119  
\bibitem[Kim et al.(2016b)]{Kim2016b} Kim, D., Jerjen, Geha, M., Chiti, A., Milone, A.P., H., Da Costa, G.~S., Mackey, D., Frebel, A.\& Conn, B.\ 2016, \apj, 833, 16  
\bibitem[King(1966)]{1966AJ.....71...64K} King, I.~R.\ 1966, \aj, 71, 64
\bibitem[Kirby et al.(2013)]{Kirby2013} Kirby, E.~N., Cohen, J.~G., Guhathakurta, P., et al.\ 2013, \apj, 779, 102
\bibitem[Kirby et al.(2015)]{Kirby2015} Kirby, E.~N., Simon, J.~D., \& Cohen, J.~G.\ 2015, \apj, 810, 56 
\bibitem[Klypin et al.(1999)]{Klypin1999} Klypin, A., Kravtsov, A.~V., Valenzuela, O., \& Prada, F.\ 1999, \apj, 522, 82
\bibitem[Koch \& Rich(2014)]{Koch2014} Koch, A., \& Rich, R.~M.\ 2014, \apj, 794, 89 
\bibitem[Koposov et al.(2007)]{Koposov2007} Koposov, S., de Jong, J.~T.~A., Belokurov, V., et al.\ 2007, \apj, 669, 337
\bibitem[Koposov et al.(2015a)]{Koposov2015a} Koposov, S.~E., Belokurov, V., Zucker, D.~B., et al. 2015a, \mnras, 446, 3110
\bibitem[Koposov et al.(2015b)]{Koposov2015b} Koposov, S.~E., Casey, A.~R., Belokurov, V., et al. 2015b, \apj, 811, 62 
\bibitem[Koposov et al.(2015)]{Koposov2015} Koposov, S.~E., Belokurov, V., Torrealba, G., \& Wyn Evans, N. 2015, \apj, 805, 130
\bibitem[Koposov et al.(2017)]{Koposov2017} Koposov, S.~E., Belokurov, V., \& Torrealba, G.\ 2017, arXiv:1702.01122 
\bibitem[Koposov et al.(2018)]{Koposov2018} Koposov, S.~E., Walker, M.~G., Belokurov, V., et al.\ 2018, \mnras
\bibitem[Kroupa(2001)]{Kroupa2001} Kroupa, P.\ 2001, \mnras, 322, 231
\bibitem[Kroupa(2002)]{Kroupa2002} Kroupa, P.\ 2002, Science, 295, 82 
\bibitem[Laevens et al.(2014)]{Laevens2014} Laevens, B.~P.~M., Martin, N.~F., Sesar, B., et al. 2014, \apjl, 786, L3 
\bibitem[Laevens et al.(2015a)]{Laevens2015a} Laevens, B.~P.~M., Martin, N.~F., Ibata, R.~A., et al. 2015a, \apj, 802, L18
\bibitem[Laevens et al.(2015b)]{Laevens2015b} Laevens, B.~P.~M., Martin, N.~F., Bernard, E.~J., et al.\ 2015b, \apj, 813, 44
\bibitem[Lavery(1990)]{Lavery1990} Lavery, R.~J.\ 1990, \iaucirc, 5139, 2 
\bibitem[Longeard et al.(2018)]{Longeard18} Longeard, N., Martin, N., Starkenburg, E., et al.\ 2018, \mnras, 480, 2609 

\bibitem[Luque et al.(2016)]{Luque2016} Luque, E., Queiroz, A., Santiago, B., et al. 2016, \mnras, 458, 603
\bibitem[Marigo et al.(2017)]{Marigo2017} Marigo et al. 2017, \apj, 835, 77
\bibitem[Martin et al.(2008)]{Martin2008} Martin, N.~F., de Jong, J.~T.~A., \& Rix, H.-W. 2008, \apj, 684, 1075
\bibitem[Martin et al.(2013a)]{Martin2013a} Martin, N.~F., Slater, C.~T., Schlafly, E.~F., et al. 2013, \apj, 772, 15 
\bibitem[Martin et al.(2013b)]{Martin2013b} Martin, N.~F., Schlafly, E.~F., Slater, C.~T., et al. 2013, \apj, 779, L10 
\bibitem[Martin et al.(2015)]{Martin2015} Martin, N.~F., Nidever, D.~L., Besla, G., et al.\ 2015, \apjl, 804, L5 
\bibitem[Martin et al.(2016a)]{Martin2016a} Martin, N.~F., Ibata, R.~A., Collins, M.~L.~M., et al.\ 2016, \apj, 818, 40 
\bibitem[Martin et al.(2016b)]{Martin2016b} Martin, N.~F., Jungbluth, V., Nidever, D.~L., et al. 2016, ApJL, 830, L10 
\bibitem[Massey(1951)]{Massey1951} Massey, F.~J. 1951, Journal of the American Statistical Association, 46, 253
\bibitem[Milone et al.(2010)]{Milone2010} Milone, A.~P., Piotto, G., King, I.~R., et al.\ 2010, \apj, 709, 1183 
\bibitem[Mu{\~n}oz et al.(2012)]{Munoz2012} Mu{\~n}oz, R.~R., Geha, M., C{\^o}t{\'e}, P., et al.\ 2012, \apjl, 753, L15
\bibitem[Mutlu-Pakdil et al.(2018)]{MutluPakdil2018} Mutlu-Pakdil, B., Sand, D.~J., Carlin, J.~L., et al.\ 2018, arXiv:1804.08627 
\bibitem[Myeong et al.(2017)]{Myeong2017} Myeong, G.~C., Jerjen, H., Mackey, D., \& Da Costa, G.~S.\ 2017, \apjl, 840, L25 
\bibitem[Navarrete et al.(2017)]{Navarrete2017} Navarrete, C., Belokurov, V., \& Koposov, S.~E.\ 2017, \apjl, 841, L23
\bibitem[Navarro et al.(1997)]{NFW1997} Navarro, J.~F., Frenk, C.~S., \& White, S.~D.~M.\ 1997, \apj, 490, 493
\bibitem[Nidever et al.(2010)]{2010ApJ...723.1618N} Nidever, D.~L., Majewski, S.~R., Butler Burton, W., \& Nigra, L. 2010, \apj, 723, 1618
\bibitem[Odenkirchen et al.(2003)]{Odenkirchen2003} Odenkirchen, M., Grebel, E.~K., Dehnen, W., et al.\ 2003, \aj, 126, 2385 
\bibitem[Paust et al.(2014)]{Paust2014} Paust, N., Wilson, D., \& van Belle, G.\ 2014, \aj, 148, 19 
\bibitem[Pieres et al.(2017)]{Pieres2017} Pieres, A., Santiago, B.~X., Drlica-Wagner, A., et al.\ 2017, \mnras, 468, 1349
\bibitem[\protect\citeauthoryear{Piotto et al.}{2012}]{Piotto2012} Piotto G., et al., 2012, ApJ, 760, 39 
\bibitem[Sales et al.(2011)]{2011MNRAS.418..648S} Sales, L.~V., Navarro, J.~F., Cooper, A.~P., et al. 2011, \mnras, 418, 648
\bibitem[Sales et al.(2017)]{2017MNRAS.465.1879S} Sales, L.~V., Navarro, J.~F., Kallivayalil, N., \& Frenk, C.~S. 2017, \mnras, 465, 1879
\bibitem[Schirmer(2013)]{2013ApJS..209...21S} Schirmer, M. 2013, \apjs, 209, 21 
\bibitem[Schlafly \& Finkbeiner(2011)]{Schlafly2011} Schlafly, E.~F., \& Finkbeiner, D.~P.\ 2011, \apj, 737, 103 
\bibitem[Schlegel et al.(1998)]{SFD1998} Schlegel, D.~J., Finkbeiner, D.~P., \& Davis, M.\ 1998, \apj, 500, 525
\bibitem[Sharma et al.(2011)]{Sharma2011} Sharma, S., Bland-Hawthorn, J., Johnston, K.~V., \& Binney, J.\ 2011, \apj, 730, 3
\bibitem[Simon(2018)]{Simon2018} Simon, J.~D.\ 2018, arXiv:1804.10230 
\bibitem[Taylor(2005)]{2005ASPC..347...29T} Taylor, M.~B. 2005, Astronomical Data Analysis Software and Systems XIV, 347, 29
\bibitem[Rockosi et al.(2002)]{Rockosi2002} Rockosi, C.~M., Odenkirchen, M., Grebel, E.~K., et al.\ 2002, \aj, 124, 349 
\bibitem[Torrealba et al.(2016a)]{Torrealba2016a} Torrealba, G., Koposov, S.~E., Belokurov, V., \& Irwin, M.\ 2016, \mnras, 459, 2370 
\bibitem[Torrealba et al.(2016b)]{Torrealba2016b} Torrealba, G., Koposov, S.~E., Belokurov, V., et al.\ 2016, \mnras, 463, 712 
\bibitem[Torrealba et al.(2018)]{Torrealba2018} Torrealba, G., Belokurov, V., Koposov, S.~E., et al.\ 2018, \mnras, 475, 5085
\bibitem[Walker et al.(2015)]{Walker2015} Walker, M.~G., Mateo, M., Olszewski, E.~W., et al.\ 2015, \apj, 808, 108
\bibitem[Wright et al.(2010)]{Wright2010} Wright, E.~L., Eisenhardt, P.~R.~M., Mainzer, A.~K., et al.\ 2010, \aj, 140, 1868-1881
\end{thebibliography}
\end{document}